\documentclass[12pt,psf,epsf]{JHEP3}
\usepackage[centertags]{amsmath}
\usepackage{amssymb}
\usepackage{graphicx}
\usepackage{epsfig}

\title{Weak Field Black Hole Formation \\
in Asymptotically $AdS$ Spacetimes}

\author{Sayantani Bhattacharyya$^a$\footnote{sayanta@theory.tifr.res.in} and  
\ Shiraz Minwalla$^a$\footnote{minwalla@theory.tifr.res.in}, \\
\small{\emph{$^{a}$Department of Theoretical Physics,Tata Institute of Fundamental Research,}} \\
\small{\emph{Homi Bhabha Rd, Mumbai 400005, India}} \\ 
}

\abstract{We use the AdS/CFT correspondence to study the 
thermalization of a strongly coupled 
conformal field theory that is forced out of its vacuum
by a source that couples to a marginal operator. The source is taken to be 
of  small amplitude and finite duration, but is otherwise
 an arbitrary function of time. When the field theory lives on 
$R^{d-1,1}$, the source sets up a translationally invariant wave in the dual 
gravitational description. This wave propagates radially 
inwards in $AdS_{d+1}$ space and collapses to form a black brane. 
Outside its horizon the bulk spacetime for this collapse process 
may systematically be constructed in an expansion in the 
amplitude of the source function, and takes the Vaidya form 
at leading order in the source amplitude. This solution is dual to a 
remarkably rapid and intriguingly scale dependent thermalization process 
in the field theory. When the field theory lives on a sphere the resultant wave either slowly scatters into a thermal gas (dual to a glueball type phase in the boundary theory) or 
rapidly collapses into a black hole (dual to a plasma type phase in the 
field theory) depending on the time scale and amplitude of the source 
function. The transition between these two behaviors is sharp and 
can be tuned to the Choptuik scaling solution in $R^{d,1}$.}

\preprint{TIFR/TH/09-10}

\begin{document}

\section{Introduction}\label{intro}

The AdS/CFT correspondence identifies asymptotically $AdS$ 
gravitational dynamics with the master field evolution of 
`large $N$' field 
theories. In particular, it relates the evolution of spacetimes with 
horizons to the non equilibrium statistical dynamics of the high temperature 
phase of the dual field theory. This connection has recently been studied 
in detail in a near equilibrium limit. It has been established that 
the spacetimes that locally (i.e. tube wise) approximate the 
black brane metric
obey the equations of boundary fluid dynamics with gravitationally determined
dissipative constants. \footnote{See 
\cite{Bhattacharyya:2008jc, Baier:2007ix, 
VanRaamsdonk:2008fp, Loganayagam:2008is, Bhattacharyya:2008xc, Dutta:2008gf, 
Bhattacharyya:2008ji, Haack:2008cp, Bhattacharyya:2008mz, Erdmenger:2008rm,  Fouxon:2008tb,  Bhattacharyya:2008kq, 
Haack:2008xx, Gupta:2008th, Hansen:2008tq, Fouxon:2008ik, 
Kanitscheider:2009as, David:2009np, Springer:2009wj, Torabian:2009qk} 
for a recent 
structural understanding of this
 connection. See the reviews \cite{Son:2007vk, Heller:2008fg, 
Ambrosetti:2008mt}for references to important earlier work. See also 
\cite{Aharony:2005bm, Lahiri:2007ae, Cheung:2007af, Bhattacharyya:2007vs, 
Evslin:2008py, Bhardwaj:2008if, Maeda:2008kj, Caldarelli:2008mv, 
Caldarelli:2008ze, Cardoso:2009bv, Bhattacharya:2009gm} 
for related line of development.} The equations of fluid dynamics are thus 
embedded in a long distance sector of asymptotically $AdS$ gravity, a 
fascinating connection that promises to prove useful in many ways.

Given the success in using gravitational physics to study near 
equilibrium field theory dynamics, it is natural to attempt to use gravitational 
dynamics to study far from equilibrium field theory processes. In this paper 
we will study the gravitational dual of the process of 
equilibration; i.e the dynamical passage of a system from a pure state in its `low temperature' phase to an approximately thermalized state in 
its high temperature phase (see \cite{Ross:1992ba, Peleg:1994wx, 
Danielsson:1998wt, Danielsson:1999zt, Giddings:1999zu, Horowitz:1999jd, Danielsson:1999fa, Balasubramanian:2000rt, Birmingham:2001dt, Giddings:2001ii, 
Alberghi:2003ce, Alberghi:2003pr, Gao:2005yq, Lin:2008rw} for closely 
related earlier work and \cite{Festuccia:2005pi, Festuccia:2006sa, 
Iizuka:2008hg, Iizuka:2008eb} for analyses of thermalization 
directly in large $N$ gauge theories). 
As has been remarked by several authors, this process is dual to 
the gravitational process of black hole formation via gravitational collapse. 
The dynamical process is fascinating in its own right, but gains additional 
interest in asymptotically $AdS$ spaces because of its link to field 
theory equilibration dynamics. In this paper we study asymptotically $AdS$ 
(and briefly asymptotically 
flat) collapse processes in a weak field limit that displays rich dynamics  
while allowing for analytic control. 

An $AdS$ collapse process that could result in black hole formation
may be set up, following Yaffe and Chesler \cite{Chesler:2008hg}, 
as follows . Consider an asymptotically locally $AdS$ spacetime, 
and let ${\cal R}$ denote a finite patch of the conformal boundary 
of this spacetime. We choose our spacetime 
to be exactly $AdS$ outside the causal future of ${\cal R}$.  On ${\cal R}$ 
we turn on the non normalizable part of a massless bulk field. 
This boundary condition sets up an ingoing shell of the corresponding 
field that collapses in $AdS$ space. Under appropriate conditions  
the subsequent dynamics can result in black hole formation.

\begin{figure}[here]
\centering
\includegraphics[scale=1.0]{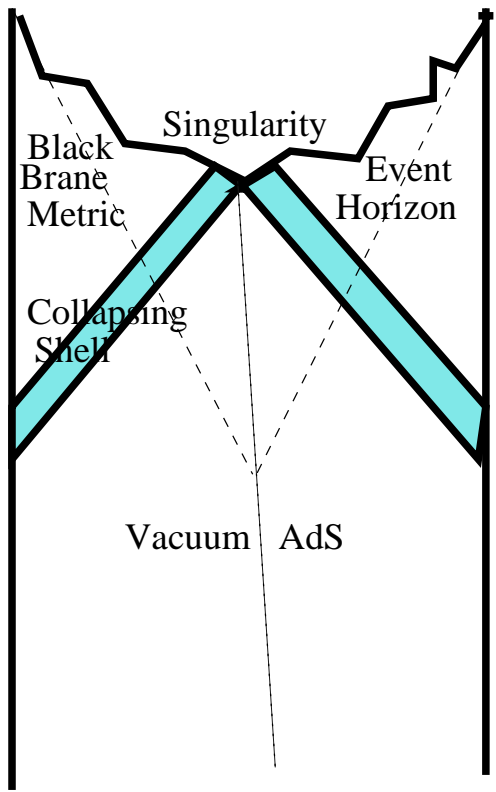}
\caption{Cross section of the causal diagram for the collapse process in an asymptotically global $AdS_{d+1}$ space. The conventional Penrose diagram for this process would include only the half of the diagram that to the right of its vertical axis of symmetry}
\label{penrose}
\end{figure}

In this paper we will study the $AdS$ collapse scenario (plus a flat space 
counterpart) outlined in the previous paragraph in a weak field limit; 
i.e. we always choose the amplitude $\epsilon$ of the non normalizable 
perturbation to be small. In the interest of simplicity we also focus on 
situations that preserve a great deal of symmetry, as we explain further 
below. In the rest of this introduction we describe the three classes of 
collapse situations we study, and the principal results of our analysis.
\footnote{In most of 
the bulk of the text of this paper we only present formulae 
for asymptotically $AdS_{d+1}$ spacetimes for the smallest nontrivial 
value of $d$ namely $d=3$. As we explain in the Appendix \ref{arb}
however, most of the qualitative results of our analysis apply to arbitrary 
odd $d$ for $d \geq 3$ and also plausibly to arbitrary even $d$ for $d\geq 4$.}

\subsection{Translationally invariant asymptotically $AdS_{d+1}$ collapse}
\label{introtrans}

In the first part of this paper we analyze spacetimes 
that asymptote to Poincare patch $AdS_{d+1}$ space and 
turn on non normalizable modes on the boundary \footnote{See 
\cite{Das:2006dz, Das:2006pw, Awad:2007fj,  Awad:2008jf} for other 
work on Poincare patch $AdS$ solutions forced by time dependent non 
normalizable data}. We choose our non normalizable data to depend 
on the boundary time but to be independent of boundary spatial 
coordinates. Moreover, our data has support only in the time interval 
$v\in (0, \delta t)$, i.e. our forcing functions are turned on only over 
a limited time interval.  Our boundary conditions   
create a translationally invariant wave of small amplitude $\epsilon$ 
near the boundary of $AdS$. This wave then propagates into the bulk of 
$AdS$ space.

In section \ref{transdil} and Appendices \ref{grav} and \ref{arbdim} 
of this paper we demonstrate that this 
wave ${\it always}$ results in black 
brane formation at small amplitude (see figure \ref{penrose} for the Penrose
diagram of the analogous process in an asymptotically global $AdS$ space). 
Outside the event horizon, 
this black brane formation process is reliably described by a perturbation 
expansion in the amplitude. At leading order in perturbation theory   
the spacetime set up by this wave takes the Vaidya form\footnote{The Vaidya 
metric is an exact solution for the propagation of a null dust - a fluid 
whose stress tensor is proportional to $\rho k_\mu k_\nu$ for a lightlike 
vector $k_\mu$ ($k_\mu=\partial_r$ in \eqref{lo}). Note that $\rho k_\mu k_\nu$ is also 
the stress tensor of a massless field in the eikonol or geometric optics 
approximation.  The AdS-Vaidya
metric has been studied before in the context of the $AdS/CFT$ correspondence
in, for instance, \cite{Hubeny:2006yu}. } (\cite{Vaidyaa, Vaidyab, Vaidyac}, see 
e.g.\cite{2007gcss.book.....J} for a review) 
\begin{equation}\label{lo} 
ds^2=2 dr dv -\left(r^2-\frac{M(v)}{r^{d-2}}\right)dv^2+r^2 dx_i^2 .
\end{equation}
This form of the metric is exact for all $r$ when $v<0$, and is 
a good approximation to the metric for 
$r \gg \frac{\epsilon^{\frac{2}{d-1}}}{\delta t}$
when $v>0$. Our perturbative procedure determines the function $M(v)$ in 
\eqref{lo} in terms of the non normalizable data at the boundary; 
$M(v)$ turns out to be of order $\frac{\epsilon^2}{(\delta t)^d}$
\footnote{More precisely, let $\phi_0(v)=\epsilon ~\chi(\frac{v}{\delta t})$ 
where $\chi$ is a function that is defined on $(0, 1)$. Then the energy of the
resultant black brane is $\frac{\epsilon^2}{(\delta t)^d} \times A[\chi]$ 
where $A[\chi]$ is a functional of $\chi(x)$ that is computed 
later in this paper.}.
 $M(v)$ reduces to constant $M$ for $v >\delta t$ in odd $d$  and 
asymptotes to that value (like a power in $\frac{\delta t}{v}$)  
in even $d$ \footnote{$M(v)$ is defined as the coefficient of the 
$\frac{dv^2}{r^{d-2}}$ term in the metric, in an expansion around small $r$. 
In even $d$ this turns out not to be equal to the mass density of the 
system, the coefficient of the same term in the metric when expanded
around large $r$.} In either case the spacetime \eqref{lo} describes 
the process of 
formation of a black brane 
of temperature $T \sim \frac{\epsilon^{\frac{2}{d}}}{\delta t}$ over the 
time scale of order $\delta t$. Note that the time scale of formation of the 
brane is much smaller than its inverse temperature. This fact 
allows us to compute the event horizon of the spacetime \eqref{lo} 
in a simple and explicit fashion in a power series in 
$ \delta t T \sim \epsilon^{\frac{2}{d}}$. To 
leading order in $\epsilon$ the event horizon  manifold is given by 
\begin{equation}\label{eventhor} \begin{split}
r_H(v)&=M^\frac{1}{d}~~~v>0 \\
r_H(v)&= \frac{M^{\frac{1}{d}}}{1-M^\frac{1}{d} v}~~~v<0\\
\end{split}
\end{equation} 
All of the spacetime outside the event horizon \eqref{eventhor} 
lies within the domain of validity of our perturbative procedure.
Of course perturbation theory does not accurately describe the process of 
singularity formation of the black brane. However the region where 
perturbation theory breaks down (and so \eqref{lo} is not reliable) 
is contained entirely within the event horizon of \eqref{lo}.
Consequently, the region outside perturbative control is 
causally disconnected from physics outside the event horizon, so 
our perturbation procedure gives a fully reliable description 
of the dynamics outside the event horizon. It follows in particular that 
any singularities that develop in our solution  are is always shielded by a 
regular event horizon, in agreement with the cosmic censorship conjecture.
 \footnote{We thank M. Rangamani for discussions on this point.} 

In section \ref{transdil} we demonstrate that the corrections to the 
Vaidya metric \eqref{lo} may be systematically computed in a power series 
in positive fractional powers of $\epsilon$. At any order 
in the perturbation expansion, the metric  may be 
determined analytically for times $v \ll T^{-1}$ ($T$ is the temperature of the eventually 
formed brane). At times of order or larger than $T^{-1}$, perturbative 
corrections to the metric are determined in terms solutions of universal(i.e. independent of the form of the  perturbation) linear 
differential equations  which we have only been able 
to solve numerically. Even at late times, however our perturbative 
procedure analytically determines the dependence  of observables
on the functional form of the non normalizable perturbation, allowing us  to 
draw conclusions that are valid for small amplitude perturbation of arbitrary form.

Let us now word our results in dual field theoretic terms. 
Our gravity solution describes a CFT initially in its 
vacuum state. Over the time period $(0, \delta t)$ the field theory is 
perturbed by a translationally invariant time dependent source, 
of amplitude $\epsilon$, that couples to a marginal operator. This 
coupling pumps energy into this system. Our perturbative 
gravitational solution gives a detailed description of the 
subsequent equilibration process; in particular it gives a 
precise formula for the temperature of the final equilibrium 
configuration as a function of the perturbation function. It also, 
very surprisingly, asserts that for some purposes\footnote{In particular, in even bulk space time dimensions, one point functions of all local operators reduce to their thermal values as soon as the perturbation is switched off. While thermalization of one point functions is not instantaneous in odd bulk space times it appears to take place over a time scale of order  $\delta t$.}
our system appears to thermalize almost instanteneously at leading order 
in $\epsilon$. We pause to explain this in detail. 
\footnote{The rest of this 
subsection was worked out in collaboration with O. Aharony, B. Kol and 
S. Raju. See also the paper \cite{Lin:2008rw}, by Lin and Shuryak, for a very similar earlier discussion. We thank E. Shuryak for bringing this paper to our attention.}

A field theorist presented with a flow towards equilibrium might choose to 
probe this flow by perturbing it with an infinitesimal source, localized 
at some time. He would then measure the subsequent change in the solution in response to 
this perturbation. However note that  the spacetime 
in \eqref{lo} is identical to the spacetime outside a static uniform black brane 
 for $v>\delta t$ when $d$ is odd 
(and for $v\gg \delta t$ for even $d$).  It follows that the response of 
our system to any boundary perturbation localized at times 
$v > \delta t$ in odd $d$ (and at $v \gg \delta t$ 
in even $d$) will be identical to the response of a thermally equilibrated
system to the same perturbation. In other words our system responds 
to perturbations at $v>\delta t$ as if it had equilibrated instanteneously. 

A field theorist could also characterize a flow towards equilibrium by 
recording the values of all observables as a function of time (in the absence 
of any further perturbation). The full set of observables consists of 
expectation values of the arbitrary product of `gauge invariant' operators,
i.e. quantities that in a gauge theory would take the form

$$\langle Tr O_1 Tr O_2 \ldots  Tr O_n \rangle.$$
In this paper we work in the strict large $N$ limit (i.e. the 
strictly classical limit from the dual bulk viewpoint). In this limit 
trace factorization (or the classical nature of the dual bulk theory) ensures
that the expectation value of products equals the product of expectation values.
 In other words our set of 
observables is given precisely by the one point functions of all gauge 
invariant operators.  

Now note that expectation values of all {\it local} boundary 
operators are determined by the bulk solution in a neighborhood 
of the boundary values. As the metric \eqref{lo} is identical to the metric 
of a uniform black brane in the neighbourhood of the boundary when
$v>\delta t$, it follows that the expectation value of all {\it local} boundary 
operators reduce instantaneously to their thermal values in odd $d$ (and when 
$v\gg \delta t$ in even $d$). Consequently, all local operators appear to 
thermalize instanteneously.

Not all gauge invariant operators are local, however. A field theorist could 
also record the values of non local observables, like circular Wilson Loops 
of radius $a$, as a function of time.  As nonlocal observables 
probe the spacetime away from the boundary, their expectation values reduce 
to thermal results only after a larger  time that depends on the 
size of the loop (this time is proportional to $a$ at small $a$). So 
a diligent infinite $N$ field theorist would be able to distinguish \eqref{lo}
from absolute thermal equilibrium at times greater than $\delta t$, but 
only by keeping track of the expectation values of non local observables. 

If one were to retreat away from the large $N$ limit one would find 
large new classes of gauge invariant observables; the connected correlators 
of, for instance, local gauge invariant operators. Such correlators also 
sample spacetime away from the boundary, the distance scale of this nonlocal 
sampling being set by the separation between the operator insertions 
(see \cite{Hubeny:2006yu} for a detailed discussion of properties of 
correlation functions in asymptotically $AdS$ Vaidya type metrics).
 As in our discussion of Wilson loops above, the time scale for thermalization 
of such connected correlators is set by their separation 
(it is proportional to their separation when this separation is small).

As we have seen, the time scale of equlibration of the solutions described
in this paper depend on the precise question you ask about it. 
We would now like to describe a concrete and possibly 
practically important experimental sense in which our system behaves as if 
it were instanteneously thermally equilibrated.  

Consider the response of a CFT in its vacuum to a forcing function 
that varies - though only slowly - with ${\vec x}$. We anticipate that at $v=\delta t$ the 
corresponding spacetime is locally (tube wise) well described by a 
black brane metric with a value of the temprature that varies with ${\vec x}$ 
(see \eqref{derexp} and the discusson arounf it in section \ref{disc}). According to 
\cite{Bhattacharyya:2008jc, Baier:2007ix, VanRaamsdonk:2008fp, 
Loganayagam:2008is, Bhattacharyya:2008xc, Dutta:2008gf, 
Bhattacharyya:2008ji, Haack:2008cp, Bhattacharyya:2008mz, Erdmenger:2008rm,  Fouxon:2008tb,  Bhattacharyya:2008kq, 
Haack:2008xx, Gupta:2008th, Hansen:2008tq, Fouxon:2008ik, 
Kanitscheider:2009as, David:2009np, Springer:2009wj, Torabian:2009qk}, 
the subsequent evolution 
of our system is governed by the equations of boundary fluid dynamics. 
The initial conditions for the relevant fluid flow are given at $v=\delta t$. 
Consequently an experimentalist who observes the subsequent fluid flow, and 
back calculates, would conclude that his system was thermalized at 
$v=\delta t$.

The thought experiment of the previous paragraph is reminiscent of 
situation at the RHIC experiment. The back calculation described in 
this paragraph, in the context of that experiment, suggests that the 
RHIC system is governed by fluid dynamics at times of order 0.5 fermi 
after the collision, much faster than suggested by naive estimates for thermalization time
(see \cite{Heinz:2009xj} and references therein). 
It is natural to wonder whether the mechanisims for rapid equilibration 
of this paper have qualitative applicability to the RHIC experiment. 
We leave a serious investigation of this question to future work.

In summary, \eqref{lo} describes a system whose response to additional 
external perturbations at $v>\delta t$ is identical to that of a thermally 
equilibrated system and whose one point functions of local 
operators also instanteneously thermalize. However expectation values 
of non local observables (or 
correlators) thermalize more slowly, over a time scale that depends on the 
smearing size of the observable (or correlator).  We find instantenous 
thermalization of expectation values local operators and 
the scale dependence in the process of equilibration 
fascinating. In fact this discussion is reminiscent of precursors in the AdS/CFT correspondence \cite{Susskind:1998vk, Polchinski:1999yd, Susskind:1999ey, 
Hubeny:2002dg}.

We emphasize that our discussion of thermalization applies only at leading order in $\epsilon$ expansion. Indeed our analysis was based on \eqref{lo} which accurately describes our spacetime only at leading order in $\epsilon$. At sub leading orders 
\eqref{lo} is corrected by perturbations that decay 
to the black brane result only over the time scale  $1/T$, in accordance with naive expextation. Consequently, the instantaneous thermalization of expectation values of local operators is corrected by sub leading equilibration  process that take place over 
the time scale $1/T$, the thermalization of 
linear fluctuations about a brane of temperature $T$. Note, in particular, 
that we have no reason so suspect that thermalization occurs over a time 
period that is faster than the naive estimate $v=\frac{1}{T}$ when 
$\epsilon$ is of order unity or larger.

\subsection{Spherically symmetric collapse in flat space}\label{introflat}

We next turn to the perturbative study of spherically symmetric collapse 
in an asymptotically flat space. Consider a spherically symmetric 
shell, propagating inwards, focused onto the origin of an asymptotically 
flat space. Such a shell may qualitatively be characterized by its thickness
and mass, or (more usefully for our purposes) by the Schwarzschild radius 
$r_H$ associated with this mass. It is a well appreciated fact that 
this collapse process may reliably be described in an amplitude 
expansion when  $y\equiv\frac{r_H}{\delta t}$ is very small. 
The starting point 
for this expansion is the propagation of a free scalar shell. This free 
motion receives weak scattering corrections at small $y$, which may be 
computed perturbatively. 

In section \ref{flat} of this paper we demonstrate that this flat space 
collapse process may also be reliably described in an amplitude expansion 
at ${\it large}$ $y$. In section \ref{flat} and Appendix \ref{arbflat} 
we study this collapse process mainly in odd $d$ (i.e. in even bulk spacetime 
dimensions). The starting point for this expansion is a Vaidya 
metric similar to \eqref{lo}, whose event horizon we are able to reliably 
compute in a power series expansion in inverse powers of $y$.
Outside this event horizon the dilaton is everywhere small and 
the Vaidya metric receives only weak 
scattering corrections that it may systematically be computed in a power 
series in $\frac{1}{y}$ at large $y$. As in the previous subsection, our 
perturbative procedure is not valid everywhere; however the breakdown 
of perturbation theory occurs entirely within the event horizon, and so 
does not impinge on our control of the solution outside the event horizon. 
 
At early times we are able to determine the perturbative corrections to the metric (order 
by order in  $\frac{1}{y}$) in an entirely analytic manner. However 
late time corrections to the metric are computed in terms 
of the solutions to relevant universal linear differential equations, which we have 
not been able to solve analytically. However our perturbative solutions 
carry a considerable amount of information, even in the absence of an explicit 
analytic solution to the relevant differential equation.  As an example, 
in section \ref{flat} we determine the fraction of energy of the incident 
pulse that is radiated back out to infinity to 
nontrivial leading order in the expansion in $\frac{1}{y}$. We are able to 
analytically determine the dependence of this fraction on the shape of 
the incident pulse upto an overall constant (see \eqref{eninpulse}). 
The determination of the value of this constant requires knowledge 
of the explicit solution of the `universal' differential equation listed in 
section \ref{flat}, and may presumably be determined numerically.

An order parameter (the 
presence of an event horizon at late times) distinguishes small $y$ from 
large $y$ behavior, so the transition between them must be sharp.
This observation was originally made about twenty years ago 
in classic paper by Christodoulou (see \cite{Christodoulou:2008nj}) 
and references therein) who rigorously demonstrated 
that collapse at arbitrarily 
large $y$ results in black hole formation, while collapse at small $y$ 
does not. As the fascinating transition between small and large $y$ 
behaviors (which has been extensively in a programme of numerical 
relativity initiated by  Choptuik
\cite{Choptuik:1992jv}) \footnote{See \cite{Gundlach:1999cu, Gundlach:2007gc}
for reviews and \cite{Birmingham:2001hc, Pretorius:2000yu, 
AlvarezGaume:2006dw, AlvarezGaume:2007rg, 
AlvarezGaume:2008qs, AlvarezGaume:2008fx} for recent work interpreting 
this transition in the context of the AdS/CFT correspondence.}) presumably 
occurs at $y$ of order unity. Consequently it cannot be studied in either 
the small $y$ or the large $y$ expansions described in our paper. 

As we do not have a holographic description of gravitational dynamics 
in an asymptotically flat space, we are unable to give a direct dual 
field theoretic interpretation of our results reviewed in this 
subsection. See however, the next subsection.

\subsection{Spherically symmetric collapse in asymptotically global $AdS$}
\label{introglobal}

The process of spherically symmetric collapse in an asymptotically 
global $AdS$ space constitutes an interesting one parameter 
interpolation between the collapse processes described in 
subsections \ref{introtrans} and \ref{introflat}. We study 
such collapse processes in section \ref{global} of this paper. 
In section \ref{global} we have studied this collapse situation in 
detail only in $d=3$. In this subsection we report the 
generalization of these results to arbitrary odd dimension, which may 
qualitatively be inferred from the results of Appendix \ref{arb}.

Consider a global $AdS$ space, whose boundary is taken to be a sphere of 
radius $R$ $\times$ time. Consider a collapse process initiated 
by radially symmetric non normalizable boundary conditions that are 
turned on, uniformly over the boundary sphere, over a time interval $\delta t$. The amplitude $\epsilon$ of this perturbation together with the 
dimensionless  ratio $x\equiv\frac{\delta t}{R}$, constitute the two 
qualitatively important parameters of this perturbation.  
In the limit $x \to 0$ it is obvious that the collapse process of this 
subsection effectively reduces to the Poincare patch collapse process 
described in 
subsection \ref{introtrans}, and results in the formation of a black hole 
that is large compared to the $AdS$ radius (and so locally well approximates
a flat black brane); quantitatively this turns out to work for 
$x \ll \epsilon^\frac{2}{d}$. When $x \gg \epsilon^{\frac{2}{d}}$ 
the most interesting part of the collapse process takes place in a bubble 
of approximately flat space. In this case the solution closely resembles 
a wave propagating in $AdS$ space at large $r$,  glued onto a flat space 
collapse process described in subsection 
\ref{introflat}. \footnote{This statement is only correct at times 
$v \ll \frac{1}{R}$. To see why recall that when a collapsing shell in 
flat space forms a black hole, some of its energy is radiated out to 
${\cal I^+}$. The resolution of the infalling  shell
 into a static black hole and plus a shell radiated
out to infinity occurs over a time scale set by $r_H$ the Schwarszshild radius 
associated with the infalling matter.  In $AdS$ space this shell eventually 
reflects off the boundary of $AdS$ space at times of order $\frac{1}{R}$ 
(note that this is a much larger length scale than $r_H$ 
when the black hole is small enough) and then is refocussed on the origin of 
space. This process repeats itself unendingly; eventaully 
all of the energy of the intitial shell is absorbed by the black hole. 
Consequently $AdS$ collapse processes always differ significantly 
from their flat space counterparts for $v\ll \frac{1}{R}$. In particular, 
while such a process can result in the formation of arbitrarily small mass 
black holes over time scale $\frac{1}{R}$, the mass of black holes created at 
long times is bounded from below (see below for an estimate). We thank 
V. Hubeny for a discussion on this point.} 
Following through the details of the gluing process, it turns 
out that the inverse of the effective flat space $y$ parameter 
(see subsection \ref{introflat}) is given by 
$\frac{x^\frac{2d-2}{d-2} }{\epsilon^\frac{2}{d-2}}$. The parameter
$y$ is of order unity when $x \sim \epsilon^{\frac{1}{d-1}} $. We conclude 
that the end point of the global $AdS$ collapse process is a black hole 
for $x \ll \epsilon^{\frac{1}{d-1}} $ but a scattering dilaton wave 
for $x \gg \epsilon^{\frac{1}{d-1}}$. 

The minimum mass of black holes formed through this process is 
$\frac{\epsilon^\frac{d-2}{d-1}}{R}$ (we work in units in which the mass of 
the black hole is simply the long time value of the parameter $M$ in 
\eqref{los}, the global analogue of \eqref{lo}).
Let us contrast this with 
the minimum mass of black holes that we expect to be produced when we 
pump energy into the more slowly (i.e. through a forcing function whose time 
variation is of order $\frac{1}{R}$) but over a long time period. As we have
described above, slow forcing deposits energy into the gravitational thermal
gas. By continually forcing the system one creates a thermal gas of increasing
energy. At a critical energy density of order $\frac{1}{R}$, however, density
fluctuations in this thermal gas become unstable \cite{Page:1985em}; the end point of this 
instability is believed to be a black hole. Clearly this slow pumping in 
of energy produces black holes of energy $\frac{1}{R}$ or greater. It follows that 
black hole production can be produced more efficiently (i.e. at lower energies)
via rapid forcing than via a slow pumping in of energy into the system.

As we have explained above, when $\epsilon \ll 1$ and when 
$x \ll \epsilon^{\frac{1}{d-1}} $, we are able to reliably establish 
black hole formation within perturbation theory (see figure \ref{penrose}
for a Penrose diagram of this process). As in the previous 
two subsections, the starting point of the perturbative expansion always 
turns out to be a metric of the Vaidya form, whose event horizon we are able
to reliably compute. Our metric receives only small scattering 
corrections outside the event horizon. Although the perturbative procedure 
breaks down badly near the black hole singularity, that is irrelevant 
for the physics outside the event horizon.

On the other hand, when $x \gg \epsilon^{\frac{1}{d-1}} $ (but at small 
$\epsilon$), the incident waves simply scatter through the origin, and 
subsequently undergo periodic motion 
in $AdS$ space. This free motion is corrected by interaction effects that 
will eventually cause this dilaton pulse to deviate significantly 
from its free motion over a time scale that we expect to scale like a 
positive power of $\frac{x^\frac{2d-2}{d-2} }{\epsilon^\frac{2}{d-2}}$ times 
the inverse radius of the sphere \footnote{We expect this pulse to 
thermalize over an even longer time scale, one that scales as a 
positive power of the larger number, $\frac{x^d}{\epsilon^2}$.}.

\begin{figure}[here]
\centering
\includegraphics[scale=1.0]{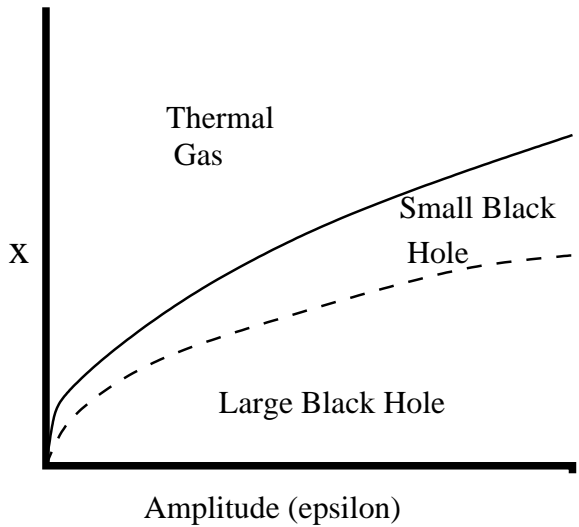}
\caption{The `Phase Diagram' for our dynamical stirring in global 
$AdS$. The final outcome is a large black hole for 
$x \ll \epsilon^\frac{2}{d}$ (below the dashed curve), a small black hole 
for $x\ll \epsilon^\frac{1}{d-1}$ (between the solid and dashed curve) 
and a thermal gas for $x \gg  \epsilon^\frac{1}{d-1}$. The solid curve 
represents non analytic behaviour (a phase transition) while the dashed 
curve is a crossover.}
\label{phased}
\end{figure}

Let us now reword our results in field theory terms. Any CFT that admits a 
two derivative gravity dual description undergoes a first order 
finite temperature phase transition when studied on $S^{d-1}$. The low 
temperature phase is a gas of 'glueballs' (dual to gravitons) while the 
high temperature phase is a strongly interacting, dissipative,  `plasma' 
(dual to the black hole). The gravitational solutions of this paper 
describe such a CFT on a sphere, initially in its vacuum state. We then 
excite the CFT over a time $\delta t$ by turning on a spherically symmetric 
source function that couples to a marginal operator. The most important 
qualitative question about the subsequent equilibration process is: in which 
phase does the system eventually settle down within classical dynamics (i.e. 
ignoring tunneling effects) ? Our gravitational solutions 
predict that the system settles in its free particle phase 
when $x \gg \epsilon^{\frac{1}{d-1}}$ but in the plasma phase
when $x \ll  \epsilon^{\frac{1}{d-1}}$. As in subsection \ref{introtrans} 
the equilibration in the high temperature phase is almost instantaneous. 
However equilibration in the low temperature phase appears to occur 
over a much longer time scale. We note also that the transition between 
these two end points appears to be singular (this is the Choptuik 
singularity) in the large $N$ limit. \footnote{See section \ref{disc} 
for a discussion of the effects of potential Gregory-Laflamme type 
instabilities near this singular surface.} This singularity is presumably 
smoothed out by fluctuations at finite $N$, a phenomenon that should 
be dual to the smoothing out of a naked gravitational singularity 
by quantum gravity fluctuations. 

In the rest of this paper and in the appendices 
we will present a detailed study of the collapse scenarios outlined in this 
introduction. In the last section of this paper we also present a discussion 
of our results.

\section{Translationally invariant collapse in $AdS$}\label{transdil}

In this section we study asymptotically planar (Poincare patch) $AdS_{d+1}$ 
solutions to negative cosmological constant Einstein gravity interacting 
with a minimally coupled massless scalar field (note that this system 
obeys the null energy condition). We focus on solutions 
in which the boundary value of the scalar field takes a given functional 
form $\phi_0(v)$ in the interval $(0,\delta t)$ but vanishes otherwise.
The amplitude of $\phi_0(v)$ (which we schematically refer to as $\epsilon$ 
below) will be taken to be small in most of this section. The boundary dual to our setup is a $d$ dimensional conformal field theory on $R^{d-1,1}$, perturbed 
by  a spatially homogeneous and isotropic source function, 
$\phi_0(v)$, multiplying a marginal scalar operator.

Note that our boundary conditions preserve an $R^{d-1} \times SO(d-1)$ 
symmetry (the $R^{d-1}$ factor is boundary spatial translations while 
the $SO(d-1)$ is boundary spatial rotations). In this section we 
study solutions on which $R^{d-1} \rtimes SO(d-1)$ lifts to an isometry of the 
full bulk spacetime.  In other words the spacetimes studied in this section 
preserve the maximal symmetry allowed
by our boundary conditions. As a consequence all 
bulk fields in our problem are functions of only two variables;  a 
radial coordinate $r$ and an Eddington Finkelstein ingoing time coordinate 
$v$. The chief results of this section are as follows:

\begin{itemize}
\item The boundary conditions described above result in black brane 
formation for an arbitrary (small amplitude) source functions $\phi_0(v)$. 
\item Outside the event horizon of our spacetime, we find an explicit 
analytic form for the metric as a function of $\phi_0(v)$. Our metric 
is accurate at leading order in the $\epsilon$ expansion, and takes 
the Vaidya form \eqref{lo} with a mass function that we determine 
explicitly as a function of time.  
\item  In particular, we find that the energy density of the resultant 
black brane is given, to leading order, by 
\begin{equation}\label{ctg}
C_2=\frac{2^{d-1}}{(d-1)} 
\left( \frac{(\frac{d-1}{2})!}{(d-1)!} \right)^2 
\int_{-\infty}^\infty \left(
\left( \partial_t^{\frac{d+1}{2}}\phi_0(t)
\right)^2 \right) 
\end{equation}
in odd $d$ and by 
\begin{equation}\label{evenmass}
C_2=-\frac{d^2}{(d-1) 2^d} \frac{1}{ \left(\frac{d}{2}!\right)^2 }
\int dt_1 dt_2 \partial_{t_1}^{\frac{d+2}{2}}\phi_0(t_1) 
 \ln (t_1-t_2) \theta(t_1-t_2) 
\partial_{t_2}^{\frac{d+2}{2}} \phi_0(t_2)
\end{equation}
in even $d$. Note that, in each case, $C_2\sim \frac{\epsilon^2}{(\delta t)^d}$.
\item We find an explicit expression for the event horizon of the resultant 
solutions, at leading order, and thereby demonstrate that singularities 
formed in the process of black brane formation are always shielded by a 
regular event horizon at small $\epsilon$.

\item  Perturbation theory in the amplitude $\epsilon$ 
yields systematic corrections to this leading order  metric.
We unravel the structure of this perturbation expansion in detail and 
compute the first corrections to the leading order result. 

\end{itemize}

While every two derivative theory of gravity that admits an $AdS$ solutions 
admits a consistent truncation to Einstein gravity with a negative cosmological 
constant, the same statement is clearly not true of gravity 
coupled to a minimally coupled massless scalar field. It is consequently 
of considerable interest to note that results closely analogous to those 
described above also apply to the study of Einstein gravity with a 
negative cosmological constant. In Appendix \ref{grav} we analyze the 
process of black brane formation by gravitational wave collapse in the 
theory of pure gravity (similar to the set up of \cite{Chesler:2008hg}), and 
find results that are qualitatively very similar 
to those reported in this section.
 The solutions of Appendix \ref{grav} yield the 
dual description of a class of thermalization processes in every 3 dimensional 
conformal field theory that admits a dual description as a two derivative 
theory of gravity. In fact, the close similarity of the results of 
Appendix \ref{grav} with those of this section, lead us to believe that the 
results reported in this section are qualitatively robust. In particular 
we think it is very likely that results of this section will qualitatively 
apply to the most general small amplitude translationally invariant 
collapse process in the systems we study.

\subsection{The set up}\label{transdilsetup}

Consider a minimally coupled massless scalar 
(the `dilaton') interacting with negative cosmological constant Einstein 
gravity in $d+1$ spacetime dimensions 
\begin{equation}\label{lagr}
S=\int d^{d+1}x \sqrt{g} \left( R -\frac{d (d-1)}{2} - 
\frac{1}{2} (\partial \phi)^2 \right) 
\end{equation}
The equations of motion that follow from the Lagrangian \eqref{lagr}
are \begin{equation}\label{eom} \begin{split}
E_{\mu\nu} & \equiv G_{\mu\nu}-\frac{1}{2} \partial_\mu \phi \partial_\nu \phi + g_{\mu \nu}\left(-\frac{d(d-1)}{2} 
+ \frac{1}{4} (\partial \phi)^2 \right)=0 \\
& \nabla^2 \phi=0 \\
\end{split}
\end{equation}
where the indices $\mu, \nu$ range over all $d+1$ spacetime coordinates.
As mentioned above, in this section we are interested in 
locally asymptotically $AdS_{d+1}$ solutions to these equations that preserve 
an $R^{d-1} \times SO(d-1)$ symmetry group. This symmetry requirement forces 
the boundary metric to be Weyl flat (i.e. Weyl equivalent to 
flat $R^{d-1,1}$); however it allows the boundary value of the 
scalar field to be an arbitrary function of boundary time $v$. 
We choose this function as 
\begin{equation} \label{bdilp} \begin{split}
\phi_0(v)&=0 ~~~(v<0)\\
\phi_0(v)&<\epsilon ~~~(0<v<\delta t)\\
\phi_0(v)&=0 ~~~(v> \delta t)
\end{split}
\end{equation}
(we also require that $\phi_0(v)$ and its first few derivatives 
are everywhere continuous.\footnote{We expect that all our 
main physical conclusions will continue to apply if we replace our 
$\phi_0$ - which is chosen to strictly vanish outside $(0, \delta t)$ - 
by any function that decays sufficiently rapidly outside this range.}).

Everywhere in this paper we adopt the 
 `Eddington Finkelstein' gauge $g_{rr}=g_{ri}=0$ and 
$g_{rv}=1$. In this gauge, and subject to our symmetry requirement, 
our spacetime takes the form 
\begin{equation} \label{metdil} \begin{split}
ds^2&= 2 dr dv -g(r,v) dv^2 +f^2(r,v) dx_i^2\\
\phi&=\phi(r,v).\\
\end{split}
\end{equation}

The mathematical problem we address in this subsection is 
to solve the equations of motion \eqref{eom} 
for the functions $\phi$, $f$ and $g$,  subject 
to the pure $AdS$ initial conditions 
\begin{equation}\label{inconp} \begin{split}
g(r,v)&=r^2 ~~~(v<0)\\
f(r,v)&=r ~~~(v<0)\\ 
\phi(r,v)&=0 ~~~(v<0)\\ 
\end{split}
\end{equation} 
and the large $r$ boundary conditions
\begin{equation} \label{bcsp} \begin{split}
&\lim_{r \to \infty} \frac{g(r, v)}{r^2}=1 \\
&\lim_{r \to \infty} \frac{f(r, v)}{r}= 1\\
&\lim_{r \to \infty} \phi(r,v)= \phi_0(v)
\end{split}
\end{equation} 
The Eddington Finkelstein gauge we adopt in this paper does not 
completely fix gauge redundancy (see \cite{Chesler:2008hg} for a related 
observation). The coordinate redefinition 
$r= {\tilde r} + h(v)$ respects both our gauge choice as well as 
our boundary conditions. In order to completely define the mathematical 
problem of this section, we must fix this ambiguity. We have assumed above that $f(r,v) = r + {\cal O}(1)$ at large $r$. It follows that under 
the unfixed diffeomorphism, $f(r,v) \rightarrow f(r,v) + h(v) + 
{\cal O}(1/r)$. Consequently we can fix this gauge redundancy by demanding 
that $f(r,v) \approx r+ {\cal O}(1/r)$ at large $r$. We make this 
choice in what follows. As we will see below, it then follows from the 
equations of motion that $g(r)=r^2+{\cal O}(1)$. Consequently, 
the boundary conditions \eqref{bcsp} on the fields $g$, $f$ and $\phi$, 
may be restated in more detail as 
\begin{equation} \label{bcsm} \begin{split}
&g(r, v)=r^2 \left(1+{\cal O}(\frac{1}{r^2}) \right) \\
&f(r, v)= r\left(1+{\cal O}(\frac{1}{r^2}) \right)\\
&\phi(r,v)= \phi_0(v) + {\cal O}(\frac{1}{r})
\end{split}
\end{equation} 
Equations \eqref{eom}, \eqref{metdil}, \eqref{inconp} and \eqref{bcsm}
together constitute a completely well defined dynamical system. 
Given a particular forcing function $\phi_0(v)$, these equations and 
boundary conditions uniquely determine the functions 
$\phi(r,v)$, $g(r,v)$ and $f(r,v)$.   

\subsection{Structure of the equations of motion}\label{transdileom}

The nonzero equations of motion \eqref{eom} consist of four nontrivial 
Einstein equations $E_{rr}$, $E_{r v}$, $E_{vv}$ and $\sum_i E_{ii}$ 
(where the index $i$ runs over the $d-1$ spatial directions) together with 
the dilaton equation of motion. For the considerations that follow below, 
we will find it convenient to study the following linear combinations of 
equations
\begin{equation}\label{eqnset} \begin{split}
E_c^1&=g^{v\mu} E_{\mu r}\\
E_c^2&=g^{v\mu} E_{\mu v}\\
E_{ec}&=g^{r \mu} E_{\mu r}\\
E_{d}&=\sum_{i=1}^{d-1} E_{ii}\\
E_\phi&= \nabla^2 \phi \\
\end{split}
\end{equation}
Note that the equations $E_c^1$ and $E_c^2$ are constraint equations from the 
point of view of $v$ evolution. 

It is possible to show that $E_d$ and $\frac {d  (r E_{ec})}{dr}$ both 
automatically vanish whenever $E_c^1 =E_c^2=E_\phi=0$. 
This implies that this last set of three independent equations - supplemented
by the condition that $r E_{ec}=0$ at any one value of $r$ - 
completely exhaust the dynamical content of \eqref{eom}. As a consequence, 
in the rest of this paper we will only bother to solve the two constraint 
equations and the dilaton equation, but take care to simultaneously ensure 
that $r E_{ec}=0$ at some value of $r$. It will often prove useful to 
impose the last equation at arbitrarily large $r$. 
This choice makes the physical content of $r E_{ec}=0$ manifest; this is 
simply the equation of energy conservation in our system. 
\footnote{It turns out that 
both $E_d$ and the dilaton equation of motion are automatically satisfied 
whenever $E_{ec}$ together with the two Einstein constraint equations are 
satisfied. Consequently $E_{ec}$ plus the two Einstein constraint equations 
form another set of independent equations. This choice of equations has 
the advantage that it  does not require the addition of 
 any additional condition analogous to energy conservation. 
However it turns out to 
be an inconvenient choice for implementing the $\epsilon$ expansion of this 
paper, and we do not adopt it in this paper.}
 
\subsubsection{Explicit form of the constraints and the dilaton equation}

With our choice of gauge and notation the 
dilaton equation takes the minimally coupled form
\begin{equation} \label{dileom}
\partial_r\left( f^{d-1} g \partial_r \phi \right) 
+ \partial_v\left( f^{d-1}  \partial_r \phi \right)
+\partial_r\left( f^{d-1} \partial_v \phi \right)
=0
\end{equation}
Appropriate linear combinations of the two constraint equations take 
the form 
\begin{equation} \label{maineqs}
\begin{split}
\left(\partial_r \phi\right)^2 &=-\frac{2(d-1) \partial^2_r f}{f} \\
& \partial_r \left( f^{d-2} g \partial_r f  + 2 f^{d-2} \partial_v f \right) 
 = f^{d-1} d \\
\end{split}
\end{equation}

Note that the equations \eqref{maineqs} (together with boundary conditions 
and the energy conservation equation) permit the unique determination 
of $f(r,v_0)$ and $g(r, v_0)$ in terms of $\phi(r, v_0)$ and 
${\dot \phi}(r,v_0)$ (where $v_0$ is any particular time). It follows
that $f$ and $g$ are not independent fields. A solution to the differential 
equation set \eqref{dileom} and \eqref{maineqs} is completely specified by 
the value of $\phi$ on a constant $v$ slice (note that the equations are 
all first order in time derivatives, so ${\dot \phi}$ on the slice 
is not part of the data of the problem) together with the boundary condition 
$\phi_0(v)$.

\subsection{Explicit form of the energy conservation equation} \label{encons}

In this section we give an explicit form for the equation $E_{ec}=0$ 
at large $r$. We specialize here to $d=3$ but see 
Appendix \ref{arbdim} for arbitrary $d$. Using the Graham Fefferman 
expansion to solve the equations of motion in a power series in $\frac{1}{r}$
we find 
\begin{equation}\label{larger}\begin{split}
f(r, v)&=r \left(1 -\frac{{\dot \phi _0}{}^2}{8 r^2}+ \frac{1}{r^4} 
\left( \frac{1}{384} 
({\dot \phi _0})^4-\frac{1}{8} L(v){\dot \phi_0}\right)  
+{\cal O}(\frac{1}{r^5}) \right) \\
g(r, v)& =r^2 \left(1 -\frac{3 ({\dot \phi_0})^2}{4 r^2}-\frac{M(v)}{r^3} 
+{\cal O}(\frac{1}{r^4})\right)\\
\phi(r,v)& =\phi_0(v)+\frac{\dot \phi_0}{r}+\frac{L(v)}{r^3} 
+{\cal O}(\frac{1}{r^4})\\
\end{split}
\end{equation}
where the functions $M(v)$ and $L(v)$ are undetermined functions of 
time that are, however,  constrained by the energy conservation equation 
$E_{ec}$, which takes the explicit form  
\begin{equation}\label{encon}
{\dot M}= {\dot \phi _0} \left(\frac{3}{8} ({\dot \phi _0})^3- 
\frac{3 L}{2}
-\frac{1}{2} {\dddot\phi _0}.
\right) 
\end{equation}
In all the equations in this subsection and in the rest of the paper, 
the symbol ${\dot P}$ denotes the derivative of $P$ with respect to 
our time coordinate $v$.  
Solving for $M(v)$ we have
\begin{equation}\label{enin}
M(v)= \frac{1}{2}\int_0^v dt \left( \left({\ddot\phi_0}\right)^2 + 
\frac{3}{4}\left({\dot \phi_0}\right)^4 -3 {\dot \phi_0} L(t) \right)
\end{equation}

\footnote{We note parenthetically that \eqref{encon} may be rewritten as 
\begin{equation}\label{feq}
{\dot T^0_0}= \frac{1}{2}{\dot \phi_0} {\cal L}
\end{equation}
where the value ${\cal L}$ of the 
operator dual to the scalar field $\phi$ and the stress tensor 
$T_{\alpha \beta}$ are given by  
\begin{equation}\label{lagst} \begin{split}
{\cal L} &\equiv \lim_{r \to \infty}  r^3 \left( 
\partial_n \phi + \partial^2 \phi \right)  \\
T^{\mu}_\nu&= \lim_{r \to \infty} r^3 \left( K^\mu_\nu- (K -2)\delta^\mu_\nu -{\cal G}^\mu_\nu +\frac{
\partial^\mu \phi \partial_\nu \phi}{2} -\frac{ (\partial \phi)^2 \delta^\mu_\nu}{4} \right).\\
\end{split}
\end{equation}
Where
\begin{equation}
\begin{split}
K^\mu_\nu &= \text{Extrinsic curvature of the constant}~r~\text{surfaces},~~~K = K^\mu_\mu\\
{\cal G}^\mu_\nu &= \text{Einstein tensor evaluated on the induced metric of the constant}~r~ \text{surfaces}
\end{split}
\end{equation}
yielding 
\begin{equation}\label{ansstl} \begin{split}
T^0_0&=-2T^x_x= -2 T^y_y= M(v)\\
{\cal L}&= \frac{3}{4} {\dot \phi}_0^3 - 3 L(v)-\partial_v^3 \phi _0
\end{split}
\end{equation}}

\subsection{The metric and event horizon at leading order}
\label{event}

Later in this section we will solve the equations of motion 
\eqref{dileom}, \eqref{maineqs} and \eqref{encon} in an expansion in powers 
of $\epsilon$, the amplitude of the forcing function $\phi_0(v)$. In this 
subsection we simply state our result for the spacetime metric 
at leading order in $\epsilon$. We then proceed to compute 
the event horizon of our spacetime to leading order in $\epsilon$. We 
present the computation of the event horizon of our spacetime before actually 
justifying the computation of the spacetime itself for the following reason. 
In the subsections below we will aim to construct the spacetime that describes
black hole formation only outside the event horizon. For this reason we will 
find it useful below to have a prior understanding of 
the location of the event horizon in the 
spacetimes that emerge out of perturbation theory.

We will show below that to leading order in $\epsilon$, 
our spacetime metric takes the Vaidya form \eqref{lo}. The mass function 
$M(v)$ that enters this Vaidya metric is also determined very simply. 
As we will show below, it turns out that $L(v) \sim {\cal O}(\epsilon^3)$ 
on our perturbative solution. It follows immediately from \eqref{enin} 
that the mass function $M(v)$ that enters the Vaidya metric, is given to 
leading order by 
\begin{equation}\label{eninp} \begin{split}
M(v)& =C_2(v)+ {\cal O}(\epsilon^4) \\
C_2(v)&= -\frac{1}{2}\int_{-\infty}^v dt {\dot \phi_0(t)} {\dddot \phi_0}(t) 
\end{split}
\end{equation}
(Here $C_2$ is the approximation to the mass density, valid to second order
in the amplitude expansion, see below). 

Note that, for $v>\delta t$, $C_2(v)$ reduces to a constant $M=C_2$ given by   
\begin{equation}\label{eninpc} 
C_2= \frac{1}{2}\int_{-\infty}^\infty dt \left({\ddot \phi_0(t)} \right)^2 
\sim \frac{\epsilon^2}{(\delta t)^3}
\end{equation}

In the rest this subsection we proceed to compute the 
event horizon of the leading order spacetime \eqref{lo} in an expansion 
in $\epsilon^\frac{2}{3}$ 
expansion. Let the event horizon manifold of our spacetime be given by 
the surface $S \equiv r-r_H(v)=0$. As the event horizon is a null manifold, 
it follows that $\partial_\mu S \partial_\nu S g^{\mu\nu}=0$, and we find  
\begin{equation} \label{eh}
\frac{dr_H(v)}{dv} = \frac{r_H^2(v)}{2}\left(1-\frac{M(v)}{r_H^3(v)}\right)
\end{equation}  
As $M(v)$ reduces to the constant $M=C_2$ for $v >\delta t$, it follows 
that the event horizon must reduce to the surface $r_H=M^{\frac{1}{3}}$
at late times. It is then easy to solve \eqref{eh} for $v<0$ and $v>\delta t$; 
we find 
\begin{equation}\label{ehs} \begin{split}
r_H(v)&=M^{\frac{1}{3}},  ~~~v \geq \delta t \\
r_H(v)&=M^{\frac{1}{3}} x(\frac{v}{\delta t}),  ~~~0<v<\delta t\\
\frac{1}{r_H(v)}&=-v + \frac{1}{M^{\frac{1}{3}}x(0)}, ~~~v \leq 0\\
\end{split}
\end{equation}
where $x(y)$ obeys the differential equation 
\begin{equation} \label{ehe} \begin{split}
\frac{dx}{dy}& = \alpha \frac{x^2}{2}\left(1-\frac{M(y \delta t)}{M x^3}\right)\\
\alpha&=M^{\frac{1}{3}}\delta t \sim \epsilon^\frac{2}{3}\\
\end{split}
\end{equation} 
and must be solved subject to the final state conditions $x=1$ for $y= 1$. 
\eqref{ehe} is easily solved in a perturbation series in $\alpha$. We set
\begin{equation}\label{psa}
x(y)= 1 +\sum_n \alpha^n x_n(y)
\end{equation}
and solve recursively for $x_n(t)$. To second order we find\footnote{In this 
section we only construct the event horizon for the Vaidya metric. The actual 
metrics of interest to this paper receive corrections away from 
the Vaidya form, in powers of $M\delta t$. Consequently, the event horizons 
for the actual metrics determined in this paper will agree with those of this 
subsection only at leading order in $M \delta t$. The determination
of the event horizon of the Vaidya metric at higher orders in $M \delta t$,
is an academic exercise that we solve in this subsection largely because it 
illustrates the procedure one could adopt on the full metric.} 
 \begin{equation}\label{xo} \begin{split}
x_1(y)&=-\int_y^1 dz~ \left( \frac{1-\frac{M(z \delta t)}{M}}{2} \right) \\
x_2(y)&=-\int_y^1 dz ~x_1(z)\left(1 +\frac{M(z \delta t)}{2 M}\right) 
\end{split}
\end{equation}
In terms of which  
\begin{equation}\label{rh}
r_H(v)=M^\frac{1}{d}\left( 1+\alpha ~x_1(\frac{v}{\delta t})
+\alpha^2 x_2(\frac{v}{\delta t}) + {\cal O}(\alpha^3) \right) ~~~
(0<v<\delta t)
\end{equation} 

Note in particular that, to leading order, $r_H(v)$ is simply given by the 
constant $M^\frac{1}{3}$ for all $v>0$.

\subsection{Formal structure of the expansion in amplitudes}
\label{transdilampexp}

In this subsection we will solve the equations \eqref{dileom}, \eqref{maineqs}
and \eqref{encon} in a perturbative expansion in the amplitude of the source 
function $\phi_0(v)$. In order to achieve this we formally replace $\phi_o(v)$ 
with $\epsilon \phi_0(v)$ and solve all equations in a power series expansion 
in $\epsilon$. At the end of this procedure we can set the formal 
parameter $\epsilon$ to unity. In other words $\epsilon$ is a formal parameter
that keeps track of the homogeneity of $\phi_0$. Our perturbative expansion 
is really justified by the fact that the amplitude of $\phi_0$ is small.

In order to proceed with our perturbative procedure, we set 
\begin{equation}\label{ampexp} \begin{split}
f(r,v)&=\sum_{n=0}^\infty \epsilon^{n} f_{n}(r,v) \\
g(r,v)&= \sum_{n=0}^\infty  \epsilon^{n} g_{n}(r, v)  \\
\phi(r,v)&= \sum_{n=0}^\infty \epsilon^{n} \phi_{n}(r,v) \\
\end{split}
\end{equation}
with 
\begin{equation}\label{zerordp}
f_0(r,v)=r, ~~~~ g_0(r,v)=r^2,~~~ \phi_0(r,v)=0.
\end{equation}
We then plug these expansions into the equations of motion, 
expand these equations in a power series in $\epsilon$, and
proceed to solve these equations recursively, order by order in 
$\epsilon$.  

The formal structure of this procedure is familiar. The coefficient of
$\epsilon^n$ in the equations of motion take the schematic form 
\begin{equation}\label{formpt}
H^i_j\chi^j_n(r, v)) = s_n^i
\end{equation} 
Here $\chi^i_N$ stands for the three dimensional `vector' of
$n^{th}$ order unknowns, i.e.  $\chi^1_n= f_n$, $\chi^2_n=g_n$ and 
$\chi^3_n=\phi_n$. The differential operator $H^i_j$ is universal (in the 
sense that it is the same at all $n$) and has a simple interpretation; it is 
simply the operator that describes linearized fluctuations about $AdS$ space. 
The source functions $s_n^i$ are linear combinations of products of 
$\chi^i_m$ $(m<n)$ ; the sum over $m$ over fields that 
appear in any particular term adds up to $n$.  

The equations \eqref{formpt} are to be solved subject to the large 
$r$ boundary conditions 
\begin{equation}\label{bcs} \begin{split}
\lim_{r \to \infty} \phi_1(r, v)&=\phi_0(r) \\  
\phi_{n}(r,v) &\leq {\cal O}(1/r), ~~~ n \geq 2 \\  
f_{n}(r,v) & \leq {\cal O}(1/r), ~~~ n \geq 1\\ 
g_{n}(r,v) & \leq {\cal O}(r),~~~ n \geq 1\\
\end{split}
\end{equation}
together with the initial conditions 
\begin{equation} \label{inconpp} 
\phi_{n}(r,v)=g_{n}(r,v) =f_{n}(r,v)=0~~~ {\rm for } ~~v<0~~(n \geq 1)
\end{equation} 
These boundary and initial conditions uniquely determine $\phi_{n}$, 
$g_{n}$ and $f_{n}$ in terms of the source functions. 

All sources vanish at first order 
in perturbation theory (i.e the functions 
$ s^i_1$ are zero). Consequently, the functions $f_1$ and $g_1$ vanish 
but $\phi_1$ is forced by its boundary condition to be nonzero. As we will 
see below, it is easy to explicitly solve for the function $\phi_1$. 
This solution, in turn, completely 
determines the source functions at  ${\cal O}(\epsilon^2)$ and so the 
equations \eqref{formpt} unambiguously determine 
$g_2$, $\phi_2$  and $f_2$. This story repeats recursively. 
The solution to perturbation theory at order $n-1$ determine the source 
functions at order $n$ and so permits the determination of the unknown 
functions at order $n$. The final answer, at every order, 
is uniquely determined in terms of $\phi_0(v)$.

To end this subsection, we note a simplifying aspect of our perturbation 
theory. It follows from the structure of the equations that  $\phi_n$ is nonzero only when $n$ is odd while $f_m$ and $g_m$ are nonzero only when $m$ is even.
We will use this fact extensively below.

\subsection{Explicit results for naive perturbation theory to fifth 
order }\label{expl}

We have implemented the naive perturbative procedure described above to 
${\cal O}(\epsilon^5)$. Before proceeding to a more structural discussion 
of the nature of the perturbative expansion, we pause here to record our 
explicit results. 

At leading (first and second) order we find
\begin{equation}\label{ampdilsol}
 \begin{split}
  \phi_1(r,v) &= \phi_0(v) + \frac{{\dot \phi_0}}{r}\\
f_2(r,v)&=-\frac{{\dot \phi_0}^2}{8 r}\\
g_2(r,v) &= -\frac{C_2(v)}{r}-\frac{3}{4}{\dot \phi_0}^2\\
\end{split}
\end{equation}
At the next order 
\begin{equation}\label{ampdilsolt} \begin{split}
\phi_3(r,v)&=\frac{1}{4 r^3}\int^v_{-\infty} B(x)~dx
\\
f_4(r,v)&=\frac{{\dot \phi_0}}{384 r^3}\left\{{\dot \phi_0}^3
-12\int^v_{-\infty} B(x)~dx \right\}\\
g_4(r,v) &= \frac{C_4(v)}{r} + \frac{{\dot \phi_0}}{24 r^2}\left\{
- {\dot \phi_0}^3 + 3\int^v_{-\infty} B(x)~dx \right\}\\
&+\frac{1}{48 r^3} \left( 3 B(v) {\dot \phi_0} - 4 {\dot \phi_0}^3 
{\ddot \phi_0} + 3 {\ddot \phi_0}\int_v^\infty B(t) dt \right) 
\end{split}
\end{equation}
while $\phi_5$ is given by 
\begin{equation}\label{ampdilsolth} \begin{split}
\phi_5(r,v) &= \frac{1}{8 r^5}\int^v_{-\infty} B_1(x)~dx\\
&+ \frac{1}{6 r^4}\int^v_{-\infty} B_3(x)~dx
 + \frac{5}{24 r^4}\int^v_{-\infty}dy\int^y_{-\infty}B_1(x)~dx \\  
&+\frac{1}{4 r^3} \int^v_{-\infty} B_2(x)~dx  + \frac{1}{6 r^3} \int^v_{-\infty}dy\int^y_{-\infty}B_3(x)~dx\\
& + \frac{5}{24 r^3}\int^v_{-\infty}dz\int^z_{-\infty}dy\int^y_{-\infty}B_1(x)~dx
\end{split}
\end{equation}
In the equations above
 
\begin{equation}\label{defins} \begin{split}
B(v) &= {\dot \phi_0} \left[-C_2(v) + 
{\dot \phi_0}{\ddot \phi_0}\right]\\
B_1(v) &= \left(-\frac{9}{4}C_2(v) + 
\frac{7}{8}{\dot \phi_0}{\ddot \phi_0}\right)\int^v_{-\infty} B(x)~dx\\ 
&+\frac{1}{2}C_2(v) {\dot \phi_0}^3
+\frac{3}{8}  {\dot \phi_0}^2 B(v) 
- \frac{1}{6}{\dot \phi_0}^4 {\ddot \phi_0}\\
B_2(v)&=C_4(v){\dot \phi_0}\\
B_3(v)&=\frac{1}{24}\left(-30 {\dot \phi_0}^2\int^v_{-\infty} B(x)~dx 
+ 7{\dot \phi_0}^5\right)
\end{split}
\end{equation}
and the energy functions $C_2(v)$ and $C_4(v)$ (obtained by integrating 
the energy conservation equation) are given by 
\begin{equation}\label{dilcon}
 \begin{split}
 {C_2}(v) =& -\int_{-\infty}^v dt \frac{1}{2}{\dot \phi_0}{\dddot \phi_0}\\
{C_4}(v) =& 
\int_{-\infty}^v dt \frac{3}{8}{\dot \phi_0}\left(-{\dot \phi_0}^3 + 
\int^t_{-\infty} B(x)~dx\right)
 \end{split}
\end{equation}

For use below, we note in particular that at $v=\delta t$ the mass of the 
black brane is given by $C_2(\delta t) - C_4(\delta t) + {\cal O}(\epsilon^6)$
while the value of the dilaton field is given by 
\begin{equation}\label{inconsp} \begin{split}
\phi(r,\delta t)&=\frac{1}{4 r^3}\int^{\delta t}_{-\infty} B(x)~dx \\
&+\frac{1}{4 r^3} \int^{\delta t}_{-\infty} B_2(x)~dx + \frac{1}{6 r^3} \int^{\delta t}_{-\infty}dy\int^y_{-\infty}B_3(x)~dx\\
&+ \frac{5}{24 r^3}\int^{\delta t}_{-\infty}dz\int^z_{-\infty}dy\int^y_{-\infty}B_1(x)~dx \\
&+ \frac{5}{24 r^4}
\int^{\delta t}_{-\infty}dy\int^y_{-\infty}B_1(x)~dx  
+ \frac{1}{6 r^4}\int^{\delta t}_{-\infty} B_3(x)~dx \\
& +\frac{1}{8 r^5}\int^{\delta t}_{-\infty} B_1(x)~dx   ~~~+ {\cal O}(\epsilon^7)\\
\end{split}
\end{equation}

\subsection{The analytic structure of the naive perturbative expansion}\label{transdilas}

In this subsection we will explore the analytic structure of the 
naive perturbation expansion in the variables $v$ (for 
$v>\delta t$) and $r$. It is possible to inductively demonstrate that 
\begin{itemize} 
\item{1.} The functions $\phi_{2n+1}$, $g_{2n+2}$ and 
$f_{2n+2}$ have the following analytic structure in the variable $r$
\begin{equation}\label{arp}\begin{split}
\phi_{2n+1}(r, v)&=\sum_{k=0}^{2n-2} \frac{\phi_{2n+1}^k(v)}{r^{2n+1-k}}, ~~~
(n \geq 2)\\
f_{2n}(r,v)&= r \sum_{k=0}^{2n-6} \frac{f_{2n}^k(v)}{r^{2n-k}},~~~~(n \geq 3)\\
g_{2n}(r,v)&= \frac{C_{2n}(\delta t)}{r} + 
r \sum_{k=0}^{2n-5} \frac{g_{2n-3}^k(v)}{r^{2n-k}},~~~~(n \geq 3)\\
\end{split}
\end{equation}
Moreover, when $v>\delta t$  $\phi_1(r,v)=f_2(r,v)=f_4(r,v)=0$ 
while $g_2(r,v)=-\frac{C_2(\delta t)}{r}$ and 
$g_4(r,v)=\frac{C_4(\delta t)}{r}$.

\item{2.} The functions $\phi_{2n+1}^k(v)$, $f_{2n}^k(v)$ and 
$g_{2n}^k(v)$ 
are each functionals of $\phi_0(v)$ that scale like $\lambda^{-2n-1+k}$, 
$\lambda^{-2n+k}$ and $\lambda^{-2n +k-1}$ respectively under 
the scaling $v \rightarrow \lambda v$.
\item{3.} For $v>\delta t$ the functions 
$\phi_{2n+1}^k(v)$ are all polynomials in $v$ of a degree 
that grows with $n$. In particular the degree of 
$\phi_{2n+1}^k$ at most $n-1+k$; the degree of $f_{2n}^k$ is 
at most $n-3+k$ and the degree of $g_{2n}^k$ is at most $n-4+k$.  
\end{itemize}

The reader may easily verify that all these properties hold for the 
explicit low order solutions of the previous subsection. 

\subsection{Infrared divergences and their cure}

The fact that $\phi_{2n+1}(v)$ are polynomials in time whose degree grows 
with $n$ immediately implies that the naive perturbation theory of the 
previous subsection fails at late 
positive times. We pause to characterize this failure in more detail. 
As we have explained above, the field $\phi(r,v)$ schematically takes the form 
$$\sum_{n, k} \frac{\epsilon^{2n+1} \phi^k_{2n+1}}{r^{2n+1-k}}$$
where $\phi_{2n+1}^k \sim \frac{v^{n-1+k}}{(\delta t)^{3n}}$ 
at large times. Let us examine this sum in the vicinity 
$r \sim \frac{\epsilon^\frac{2}{3}}{\delta t}$, a surface that will turn  
out to be the event horizon of our solution.  The term with labels 
$n, k$ scales like $\epsilon \times (\epsilon^\frac{2}{3} \frac{v}{\delta t})
^{n-1+k}$. Now $\frac{\epsilon^{\frac{2}{3}}}{\delta t} =T$ is approximately 
the temperature of a black brane of event horizon $r_H$. We conclude that 
the term with labels $n, k$ scales like $(v T)^{n-1+k}$. It follows that, 
at least in  the vicinity of the horizon, the naive expansion for $\phi$ 
is dominated by the smallest values of $n$ and $k$ when 
$\delta t T  \ll 1$. On the other hand, at times large compared to the 
inverse temperature, this sum is dominated by the largest values 
of $k$ and $n$. 
As the sum over $n$ runs to infinity, it follows that naive perturbation 
theory breaks down at time scales of order $T^{-1}$.

A long time or  IR divergence in perturbation theory usually signals 
the fact that the perturbation expansion has been carried out about the 
wrong expansion point; i.e. the zero order `guess' with which 
we started perturbation (empty $AdS$ space) does not everywhere 
approximate the true solution even at arbitrarily small $\epsilon$. 
Recall that naive perturbation theory is perfectly satisfactory for times 
of order $\delta v$ so long as $r \gg \frac{\epsilon}{\delta t}$. 
Consequently this perturbation theory may be used to check if our spacetime 
metric deviates significantly from the pure $AdS$ in this range 
of $r$ and at these early times. The answer is that it does, even in the limit 
$\epsilon \to 0$. In order to
see precisely how this comes about, note that the most singular term in 
$g_{2n}$ is of order $r \times \frac{1}{r^{2n}}$ for $n \geq 1$, the 
exact value of $g_0=r^2= (r \times \frac{1}{r^{0}} \times r$). In other 
words $g_0$ happens to be less singular, near $r=0$, than one would expect 
from an extrapolation of the singularity structure of $g_n$ at finite 
$n$ down to $n=0$. As a consequence, even though $g_0$ is of lowest order 
in $\epsilon$, at small enough $r$ it is dominated by the most singular 
term in $g_2(r,v)$. Moreover this crossover in dominance occurs at 
$r\sim \frac{\epsilon^\frac{2}{3}}{\delta t} \gg \frac{\epsilon}{\delta t}$ 
and so occurs well within the domain of applicability of perturbation theory. 
In other words,  in the variable range 
$r \gg \frac{\epsilon}{\delta t}$, $g(r,v)$ is not uniformly well 
approximated  by $g_0=r^2$ at small $\epsilon$ but instead by   
$$ g(r,v) \approx r^2-\frac{C_2(v)}{r}.$$ 
This implies that, in the appropriate parameter 
range,  the true metric of the spacetime is everywhere well approximated 
by the Vaidya metric  \eqref{lo}, with $M(v)$ given by \eqref{eninp}
in the limit $\epsilon \to 0$. 

Of course even this corrected estimate for $g(r,v)$ breaks down at 
$r \sim \frac{\epsilon}{\delta t}$. However, as we have indicated above, 
this will turn out to be irrelevant for our purposes as our spacetime 
develops an event horizon at $r\sim \frac{\epsilon^{\frac{2}{3}}}{\delta t}$.

We will now proceed to argue that the metric is well approximated by 
the Vaidya form at all times (not just at early times) outside its event
horizon, so that the Vaidya metric \eqref{lo} rather than empty $AdS$ 
space, constitutes the correct starting point for the perturbative expansion
of our solution.

\subsection{The metric to leading order at all times}

The dilaton field and spacetime metric begin a new stage in their 
evolution at $v= \delta t$. At later times the solution is 
a normalizable, asymptotically $AdS$ solution to the equations
of motion. This late time motion is unforced and so is 
completely determined by two pieces 
of initial data; the mass density  $M(\delta t)$ and the dilaton function 
$\phi(r, \delta t)$. As the naive perturbation expansion described in
subsection \ref{transdilas} is valid at times of order $\delta t$, it 
determines both these quantities perturbatively in $\epsilon$. The explicit
results for these quantities, to first two nontrivial orders in $\epsilon$, 
are listed in \eqref{inconsp}. 

The leading order expression for the mass density is simply given by 
$C_2$ in \eqref{eninp}. Now if one could ignore $\phi(r, \delta t)$ (i.e. 
if this function were zero) this initial condition would define a  unique, 
simple subsequent solution to Einstein's equations; the uniform black brane with 
mass density $C_2$. While $\phi(r,\delta t)$ is not zero, 
we will now show it induces only 
a small perturbation about the black brane background. 

In order to see this it is useful to move to a rescaled variable 
${\tilde r}=\frac{r}{C_2^{\frac{1}{3}}}$. In terms of this rescaled variable, 
our solution at $v=\delta t$ is a black brane of unit energy density, 
perturbed by $\phi(r, \delta t)$. With this choice of variable the background
metric is independent of $\epsilon$, so that all $\epsilon$ dependence 
in our problem lies in the perturbation. It follows that, to leading order
in $\epsilon$ ( recall $\phi_1(r, \delta t)=0$)  
\begin{equation}\label{incondp}
\phi(r, \delta t)= \frac{\phi^0_3(\delta t)}{r^3}\left(1+{\cal O}
(\epsilon^{\frac{2}{3}}) \right)
= \frac{1}{{\tilde r}^3} \times \frac{\phi^0_3(\delta t)}{M}\left(1+{\cal O}
(\epsilon^{\frac{2}{3}}) \right)
\sim \frac{\epsilon}{{\tilde r}^3}
\end{equation}
where, from subsection \ref{expl} \begin{equation}\label{inconspsp} 
\phi^0_3(\delta t) =\frac{1}{4}\int^{\delta t}_{-\infty} B(x)~dx 
\end{equation}
  
The important point here is that the perturbation is proportional to 
$\epsilon$ and so represents a small deformation of the 
dilaton field about the unit energy density black brane initial 
condition. Moreover, any regular linearized perturbation about the 
black brane may be re expressed as a linear sum of quasinormal modes 
about the black brane and so decays exponentially over a time scale of 
order the inverse temperature.
It follows that the initialy small dilaton perturbation remains small at all 
future times and in fact decays exponentially to zero over a finite time.  
The fact that perturbations about the Vaidya metric 
\eqref{lo} are bounded both in amplitude as well as in temporal duration
allows us to conclude that the event horizon 
of the true spacetime is well approximated by the 
event horizon of the Vaidya metric at small $\epsilon$, as described in 
subsection \ref{event}.

\subsection{Resummed versus naive perturbation theory}

Let us define a resummed perturbation theory which uses the corrected 
metric \eqref{lo} (rather than the unperturbed $AdS$ metric) as the 
starting point of an amplitude expansion. This amounts 
to correcting the naive perturbative expansion by 
working to all orders in $M \sim \epsilon^2$, 
while working perturbatively in all other sources of 
$\epsilon$ dependence. 
\footnote{This 
is conceptually similar to the coupling constant expansion in finite 
temperature weak coupling 
QED. There, as in our situation, naive perturbation theory leads to IR 
divergences, which are cured upon exactly accounting for the photon mass 
(which is of order $g^2_{YM}$). Resummed perturbation theory in that context
corresponds to working with a modified propagator which effectively 
includes all order effects in the photon mass, while working perturbatively
in all other sources of the fine structure constant 
$\alpha$. } As we have argued above, resummed 
perturbation theory (unlike its naive counterpart) is valid at all times.

We have seen above that the naive perturbation theory gives reliable results
when $vT \ll 1$. This fact has a simple `explanation'; we will now 
argue that the resummed perturbation theory (which is always reliable
at small $\epsilon$) agrees qualitatively with naive perturbation theory 
$v T \ll 1$.
 
At each order, resummed perturbation theory involves solving the equation 
\begin{equation}\label{rpt}
\partial_r \left[r^4\left(1 - \frac{M(v)}{r^3} \right) \partial_r \phi \right]
+2 r \partial_v \partial_r (r \phi)= {\rm source}
\end{equation}
The naive perturbation  procedure requires us to solve an equation of the 
same form but with $M$ set to zero. In the vicinity of the horizon, 
the two terms in the expression $(1 - \frac{M(v)}{r^3})$ are comparable, so 
that the resummed and naive perturbative expansions can agree 
only when the entire first term on the LHS of \eqref{rpt} is 
negligible compared to the second term on the LHS of the same equation. 
The ratio of the first 
term to the second may be approximated by $r v$ where $v$ is the time 
scale for the process in question. Now the term multiplying the mass 
in \eqref{rpt} is only important in the neighborhood of the horizon, 
where $r\sim M^\frac{1}{3} \sim T$ where $T$ is the temperature 
of the black brane. It follows that resummed and naive perturbation 
expansions will differ substantially from each other only at time scales
of order and larger than the inverse temperature.  

Let us restate the point in a less technical manner. 
The evolution of a field $\phi$, outside the horizon of a black 
brane of temperature $T$, is not very different from the evolution of the 
same field in Poincare patch $AdS$ space, over time scales $v$ where 
$v T \ll 1$. However the two motions differ significantly over time scales 
of order or greater than the inverse temperature. In particular, in 
the background of the black brane, the field $\phi$ outside the horizon 
decays exponentially with time over a time scale set by the inverse 
temperature; i.e. the solution involves factors like $e^{-v T}$. As the 
temperature is itself of order $\epsilon^{\frac{2}{3}}$, naive perturbation 
theory deals with these exponentials by power expanding them. Truncating 
to any finite order then gives apparently divergent behavior at large times. 
Resummed perturbation theory makes it apparent that these divergences actually 
resum into completely convergent, decaying, exponentials.

\subsection{Resummed perturbation theory at third order}

In the previous subsection we have presented explicit results for the 
behavior of the dilaton and metric fields, at small $\epsilon$ and 
for early times $v M^{\frac{1}{3}} \ll 1$. The resummed perturbation theory 
outlined in this section may be used to systematically correct the 
leading order spacetime \eqref{lo} at all times, in a power series in 
$\epsilon^{\frac{2}{3}}$. In this section we explicitly evaluate 
the leading order correction in terms of a universal (i.e. $\phi_0$ independent)
function $\psi(x,y)$, whose explicit form we are able to determine only 
numerically.  

Let us define the function $\psi(x,y)$
as the unique solution of the differential equation
\begin{equation}\label{pte}
\partial_x\left(x^4\left(1-\frac{1}{x^3}\right) \partial_x \psi 
\right) + 2 x \partial_y \partial_x \left(x \psi \right) = 0
\end{equation}
subject to the boundary condition $\psi \sim {\cal O}({\frac{1}{x^3}})$ 
at large $x$ and the initial condition $\psi(x,0)=\frac{1}{x^3}$. 
The leading order solution to the resummed perturbation theory for $\phi$, 
for $v>\delta t$, is given by 
\begin{equation}\label{logo}
\phi= \frac{\phi^0_3(\delta t)}{M} \psi( \frac{r}{M^{\frac{1}{3}}}, 
(v-\delta t)M^{\frac{1}{3}})
\end{equation}

Unfortunately, the linear differential equation 
\eqref{pte} - appears to be difficult to solve analytically. 
In this section we  present a numerical solution of \eqref{pte}. 
Although we are forced to resort to numerics to determine $\psi(x,y)$, 
we emphasize that a single numerical evaluation suffices to determine the 
leading order solution at all values of the forcing function $\phi_0(v)$. 
This may be contrasted with an ab initio numerical approach to the full 
nonlinear differential equations, which require the re running of the full numerical code for every initial function $\phi_0$. In particular 
the ab initio numerical method cannot be used to prove general statements about a wide class of forcing functions $\phi_0$.

\begin{figure}[htbp]
\centering
\includegraphics[scale=1.0]{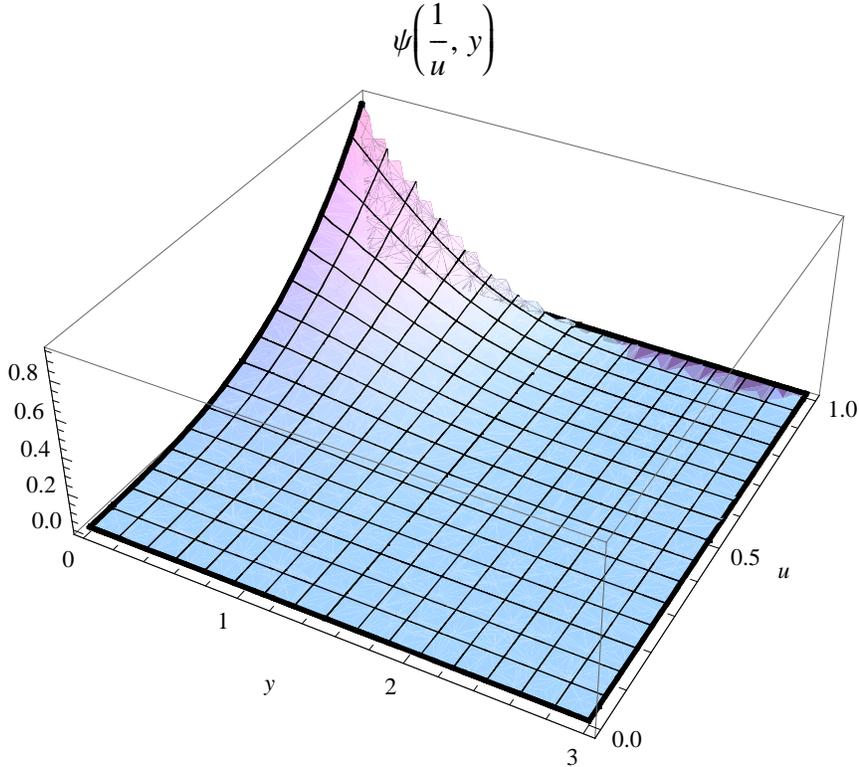}
\caption{Numerical solution for dilaton to the leading order in amplitude at late time}
\label{fig:plot4}
\end{figure}

In Figure \ref{fig:plot4} we present a plot of $\psi(\frac{1}{u}, y)$ 
against 
the variables $u$ and $y$. The exterior of the event horizon lives in the compact interval $\frac{1}{x} = u\in (0,1)$, and in our figure $y$ 
runs from zero to three.  
 
\begin{figure}[htbp]
\centering
\includegraphics[scale=1.0]{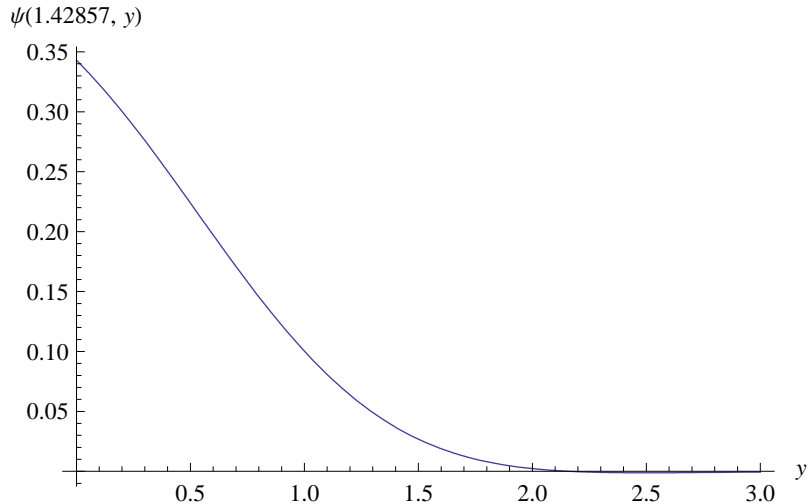}
\caption{A plot of $\psi(\frac{1}{0.7}, y)$ as a function of $y$}
\label{fig:pt}
\end{figure}

In order to obtain this plot we rewrote the differential equation \eqref{pte}
in terms of the variable $u=\frac{1}{x}$ (as explained above) and worked 
with the field variable $\chi(u, y)= (1-u)\psi(\frac{1}{u}, y)$. Recall 
that our original 
field $\psi$ is expected
to be regular at the horizon $u=1$ at all times. This expectation imposes 
the boundary condition  $\chi(.999999 , y)=0$. We further imposed the
condition of normalizability $\chi(0, y)=0$ and the initial condition 
$\chi(u, 0)=(0.999999-u)u^3$. Of course $0.999999$ above is simply a good 
approximation to $1$ that avoids numerical difficulties at unity. 
The partial differential equation solving routine of Mathematica-6 was able 
to solve our equation subject to these boundary and initial conditions,
with a step size of 0.0005 and an accuracy goal of 0.001; we have displayed 
this Mathematica output in figure~\ref{fig:plot4}. In order to give a better 
feeling for the function $\psi(x,y)$ in figure~\ref{fig:pt} we present a 
graph of $\psi(\frac{1}{0.7}, y)$ 
(i.e. as a function of time at a fixed radial location).
Notice that this graph decays, roughly exponentially for $v>0.5$ and that 
this exponential decay is dressed with a sinusodial osciallation, as
expected for quasinormal type behavior. A very 
very rough estimate of this decay constant $\omega_I$ may be obtained 
from equation 
$\frac{\psi(\frac{1}{0.7}, 1.5)}{\psi(\frac{1}{0.7}, .5)}= e^{-\omega_I}$ 
which gives $\omega_I \approx 8.9 T$ (here $T$ is the temperature of our 
black brane given by $T=\frac{4\pi}{3}$). This number is the same ballpark 
as the decay constant for the first quasi normal mode of the uniform
black brane, $\omega_I=11.16 T$, quoted in \cite{Horowitz:1999jd}.

\section{Spherically symmetric asymptotically flat collapse}
\label{flat}

\subsection{The Set Up}\label{flatsetup}

In this section\footnote{We thank B. Kol and O. Aharony for discussions 
that led us to separately study collapse in flat space.} 
we study spherically symmetric asymptotically flat solutions 
to Einstein gravity (with no cosmological constant) interacting with a 
minimally coupled massless scalar field, in 4 bulk dimensions. The Lagrangian 
for our system is 
\begin{equation}\label{lagrangianf}
S=\int d^4 x \sqrt{g} \left(R - 
\frac{1}{2} (\partial \phi)^2 \right) 
\end{equation}
We choose a gauge so that our metric and dilaton take the form
\begin{equation} \label{metdilf} \begin{split}
ds^2&= 2 dr dv -g(r,v) dv^2 +f^2(r,v) d \Omega_2^2\\
\phi&=\phi(r,v).\\
\end{split}
\end{equation}
where $d \Omega_2^2$ is the line element on a unit two sphere. We will 
explore solutions to the equations of motion of this system subject to the 
pure flat space initial conditions  
\begin{equation}\label{inconf} \begin{split}
g(r,v)&=1,~~~(v<0)\\
f(r,v)&=r, ~~~(v<0)\\ 
\phi(r,v)&=0, ~~~(v<0)\\ 
\end{split}
\end{equation} 
and the large $r$ boundary conditions
\begin{equation} \label{bcsmf} \begin{split}
&g(r, v)=1+{\cal O}(\frac{1}{r})  \\
&f(r, v)= r\left(1+{\cal O}(\frac{1}{r^2}) \right)\\
&\phi(r,v)= \frac{\psi(v)}{r} + {\cal O}(\frac{1}{r^2})
\end{split}
\end{equation}
where $\psi(v)$ takes the form 
\begin{equation} \label{bdilf} \begin{split}
\psi(v)&=0, ~~~(v<0)\\
\psi(v)&<\epsilon_f \delta t,~~~(0<v<\delta t)\\
\psi(v)&=0 ~~~(v> \delta t),
\end{split}
\end{equation}
In other words our spacetime starts out in its vacuum, but has 
a massless pulse of limited duration focused to converge at the origin at 
$v=0$. This pulse could lead to interesting behavior - like black hole 
formation, as we explore in this section.

The structure of the equations of motion of our system 
was described in subsection \ref{transdileom}. As in 
that subsection, the independent dynamical equations for our system 
may be chosen to be the dilaton equation of motion plus the two constraint 
equations, supplemented by an energy conservation equation. The explicit 
form of the dilaton and constraint equations is given by 
\begin{equation}\label{dilconstf}\begin{split}
& \partial_r\left( f^2 g \partial_r \phi \right) 
+ \partial_v\left( f^2  \partial_r \phi \right)
+\partial_r\left( f^2  \partial_v \phi \right)
=0\\
\left(\partial_r \phi\right)^2 &=-\frac{4 \partial^2_r f}{f} \\
& \partial_r \left( f g \partial_r f  + 2 f \partial_v f \right) 
 = 1 \\
\end{split}
\end{equation}
As in the previous section, we may choose to evaluate the energy conservation 
equation at large $r$. As we have explained, the large $r$ behavior of 
the function $g$ is given by  
\begin{equation}\label{masscons}
g(r,v)=1-\frac{M(v)}{r} +{\cal O}(\frac{1}{r^2})
\end{equation}
The energy conservation equation, evaluated at large $r$, yields
\begin{equation}\label{econf}
{\dot M}= - \frac{\psi {\ddot \psi}}{2}
\end{equation}
The equations\eqref{dilconstf} together with \eqref{econf} constitute 
the full set of dynamical equations for our problem. 

By integrating \eqref{econf} we find an exact expression for $M(v)$ 
\begin{equation}\label{mvf}
M(v)= \frac{-\psi {\dot \psi} +\int_{-\infty}^v {\dot \psi}^2 }{2}
\end{equation}
Note in particular that $M(v)$ reduces to a constant $M$ for $v>\delta t$ 
where 
\begin{equation}\label{mvff}
M= \frac{\int_{-\infty}^{\delta t} {\dot \psi}^2 }{2} \sim 
\epsilon_f^2 \delta t
\end{equation}

\subsection{Regular Amplitude Expansion}\label{se}

Our equations may be solved in the amplitude expansion
formally described in 
\eqref{transdilampexp}, i.e. in an expansion in powers of the function 
$\psi(v)$. As we will argue in this paper, there are two inequivalent 
valid amplitude expansions of these equations. In the first, 
the spacetime is everywhere regular and the dilaton is everywhere small.  
In the second, the spacetime is singular at small $r$ 
but this singularity is shielded from asymptotic infinity by a regular 
event horizon. The second amplitude expansion reliably describes the 
spacetime only outside the event horizon; this expansion works because 
the dilaton is uniformly small outside the event horizon. 
As we will see two amplitude expansions described above have 
non overlapping regimes of validity, and so describe dynamics in different 
regimes of parameter space. 

In this subsection we briefly comment on the more straightforward 
fully regular expansion. At every order
in perturbation theory, the requirement or regularity uniquely determines the 
solution. Explicitly at first order we have
\begin{equation} \label{dilval}
\phi_1(r,v)= \frac{\psi(v)-\psi(v-2r)}{r}
\end{equation}
The perturbation expansion that starts with this solution is valid only 
when $\phi(r)$ is everywhere small. $\phi(r)$ reaches its maximum value 
near the origin, and $ \phi_1(0,v) \sim 2 {\dot \psi(v)} 
\sim {\epsilon_f}$. Consequently the regular perturbation expansion, 
 sketched in this section, is valid only when $\epsilon_f \ll 1$ 
i.e. when $\frac{\delta t}{M} \gg 1$. 

At next order in the amplitude expansion we find 
\begin{equation}\label{flatsol}
 \begin{split} 
f_2(r,v) &= \frac{1}{4}\left(r\int_r^\infty \rho \left[\partial_\rho\phi_1(\rho,v)\right]^2~d\rho - \int_r^\infty \rho^2 \left[\partial_\rho\phi_1(\rho,v)\right]^2~d\rho\right)\\
g_2(r,v) &= -2\partial_v f_2(r,v) - \frac{f_2(r,v) - f_2(0,v)}{r} - \partial_r f_2(r,v)
 \end{split}
\end{equation}
The integration limits in the expression for $f_2(r,v)$ 
in \ref{flatsol} are fixed such that at large $r$ 
$f(r,v)$ decays like $\frac{1}{r}$. The integration constant in 
$g_2(r,v)$ is fixed  
by the requirement that the solution be regular at $r = 0$.

\subsubsection{Regularity implies energy conservation}

In this subsection we pause to explain an interesting technical subtlety 
that arises in carrying out the regular amplitude expansion. The discussion 
of this subsection will play no role in the analysis of spacetimes that 
describe black hole formation, so the reader who happens to be uninterested 
in the regular expansion could skip to the next section. 

Note that in order to obtain \eqref{flatsol} we did not make any use of 
the energy conservation equation. We will now verify (first in terms of the 
answer, and then more abstractly) that \eqref{flatsol} automatically obeys 
the energy conservation equation. 
At large $r$,  these functions have the following expansion
\begin{equation}\label{flargerex}
 \begin{split}
\phi_1(r,v) &= \frac{\psi(v)}{r}\\
f_2(r,v) &= -\frac{\psi(v)^2}{8 r}\\
g_2(r,v) &= -\frac{C_2(v)}{r}, ~~~\text{where}\\
C_2(v)&= -\frac{\psi(v)\dot \psi(v)}{2} - f_2(0,v)\\
 \end{split}
\end{equation}
If our solution does indeed obey the energy conservation relation, we 
should find that $C_2(v)$ is equal to $M(v)$ in \eqref{mvff}. We will now 
proceed to directly verify that this is the case.

The first term in $C_2(v)$ comes from the coefficient of 
$\frac{1}{r}$ in $\partial_v f_2(r,v)$.
For the  second term in the expression for $C_2(v)$, $f_2(0)$, is given by 
$$ f_2(0, v)=-\frac{1}{4}\int_0^\infty\rho^2 \left[\partial_\rho\phi_1(\rho,v)\right]^2~d\rho$$
The integrand in this expression may be split into 
four terms in the following way.
\begin{equation}\label{spliteq}
 \begin{split}
  r^2 \left[\partial_r\phi_1(r,v)\right]^2 &= 2 \psi(v)\partial_r\left[\frac{\psi(v - 2r)}{r}\right] +\frac{\psi(v)^2}{r^2} + r^2\left[\partial_r\left(\frac{\psi(v-2r)}{r}\right)\right]^2 \\
&= 2 \psi(v)\partial_r\left[\frac{\psi(v - 2r)}{r}\right] +\frac{\psi(v)^2}{r^2} + 4\left[{\dot\psi}(v - 2r)\right]^2 - \partial_r\left[\frac{\psi^2(v - 2r)}{r}\right]
\end{split}
\end{equation}
Now each of the terms can be integrated.
\begin{equation}\label{termint}
 \begin{split}
  &\int_0^r  2 \psi(v)\partial_\rho\left[\frac{\psi(v - 2\rho)}{\rho}\right]d\rho = -2\lim_{r\to 0}\frac{\psi(v)\psi(v - 2r)}{r} = -2\lim_{r\to 0}\frac{\psi(v)^2}{r}\\
&\int_0^r \frac{\psi(v)^2}{\rho^2} d\rho = \lim_{r\to 0}\frac{\psi(v)^2}{r}\\
&\int_0^r 4\left[{\dot\psi}(v - 2\rho)\right]^2 d\rho =2\int_{-\infty}^v {\dot\psi}(t)^2~dt\\
&-\int_0^r\partial_\rho\left[\frac{\psi^2(v - 2\rho)}{\rho}\right]d\rho = \lim_{r\to 0}\frac{\psi(v -2r)^2}{r} = \lim_{r\to 0}\frac{\psi(v)^2}{r}
 \end{split}
\end{equation}

Adding all the terms one finally finds
\begin{equation}\label{secondterm}
 -f_2(0,v) =  \frac{1}{2}\int_{-\infty}^v{\dot \psi(t)}^2~dt
\end{equation}

This implies
\begin{equation}\label{fmass}
 C_2(v) = -\frac{\psi(v)\dot \psi(v)}{2} + \frac{1}{2}\int_{-\infty}^v{\dot \psi(t)}^2~dt = M(v)
\end{equation}
as expected from energy conservation.

Let us summarize In order to obtain our result for 
$g_2$ above, we were required to fix the 
value of an integration constant. The value of this constant may determined
in two equally valid ways
\begin{itemize}
\item By imposing the energy conservation equation $E_{ec}$ 
\item By demanding regularity of the solution at $r=0$
\end{itemize}
In fact these two conditions are secretly the same, as we now argue. 
As we have explained in subsection \ref{transdileom}, 
$\partial_r(r E_{ec})$ automatically vanishes whenever the three equations 
\eqref{dilconstf} are obeyed. Consequently, if $r E_{ec}$ vanishes at any one 
value of $r$ it automatically vanishes at every $r$. Now the equation 
$E_{ec}$ evaluates to a finite value at $r=0$ provided our solution is 
regular at $r=0$. It follows that the regular solution automatically 
has $r E_{ec}=0$ everywhere.

Configurations in the amplitude expansion of the previous section 
(or the singular amplitude expansion we will describe shortly below), on the 
other hand, are all singular at $r=0$. $r E_{ec}$ does not automatically 
vanish on these solutions, and the energy conservation equation $E_{ec}$ is
not automatic but must be imposed as an additional constraint on solutions. 

It would be a straightforward - if cumbersome - 
exercise to explicitly implement the perturbation theory, described in this 
subsection, to higher orders in $\epsilon_f$. As our main interest is 
black hole formation, we do not pause to do that. 

\subsection{Leading order metric and event horizon for black hole formation
}\label{flathor}

In the rest of this section we will describe the formation of black holes 
in flat space in an amplitude expansion. In contrast with the previous 
subsection, our amplitude expansion will be justified by the small parameter 
$\frac{1}{\epsilon_f}$. Our analysis will reveal that 
our spacetime takes the Vaidya form to leading order in $\frac{1}{\epsilon_f^2}$, 
\begin{equation} \label{vff}
ds^2=2 dr dv -\left(1-\frac{M(v)}{r} \right) dv^2 + r^2 d\Omega_2^2
\end{equation}
where $M(v)$ is given by \eqref{mvf}.

In this subsection we will compute the event horizon of the spacetime 
\eqref{vff} at large $\epsilon_f$. We present the computation of this 
event horizon even before we have justified the form \eqref{vff},
as our aim in subsequent subsections is to have a good perturbative 
expansion of the true solution only outside the event horizon; consequently
the results of this subsection will guide the construction of the amplitude 
expansion in subsequent subsections. 

As in the previous section 
the event horizon takes the form 
\begin{equation}\label{ehflat} \begin{split}
r_H(v)&=M, ~~~(v>\delta t)\\
r_H(v)&=M x(\frac{v}{\delta t}),  ~~~(0<v<\delta t) \\
r_H(v)&=M x(0)+v,~~~(-x(0)<v<0) \\
\end{split}
\end{equation}
where the function $x(t)$ may easily be evaluated in a power series in 
$\frac{\delta t}{M} \sim \frac{1}{\epsilon_f^2}$. 
We find 
\begin{equation}\label{ehfs}\begin{split}
x(t)&= 1+ \left(\frac{\delta t}{M}\right) x_1(t) + \left(\frac{\delta t}{M}\right)^2 x_2(t) + \ldots\\
x_1(t)&=-\int_t^1 dy 
\left( \frac{1-\frac{M(y\delta t)}{M}}{2}\right)\\
x_2(t)&=-\int_t^1 dy x_1(y)\frac{M(y\delta t)}{M}.\\
\end{split}
\end{equation}
In particular $r_H=M$ for all $v>0$ at leading order.

\subsection{Amplitude expansion for black hole formation}\label{le}

Let us now construct an amplitude expansion (i.e. expansion 
in powers of $\psi(v)$) of our solution in the opposite limit to that 
of the previous subsection, namely $\frac{M}{\delta t} \sim 
\epsilon_f^2 \gg 1$. It is intuitively clear that such a dilaton shell 
will propagate into its own Schwarzschild radius and then cannot 
expand back out to infinity. 
In other words the second term in \eqref{dilval} cannot form a good 
approximation to the leading order solution for the collapse of such 
a shell. Now 
\eqref{dilval} deviates from 
\begin{equation} \label{dilvaln}
\phi_1(r,v)= \frac{\psi(v)}{r};
\end{equation}
only at spacetime points that feel the back scattered expanding wave in 
\eqref{dilval}. This observation suggests that \eqref{dilvaln} itself   
is the appropriate starting point for the amplitude expansion at large 
$\epsilon_f$, and this is indeed the case.

The incident dilaton pulse \eqref{dilvaln}
 will back react on the metric; above we have 
derived an exact expression for one term - roughly the Newtonian potential -
 (see \eqref{mvff}) of this back 
reacted metric. Including this backreaction (all others turn out to be 
negligible at large $\epsilon_f$) the spacetime metric takes the form 
\begin{equation}\label{backmet}
ds^2=2 dv dr -dv^2(1-\frac{M(v)}{r} )+ r^2 d\Omega_2^2
\end{equation}
As we have explained in the previous subsection, this 
solution has an event horizon located at 
$r_H\sim M \sim \epsilon_f^2 \delta t$ for $v>0$ (see below).
Consequently, $\phi_1(r,v)$ outside the 
event horizon $ \leq \frac{\psi}{r_H} \sim 
\frac{1}{\epsilon_f} \sim \sqrt{ \frac{\delta t}{r_H}}$, i.e. is 
parametrically small at large $\epsilon_f$. This fact allows us 
to construct a large $\epsilon_f$ amplitude expansion for the solution 
outside its event horizon.

The perturbation expansion of our solutions in $\frac{\delta t}{M}$ 
is similar in many ways to the perturbation theory described in 
detail in section \ref{transdil}. As in that section, the true (resummed) 
expansion (built around the starting metric \eqref{backmet}) is well 
approximated at early times  by a naive expansion built 
around unperturbed flat space. Naive and resummed expansions agree 
 whenever the first term in the first equation of 
\eqref{dilconstf} is negligible compared to the other terms in that equation, 
i.e. for $v \ll M \sim \epsilon_f^2 \delta t$. 
As $\epsilon_f$ is large  in this subsection, naive and resummed perturbation 
theory are simultaneously valid for times that are 
of order $\delta t$. However we expect the naive expansion to break down 
at $v\gg M$. We will now study the naive expansion in more detail and 
confirm these expectations. 

\subsection{Analytic structure of the naive perturbation expansion}
\label{flatas}

In this subsection we describe the structure of a perturbative expansion built 
starting from the  flat space metric.
We expand the full solution as 
\begin{equation}\label{fullsolf}\begin{split}
\phi(r,v)&= \sum_{n=0}^\infty \Phi_{2n+1}\\
f(r,v)&=r +\sum_{n=1}^\infty F_{2n}(r,v)\\
g(r,v)&=1+\sum_{n=1}^\infty G_{2n}(r,v)\\
\end{split}
\end{equation}
where, by definition, the functions $\Phi_m$ $F_m$ and $G_m$ are each of 
homogeneity $m$ in the source function $\psi(v)$. As explained above 
we take 
\begin{equation}\label{phiof}
\Phi_1(r,v)=\frac{\psi(v)}{r}
\end{equation}
 By studying the formal structure of the perturbation
expansion, it is not difficult to inductively establish that 
\begin{itemize} 
\item{1.} The functions $\Phi_{2n+1}$, $F_{2n}$ and 
$G_{2n}$ have the following analytic structure in the variable $r$
\begin{equation}\label{arf}\begin{split}
\Phi_{2n+1}(r, v)&=\sum_{m=0}^{\infty} \frac{\Phi_{2n+1}^m(v)}{r^{2n+m+1}}\\
F_{2n}(r,v)&=r \sum_{m=0}^{\infty} \frac{F_{2n}^m(v)}{r^{2n+m}}\\
G_{2n}(r,v)&=- \delta_{n, 1} \frac{ M(v)}{r} + 
r \sum_{m=0}^\infty \frac{G_{2n}^m(v)}{r^{2n+m}}\\
\end{split}
\end{equation}
\item{2.} The functions $\Phi_{2n+1}^m(v)$, $F_{2n}^m(v)$ and 
$G_{2n}^m(v)$ 
are each functionals of $\psi(v)$ that scale like $\lambda^{m}$ 
$\lambda^{m}$ and $\lambda^{m-1}$ under the the scaling 
$v \rightarrow \lambda v$.
\item{3.} For $v>\delta t$ the 
$\Phi_{2n+1}^m(v)$ are polynomials in $v$ of degree $\leq n+m-1$; 
$F_{2n}^m(v)$ and $G_{2n}^m$ are polynomials in $v$ of degree 
$\leq n+m-3$ and $n+m-4$ respectively.    
\end{itemize}

It follows that, say, $\phi(r,v)$, is given by a double sum
$$\phi(r,v)=\sum_{n}\Phi_{2n+1}(r, v)=\sum_{n,m=0}^{\infty} 
\frac{\Phi_{2n+1}^m(v)}{r^{2n+m+1}}.$$
Now sums over $m$ and $n$  are controlled by the effective 
expansion parameters $\sim \frac{v}{r}$ (for $m$) and 
$\frac{\psi^2 v  }{ \delta t r^2} 
\sim \frac{v}{\delta t \epsilon_f^2} \sim \frac{v}{M}$ 
(for $n$; recall that in the neighborhood of the horizon 
$r_H\sim \delta t \epsilon_f^2$).

It follows that the sum over $m$ is well approximated by its first few 
terms if only $v \ll M$ (recall we are interested in the solution only for 
$r > M$). The sum over $n$ may also be truncated to leading order 
only for $v \ll M$. As anticipated above, therefore, our naive perturbation 
expansion breaks over time scales $v$ of order and larger than $M$. 

Let us now focus on times $v$ of order $\delta t$. Over these time scales 
naive perturbation theory is valid for $r \gg \epsilon_f \delta t$ (recall 
that this domain of validity includes the event horizon surface $r_H \sim 
\epsilon_f^2 \delta t)$. Focusing on the region of interest, $r \geq r_H$, 
$\frac{\Phi_{2n+1}^m}{r_H^{2n+m+1}}$ scales like 
$\frac{1}{\epsilon_f^{2n+2m+1}}$. It follows that 
$\Phi^m_{2n+1}$, with equal values of $n+m$ are comparable 
at times of order $\delta t$. For this reason we find it useful to define 
the resummed fields 
\begin{equation}\label{resum} \begin{split}
\phi_{2n+1}(r,v)&= 
\sum_{k=0}^{n-1} \frac{\Phi_{2n+1-2k}^k(r,v)}{r^{2n+1-k}}\\
f_{2n}(r,v)&=r \delta_{n,2} F_2^0 + 
r \sum_{k=0}^{n-2}\frac{F_{2n-2k}^k(r,v)}{r^{2n-k}}\\
g_{2n}(r,v)&=r \delta_{n,2} G_2^0 +
r \sum_{k=0}^{n-2}\frac{G_{2n-2k}^k(r,v)}{r^{2n-k}}\\
\end{split}
\end{equation}
$\phi_{2n-1}$, unlike $\Phi_{2n-1}$, receives contributions from only a finite 
number of terms at any fixed $n$, and so is effectively computable at low 
orders. According to our definitions, $\phi_m$, $f_m$ and $g_m$ 
capture all contributions to our solutions of order $\frac{1}{\epsilon_f^m}$, at time scales of order $\delta t$. 

We now present explicit computations of the fields $\phi_m$, $f_m$ and 
$g_m$ up to 5th order.  We find 
\begin{equation} \label{flatexp}
 \begin{split}
  f_2(r,v) &= -\frac{\psi (v)^2}{8 r}\\
g_2(r,v) &= -\frac{M(v)}{r}\\
f_4(r,v) &= \frac{\psi (v)^4}{384 r^3}-\frac{\psi (v) B(v)}{32
   r^3}\\
g_4(r,v) &= -\frac{{\dot \psi}(v) \psi (v)^3}{48 r^3} 
-\frac{M(v) \psi (v)^2}{16
   r^3}+\frac{{\dot \psi} (v) B(v) }{16 r^3}\\
\phi_3(r,v) &= \frac{B(v)}{4
   r^3}\\
\phi_5(r,v) &= \frac{\int_{-\infty}^v \left(48 B(t)-16 \psi (t)^3\right) \, dt}{192 r^4}\\
& + \frac{\int_{-\infty}^v \left[\psi (t) {\dot \psi} (t)\left\{5 
\psi (t)^3+21 B(t)\right\} +3 M(t) \left\{\psi (t)^3-18 B(t)\right\}\right] \, dt}{192 r^5}\\
\end{split}
\end{equation}
Where
\begin{equation*}
B(v) = \int_{-\infty}^v \psi (t) \left(-M(t)+\psi (t) {\dot \psi} (t)\right) \, dt 
\end{equation*}

\subsection{Resummed perturbation theory at third order}

As in the previous subsection, even at times of order $\delta t$ (where 
naive perturbation theory is valid) naive perturbation theory 
yields a spacetime metric that is 
not uniformly well approximated by empty flat space over its region 
of validity $r \gg \delta t \epsilon_f$. The technical reason for this fact 
is very similar to that outlined in the previous section; 
$g_0$ is a constant, so is smaller at $r \sim r_H$ 
than one would have guessed from the naive extrapolation of \eqref{arf} to 
$n=0$. It follows that, in the previous section that, even at arbitrarily small 
$\epsilon$, the resultant solution is well approximated by 
$$ g(r,v) \approx 1-\frac{M(v)}{r}$$
rather than the flat space result $g(r,v)=1$, over the full domain of 
the amplitude expansion. It follows that the correct (resummed) amplitude
expansion should start with the Vaidya solution \eqref{backmet}
rather than the empty flat space. The IR divergences of the naive expansion
are a consequence of the incorrect choice of starting point for the 
perturbative expansion. 

At $v=\delta t$ our metric, to leading order, is the Schwarzschild metric 
of a black hole Schwarzschild radius $M$ with a superposed dilaton (and 
consequently metric) perturbation. We will now demonstrate that these 
pertubrations are small.  As in the previous section, 
it is useful to define rescaled radial and time variables $x=\frac{r}{M}$
and $y=\frac{v}{M}$. In terms of the rescaled variables, the leading order
metric takes the form 
\begin{equation}\label{resm}
ds^2=M^2 \left( 2 dx dy - dy^2 \left(1-\frac{1}{x} \right) 
+ x^2 d \Omega_2^2 \right) 
\end{equation}
while the $\phi$ perturbation is given to leading order by 
\begin{equation}\label{ptm}
\frac{\phi^0_3(\delta t)}{r^3}= \frac{\phi^0_3(\delta t)}{M^3 x^3}
\sim \frac{1}{\epsilon_f^3 x^3}
\end{equation}
(recall from \eqref{flatexp} that 
\begin{equation}\label{ptt}
\phi^0_3(\delta t) = \frac{1}{4} 
\int_0^{\delta t} \psi (v) \left[-M(v)+\psi (v) {\dot \psi}(v)
\right] \, dv
\end{equation}
and $M(v)$ is given in \eqref{mvf}).

As a constant rescaling of the metric is an invariance of the 
equations of motion of the Einstein dilaton system, 
the factor of $M^2$ in \eqref{resm} is irrelevant for dynamics. As 
the dilaton perturbation above is parametrically small 
(${\cal O}(1/\epsilon_f^3))$ the subsequent evolution of the dilaton field is linear to leading 
order in the $\frac{1}{\epsilon_f}$ expansion. 

Let $\chi(x,y)$ denote the 
unique solution to 
\begin{equation}\label{ptef}
\partial_x\left(x^2\left(1-\frac{1}{x}\right) \partial_x \chi 
\right) + 2 x \partial_y \partial_x \left(x \chi \right) = 0
\end{equation}
subject to the boundary condition $\chi \sim {\cal O}({\frac{1}{x^3}})$ 
at large $x$ and the initial condition $\chi(x,0)=\frac{1}{x^3}$. 
The leading order solution to the resummed perturbation theory for $\phi$, 
for $v>\delta t$, is given by 
\begin{equation}\label{log}
\phi= \frac{\phi_3^0(\delta t)}{M^3} \chi \left( \frac{r}{M}, 
\frac{(v-\delta t)}{M}\right)
\end{equation}

Unfortunately, the function $\chi(x,y)$ appears to be difficult to 
determine analytically. As in section \ref{transdil} this 
solution may presumably be determined numerically with a little effort. 
We will not attempt the requisite numerical calculation here. In the rest 
of this subsection we will explain in an example 
how the general analysis of this subsection yields useful precise information 
about the subleading solution even in the absence of detailed knowledge of the function $\chi(x,y)$.

Consider a spherically symmetric shell, of the form discussed in this section, imploding inwards to form a black hole. On general grounds we expect some of 
the energy of the incident shell to make up the 
mass of the black hole, while the remaining energy is reflected back out
in the form of an outgoing wave that reaches ${\cal I}^+$. Let the fraction of 
the mass that is reflected out to ${\cal I}^+$ be denoted by $f$. \footnote{In fancy parlance $f=\frac{A-B}{A}$ where $A$ is the ADM mass of the spacetime 
and $B$ is the late time Bondi mass.}. $f$ is one of the most interesting and 
easily measured observables that characterize black hole formation. 

At leading order in the expansion in $\frac{1}{\epsilon_f}$ our spacetime metric 
takes the Vaidya form with no outgoing wave, and so $f=0$. This prediction is corrected at first subleading 
order, as we now explain. It follows on general grounds that, at late times   
$$\chi(x,y) \approx \frac{\zeta(y-2x)}{x}$$
for some function $\zeta(v)$. Note that $\zeta$, like the function $\chi$, 
is universal (i.e. independent of the initial condition $\psi(v)$). 
It follows that at late times (and to leading order) \begin{equation}\label{logt}
\phi= M \frac{\phi_3^0(\delta t)}{M^3} \frac{\zeta\left( 
\frac{v-2r}{M} \right)}{r} .
\end{equation}
It then follows from \eqref{mvff} (but now applied to an outgoing rather than 
an ingoing wave) that the energy\footnote{We have chosen our units of energy 
so that a black hole with horizon radius $r_H$ has energy $M$.} 
carried by this pulse is
 
\begin{equation}\label{eninpulse}
 \left( M \frac{\phi_3^0(\delta t) } {M^3} \right)^2 
\times \frac{1}{2} 
\int dt \left(\partial_t \zeta(\frac{t}{M}) \right)^2=
M \times  
\left(\frac{\phi_3^0(\delta t)}{M^3} \right)^2 \times \frac{1}{2}\int_{-\infty}^{\infty} dy \left({\dot \zeta}(y) \right)^2 
\end{equation}
It follows that
\begin{equation}\label{pf} \begin{split}
f&=A \left(\frac{\phi_3^0(\delta t)}{M^3} \right)^2\\
&A= \frac{1}{2}\int_{-\infty}^{\infty} {\dot \zeta}^2 
\end{split}
\end{equation}
\eqref{pf} analytically determines the dependence of $f$ on the shape of 
the incident wave packet, $\psi(v)$ (recall that $\phi_3^0(\delta t)$ 
and $M$ are determined in terms of $\psi(v)$ by \eqref{ptt} and 
\eqref{mvff}). Detailed knowledge of function $\chi(x,y)$ 
is required only to determine the precise value of 
universal dimensionless number $A$.

\section{Spherically symmetric collapse in global $AdS$}\label{global}

We now turn to the study of black hole formation induced
by an ingoing spherically symmetric dilaton pulse in an asymptotically 
$AdS_{d+1}$ space in global coordinates. As in section \ref{transdil}
our bulk dynamics is described by the Einstein Lagrangian with a negative 
cosmological constant and a minimally coupled dilaton. However as in 
section \ref{flat} we study solutions that preserve an $SO(d)$ invariance; 
this $SO(d)$ may be thought of as the group of rotations of the boundary 
$S^{d-1}$. As in both sections \ref{transdil} and \ref{flat} our symmetry 
requirement determines our metric up to three unknown functions of the
two variables; the  time coordinate $v$ and the radial coordinate $r$. 
Our solutions 
are completely determined by the boundary value, $\phi_0(v)$ of the dilaton 
field. As in section \ref{transdil} we assume that $\phi_0(v)$ is everywhere 
bounded by $\epsilon$ and vanishes outside the interval $(0, \delta t)$. 
Through out this section we will focus on the regime $\delta t \ll R$ (where 
$R$ is the radius of the boundary sphere) and $\epsilon \ll 1$. The complementary 
regime $\delta t \gg R$ and arbitrary $\epsilon$ is under independent current 
investigation \cite{SanSum}.

The collapse process studied in this section  
depends crucially on two independent dynamical parameters; 
$x=\frac{\delta t}{R}$ together with $\epsilon$ of previous subsections. 
We study the 
evolution of our systems in a limit in which $x$ and $\epsilon$ are both 
small. The problem of asymptotically $AdS$ spherically symmetric collapse is 
dynamically richer than the collapse scenarios studied in sections 
\ref{transdil} and \ref{flat}, and indeed reduces to those two special cases 
in appropriate limits. 

\subsection{Set up and equations}
 The equations of motion 
for our system are given by \eqref{eom}. The form of our metric and 
dilaton is a slight modification of \eqref{metdil} 
\begin{equation} \label{metdils} \begin{split}
ds^2&= 2 dr dv -g(r,v) dv^2 +f^2(r,v) d\Omega_{d-1}^2\\
\phi&=\phi(r,v).\\
\end{split}
\end{equation}
where $d\Omega_{d-1}^2$ represents the metric of a unit $d-1$ sphere. 
Our fields are subject to the pure global $AdS$ initial conditions 
\begin{equation}\label{incong} \begin{split}
g(r,v)&=r^2+\frac{1}{R^2}, ~~~(v<0)\\
f(r,v)&= r R, ~~~(v<0)\\ 
\phi(r,v)&=0, ~~~(v<0)\\ 
\end{split}
\end{equation} 
and the large $r$ boundary conditions
\begin{equation} \label{bcsms} \begin{split}
&g(r, v)=r^2 \left(1+{\cal O}(\frac{1}{r^2}) \right) \\
&f(r, v)= r\left(R+{\cal O}(\frac{1}{r^2}) \right)\\
&\phi(r,v)= \phi_0(v) + {\cal O}(\frac{1}{r})
\end{split}
\end{equation} 
Equations \eqref{eom}, \eqref{metdils}, \eqref{incong} and \eqref{bcsms}
together constitute a completely well defined dynamical system. 
Given a particular forcing function $\phi_0(v)$, these equations and 
boundary conditions uniquely determine the functions 
$\phi(r,v)$, $g(r,v)$ and $f(r,v)$.   

The structure of the equations of motion of our system was described in 
subsection \ref{transdileom}. In particular, we may choose the dilaton 
equation of motion, together with the two constraint equations, as our 
independent equations of motion;
this set is supplemented by the 
energy conservation relation. With our choice of gauge and notation, the 
dilaton equation of motion and constraint equations take the explicit form    
\begin{equation}\label{sphereeq}
 \begin{split}
&\partial_r\left( f^{d-1} g \partial_r \phi \right) 
+ \partial_v\left( f^{d-1}  \partial_r \phi \right)
+\partial_r\left( f^{d-1} g \partial_v \phi \right)
=0~~\\
&\left(\partial_r \phi\right)^2 +\frac{2(d-1) \partial^2_r f}{f} = 0~~\\
& \partial_r \left( f^{d-2} g \partial_r f  + 2 f^{d-2} \partial_v f \right) 
 - d~ f^{d-1} - (d-2) f^{d-3}=0 ~~ \\
 \end{split}
\end{equation}
As in section 2, the initial data needed to specify a solution to these 
equations is given by the value of $\phi(r)$ on a given time slice, 
supplemented by the initial value of the mass, and boundary conditions at 
infinity. In order to obtain an explicit form for the energy conservation 
equation we specialize to $d=3$ and explicitly `solve' our system at 
large $r$ a la Graham and Fefferman. We find 

\begin{equation}\label{onebyr}
\begin{split}
 f(r,v)&= R r\left(1 - \frac{ {\dot \phi_0}^2}{8 r^2} 
+ {\cal O}(\frac{1}{r^4}) \right)\\
g(r,v)&= r^2\left(\frac{1}{R^2} + 
\frac{1 - \frac{3 {\dot \phi_0}^2}{4}} {r^2} 
-\frac{M(v)}{r^3}  + {\cal O}(\frac{1}{r^4}) \right)\\
 \phi(r,v)&= \phi_0(v) + \frac{{\dot \phi_0}}{r} 
+ \frac{L(v)}{r^3}  + {\cal O}(\frac{1}{r^4}) \\
\end{split}
\end{equation}
The energy conservation equation constrains the (otherwise arbitrary) 
functions $M(v)$ and $L(v)$ to obey 
\begin{equation}\label{sphereeom}
  {\dot M} = -\frac{\dot \phi_0}{8}\left(12 L(v) + 4\frac{{\dot \phi_0}}{R^2} 
-3 \left({\dot \phi_0}\right)^3 + 4 {\dddot \phi_0}(v)\right)
\end{equation}
\footnote{Note that the stress tensor and Lagrangian ${\cal L}$ of our system 
are given by  
\begin{equation}\label{spherestres}
\begin{split}
T_v^v &= M(v)\\
T_\theta^\theta &= T_\phi^\phi = -\frac{M(v)}{2}\\
{\cal L} &= -3 L(v) - \frac{{\dot \phi_0}}{R^2} 
+ \frac{3}{4}\left({\dot \phi_0}\right)^3 - {\dddot \phi_0}(v)
\end{split}
\end{equation}
It follows that \eqref{sphereeom} may be rewritten as ${\dot M}= 
\frac{{\dot \phi_0} {\cal L} }{2}$. }

\subsection{Regular small amplitude expansion}\label{seg}

As in section \ref{flat} there are two legitimate amplitude expansions 
of spacetime we wish to determine. In this subsection we discuss the 
expansion analogous to the expansion of subsection \ref{se}. That is we 
expand all our fields as in \eqref{ampexp} (where the 
functions $f_n$, $g_n$ and $\phi_n$ are all defined to be of 
homogeneity $n$ in the boundary field $\phi_0$) and demand that all 
functions are everywhere regular. The requirement of regularity, together with 
our boundary and initial conditions, uniquely specifies 
all functions in \eqref{ampexp}. Explicitly, to second order we find  

\begin{equation}\label{solsph}
 \begin{split}
  \phi_1(r,v) &= \sum_{m=0}^\infty(-1)^m\bigg[\phi_0(v - m \pi R) 
+ \frac{{\dot \phi_0}(v-  m \pi R )}{r}+\phi_0(v -  R m\pi - 2 R \tan^{-1}(r R) )\\
& ~~~~~~ - \frac{\dot \phi_0(v - m\pi R - 2 R \tan^{-1}(r R))}{r} \bigg]\\
f_2(r,v) &= \frac{R}{4} \left(r \int_r^{\infty } \rho  K(\rho,v ) \, d\rho -\int_r^{\infty } \rho ^2 K(\rho,v ) \, d\rho\right)\\ 
g_2(r,v)&=-\frac{1}{4r}\bigg[\frac{2 r}{R^2} 
\int_r^\infty  \rho K(\rho,v ) \,d\rho + 2 r^2\int_r^\infty \rho^2 K(\rho,v ) \, d\rho \\
&~~~~~~+ \int_0^r\rho^2(\frac{1}{R^2} + \rho^2)K(\rho,v) \, 
d\rho\bigg] -\frac{2}{R} \partial_vf_2(r,v)\\
\text{where}&\\
K(\rho,v ) &= \left(\partial_r\phi_1(r,v)\right)^2
 \end{split}
\end{equation}

The perturbation expansion in this section is valid only if $\phi(r, v)$ 
is everywhere small on the solution. $\phi_1(r,v)$ reaches its maximum 
value in the neighborhood of the origin where it is 
given approximately by $\phi_0+ {\ddot \phi_0} \sim \epsilon +
\frac{\epsilon}{x^2}$. Consequently the validity of the amplitude 
expansion sketched in this section requires both that $\epsilon \ll 1$ 
and that $ x^2 \gg \epsilon$. 

We have chosen integration constants to ensure that the solution in 
\eqref{solsph} is regular at $r=0$. In particular 
$$g_2(0,v)= \frac{1}{2}\left(\int_0^\infty \rho^2 \partial_v 
K(\rho,v) \, d\rho - \frac{1}{R^2} \int_0^\infty \rho K(\rho,v) \, d\rho\right).$$
As in subsection \ref{se}, this choice automatically implies the 
energy conservation equation. In particular, expanding $g_2(r,v)$ at 
large $r$ we find 
\begin{equation}
 -M(v) = -\frac{1}{4}\left(\int_0^\infty\rho^2(\frac{1}{R^2} + \rho^2)
K(\rho,v) \, d\rho - {\dot\phi_0}(v)^2 -2 {\dot\phi_0} (v)
{\ddot \phi_0}(v)\right)
\end{equation}
(this equation is valid only for $v<\pi R$; it turns out that 
$M(v)$ is constant for $v > \delta t$) 
in agreement with the energy conservation equation.

Finally, let us focus on the coordinate range $r R, \frac{v}{R} 
\ll 1$ and also 
require that $x$ is small so that the time scale in $\phi_0$ is also 
smaller than the $AdS$ radius.  In this parameter and coordinate range 
\eqref{solsph} should reduce to a solution of the flat 
space propagation equation \eqref{dilval}; this is easily verified 
to be the case. In the given variable and parameter regime, 
all terms with \eqref{solsph} with $m\neq 0$ vanish; $\tan^{-1}( R r) 
\approx r R$  and the first and the third terms in \eqref{solsph} 
are negligible compared to the second and fourth as $x$ is small. 
Putting all this together, \eqref{solsph} reduces to 
\eqref{dilval} under the identification 
$\psi(v) = R^2 {\dot \phi_0(v)}$, once we also identify the coordinate 
$r$ of section \ref{flat} with $R^2 r$ in this section. 
Notice that this replacement implies that $\epsilon_f = 
\frac{\epsilon}{x^2}$ (where $\epsilon_f$ was the perturbative 
expansion of section \ref{flat}). This identification of parameters 
is consistent with the fact that the expansion of this 
section breaks down when $\frac{\epsilon}{x^2}$ becomes large, while the 
expansion of subsection \ref{se} breaks down at large $\epsilon_f$. 

\subsection{Spacetime and event horizon for black hole formation}

In the rest of this section we will describe the process of black hole 
formation via collapse in an amplitude expansion. 
As in earlier sections, the spacetime that describes this collapse process
will turn out to be given, to leading order, by the Vaidya form  
\begin{equation}\label{los} \begin{split}
ds^2&=2 dr dv -\left(\frac{1}{R^2}+ r^2-\frac{M(v)}{r}\right)dv^2
+R^2 r^2 d \Omega_2^2\\
\phi(r,v)&=\phi_0(v)+ \frac{{\dot \phi_0}}{r}\\
\end{split} 
\end{equation}
where $M(v)$ is approximated by $C_2(v)$, the order $\epsilon^2$ piece of 
\eqref{sphereeom}
\begin{equation}\label{set}
C_2(v)=-\frac{1}{2} \int_{-\infty}^v dt {\dot \phi_0}(t) \left( 
{\dddot \phi_0}(t) +\frac{{\dot \phi_0}(t)}{R^2} \right)
\end{equation}  

In this subsection we will compute the event horizon of the spacetime 
\eqref{los} in a perturbation expansion in a small parameter, whose nature 
we describe below. The horizon is 
determined by the differential equation 
\begin{equation}\label{eheg}
2 \frac{dr_H}{dv}=\frac{1}{R^2}+r_H^2-\frac{M(v)}{r_H}
\end{equation}
where $M(v)$ reduces to a constant $M$ for $t>\delta t$. At late times the
event horizon surface must reduce to the largest real solution of the equation
$$\frac{1}{R^2}+(r^0_H)^2-\frac{M}{r^0_H}=0.$$ It then follows 
from \eqref{eheg} that 
\begin{equation}\label{ehsg} \begin{split}
r_H(v)&= r_H^0,    ~~~~(v > \delta t)\\
r_H(v)&= r_H^0 x(\frac{v}{\delta t}),~~~(0<v<\delta t)\\
\tan^{-1}\left(r_H(v) \right)&= \tan^{-1} \left( r_H^0 x(0) \right)+v
~~~(v<0), ~~~ \tan^{-1}(r_H(v))>0
\end{split}
\end{equation}

As in previous subsections, the function $x(t)$ is easily generated in a 
perturbation expansion
\begin{equation}\label{xexp}
x(t)=1+\left( \frac{M \delta t}{(r^0_H)^2} \right) x_1(t) 
+\left( \frac{M \delta t}{(r^0_H)^2} \right)^2 x_2(t) + \ldots 
\end{equation}
The small parameter for this expansion is 
$\frac{M \delta t}{(r^0_H)^2}$. This parameter varies from approximately
$\epsilon^{\frac{2}{3}}$ when $ x \ll \epsilon^\frac{2}{3}$ to 
$\frac{x^4}{\epsilon^2}$ when $x \gg \epsilon^\frac{2}{3}$ and is 
always small provided $x\ll \sqrt{\epsilon}$ and $\epsilon \ll 1$.
These conditions will always be met in our amplitude constructions 
below.  Note that the event horizon of our solution is created (at $r=0$) 
at the time $v=-\tan^{-1}(r_H^0) +$ subleading. 

Explicitly working out the perturbation series we find 
\begin{equation}\label{horbor} \begin{split}
x_1(t)&=- \int_t^1 dt \frac{1-\frac{M(y \delta t)}{M}}{2}\\
x_2(t)&=-\int_t^1 dy \left( \frac{2 (r^0_H)^3}{M}+ 
\frac{M(y \delta t)}{M} \right)\\
\end{split}
\end{equation}

\subsection{Amplitude expansion for black hole formation}

The amplitude expansion of the previous subsection breaks down 
for $x^2 \ll \epsilon$. As in section \ref{flat}, we have a new 
amplitude expansion in this regime. As in section \ref{flat}, the 
starting point for this expansion is the Vaidya metric and dilaton 
field \eqref{los}. 
As in sections \ref{transdil} and \ref{flat}, the perturbation expansion 
based on \eqref{los} is technically difficult to implement at late times. 
However as in earlier sections, at early times - i.e. times of order 
$\delta t$ - the perturbative expansion is well approximated by the 
naive expansion based on the solution \eqref{los} with $M(v)$ set
equal to zero. Following the terminology of previous sections
we refer to this simplified expansion as the naive expansion. 
In the rest of this subsection we will elaborate on the analytic structure 
of the naive perturbative expansion. 

In order to build the naive expansion, we expand the fields $f(r,v)$, 
$g(r,v)$ and $\phi(r,v)$ in the form \eqref{ampexp}. It is not too difficult 
to inductively demonstrate that 
\begin{itemize} 
\item{1.} The functions $\phi_{2n+1}$, $g_{2n}$ and 
$f_{2n}$ have the following analytic structure in the variable $r$
\begin{equation}\label{aro}\begin{split}
\phi_{2n+1}(r, v)&=\sum_{m=0}^\infty  \frac{1}{R^{2m}}
\sum_{k=0}^{2n+m -2}
 \frac{\phi_{2n+1}^{k,m}(v)}{r^{2n+1-k+m}}~~~(n \geq 1)\\
f_{2n}(r,v)&= r R\sum_{m=0}^\infty  \frac{1}{R^{2m}}
\sum_{k=0}^{2n-4} \frac{f_{2n}^{k, m}(v)}{r^{2n-k+m}}~~~(n \geq 2)\\
g_{2n}(r,v)&= -\frac{C_{2n}(v)}{r} + r \sum_{m=0}^\infty \frac{1}{R^{2m}}
\sum_{k=0}^{2n-3} \frac{g_{2n}^{k,m}(v)}{r^{2n-k+m}}~~~(n \geq 2)\\
\end{split}
\end{equation}
\item{2.} The functions $\phi_{2n+1}^{k,m}(v)$, $f_{2n}^{k,m}(v)$ and 
$g_{2n}^{k,m}(v)$ 
are functionals of $\phi_0(v)$ that scale like $\lambda^{-2n-1+m+k}$, 
$\lambda^{-2n+m+k}$ and $\lambda^{-2n +m+k-1}$ respectively under 
the scaling $v \rightarrow \lambda v$.
\item{3.} For $v>\delta t$ we have some additional simplifications 
in structure. At these times $f_4(r,v)=0$ and $g_4(r,v)=-\frac{C_4(v)}{r}$. 
Further, the sums over $k$ in the second and third of the 
equations above run from $0$ to $ 2n-6+m$ and $2n-5 +m$ respectively. 
Finally, functions 
$\phi_{2n+1}^{k,m}(v)$ are all polynomials in $v$ of a degree 
that grows with $n$. In particular the degree of 
$\phi_{2n+1}^{k,m}$ is at most $n-1+k+m$; the degree of $f_{2n}^{k, m}$ is 
at most $n-3+k+m$ and the degree of $g_{2n}^{k, m}$ is at most $n-4+k+m$.  
\end{itemize}

As we have explained above, 
$$\phi(r, v)=\sum_{n=1}^\infty
\sum_{m=0}^\infty \frac{1}{R^{2m}} \
\sum_{k=0}^{2n-2+m} \frac{\phi^{k,m}_{2n+1}(v)}{r^{2n+1-k+m}}
.$$
We will now discuss the relative orders of magnitude of different terms 
in this summation. Abstractly, at times that are larger than or of order 
$\delta t$, the effective weighting factor for the sum over $n, m, k$ 
respectively are approximately given by 
$\frac{\epsilon^2 v}{r^2 (\delta t)^3}$, $\frac{v}{R^2 r}$ and ${v r}$ 
respectively. We will try to understand the implications of these estimates
 in more detail. 

Let us first suppose that 
$x \ll \epsilon^\frac{2}{3}$. In this case the black hole that is formed has a 
horizon radius of order $\frac{\epsilon^\frac{2}{3}}{\delta t}\gg 
\frac{1}{R}$ 
( this estimate is corrected in a power series in 
$\frac{x^2}{\epsilon^{\frac{4}{3}}}$). Consequently, the resultant black hole is large compared to the $AdS$ radius. At $m=0$,  this regime,  the summation 
over $k$ and $n$ simply reproduce the solution of section \ref{transdil}. 
As in section \ref{transdil} these summations are dominated by the 
smallest values of $k$ and $n$ for $v r_H \sim v T \ll 1 $, in the 
neighborhood of the horizon. As in section \ref{transdil} the sum over 
$k$ is dominated by the largest value of $k$ at large enough $r$. The new 
element here is the sum over $m$; this summation is dominated 
by small $m$  when $v T \ll \frac{\epsilon^{\frac{4}{3}}}{x^2} $. 
When $x \ll \epsilon^{\frac{2}{3}}$, this condition automatically follows 
whenever $v T  \ll 1$. Consequently, naive perturbation theory is 
always good for times small compared to the inverse black hole temperature, 
in this regime. 

We emphasize that naive perturbation theory is always good at times  
of order  $\delta t$. Over such time scales 
(and for $r \sim r_H$) we note that the sum over $n$ and $k$ are weighted 
by $\epsilon^{\frac{2}{3}}$ (this is as in section \ref{transdil}) 
while the sum over $m$ is weighted  by $\epsilon^\frac{2}{3} (\frac{x}{\epsilon^\frac{2}{3}})^2$.
Note that the weighting factor for the sum over $m$ is smaller than the 
weighting factor for the sum over, for instance $n$, provided $x \ll 
\epsilon^\frac{2}{3}$. It follows that our naive perturbation theory 
represents a weak departure from the black brane formation solution of 
section \ref{transdil} when $x\ll \epsilon^{\frac{2}{3}}$.

Now let us turn to the the parameter regime $x \gg \epsilon^{\frac{2}{3}}$.
In this regime ${r_H R} \sim \frac{\epsilon^2}{x^3}$, so that 
black holes that are formed in the collapse process are always small in 
units of the $AdS$ radius. At times that are larger or of order 
$\delta t$, the sum over $m$ and $n$ are dominated by their smallest 
values provided $\frac{v}{R} \ll \frac{\epsilon^2}{x^3}$. Making the 
replacement $\epsilon= x^2 \epsilon^f$, this condition reduces to 
$v  \ll \epsilon_f^2 \delta t$ which was exactly the condition for 
applicability of naive perturbation theory in flat space in section 
\ref{flat}. The new element here is the sum over $k$. $k$ is zero in 
section \ref{flat}, and the sum over $k$ here 
is dominated by $k=0$ near $r=r_H$ for 
$\frac{v}{R} \ll \frac{x^3}{\epsilon^2}$, a condition 
that is automatically implied by $\frac{v}{R} \ll \frac{\epsilon^2}{x^3}$. 
Note, however,  that, as in the previous paragraph, the sum over $k$ is always 
dominated by the largest value of $k$ at sufficiently large $r$. This reflects
the fact that $AdS$ space is never well approximated by a flat bubble at 
large $r$. Finally, specializing  
to $v$ of order $\delta t$ and $r \sim r_H$, the sum over $n$ and  $m$ 
are each weighted by $\frac{x^4}{\epsilon^2} \sim \frac{1}{\epsilon_f^2}$ 
while the sum over $k$ is weighted by $\epsilon \frac{\epsilon}{x^2}$. 
In particular naive perturbation theory is good at times of order $\delta t$
provided $x \ll \sqrt{\epsilon}$. 

Let us summarize in broad qualitative terms. Naive perturbation theory is 
a good expansion to the true solution when $v T \ll 1$ for 
$\frac{v}{R} \ll \frac{\epsilon^2}{x^3}$. In particular, this condition is 
always obeyed for times of order $\delta t$ when $ x \ll \sqrt{\epsilon}$.

\subsection{Explicit results for naive perturbation theory}
As we have explained above, the functions $\phi$, $f$ and $g$ may be 
expanded in an expansion in $\epsilon$ as 
\begin{equation}\label{globalexp}
\begin{split}
 \phi(r,v) &= \epsilon \phi_1(r,v) + \epsilon^3 \phi_3(r,v) +{\cal O}(\epsilon)^5\\
f(r,v) &= r R\left(1  + \epsilon^2 f_2(r,v) + \epsilon^4 f_4(r,v) +{\cal O}(\epsilon)^6\right)\\
g(r,v)&= r^2 + \frac{1}{R^2} + \epsilon^2 g_2(r,v) + \epsilon^4 g_4(r,v) +{\cal O}(\epsilon)^6\\
\end{split}
\end{equation}
Moreover the functions $\phi_{2n+1}$, $f_n$ and $g_n$ may themselves each 
be expanded as a sum over two integer series (see \eqref{aro}). The sum 
over $k$ runs over a finite number of values in \eqref{aro} and we will deal 
with this summation exactly below. However the sum over 
$m$ runs over all integers, and is computatble only after truncation
to some finite order.  This truncation is justified as the 
sum over $m$ is effectively weighted by a small parameter as explained in the section above. In this section we present exact expressions 
for the functions $\phi_1$, $g_2$ and $f_2$, and expressions for $\phi_3$,
$f_4$ and $g_4$ to the first two orders in the expansion over the integer 
$m$ (this summation is formally weighted by $\frac{1}{R}$); 

The solutions are given as
\begin{equation}\label{expsolglo}
 \begin{split}
  \phi_1(r,v) &= \phi_0(v) + \frac{\dot \phi_0(v)}{r}\\
f_2(r,v) & = -\frac{\dot \phi_0^2}{8 r^2}\\
g_2(r,v) &= -\frac{3\dot \phi_0^2}{4} - \frac{C_2(v)}{r}\\
\phi_3(r,v) &= \frac{K(v)}{r^3}\\
 &+ \frac{1}{R^2}\bigg[\frac{\int_{-\infty}^v \left(3 K(t) - {\dot\phi_0}(t)^3\right)dt}{12r^4}
+\frac{\int_{-\infty}^v dt_1\int_{-\infty}^{t_1}dt_2 \left(3 K(t_2) - {\dot\phi_0}(t_2)^3\right)}{12r^3}\bigg]\\
& + {\cal O}\left(\frac{1}{R}\right)^4\\
f_4(r,v) &= \left(\frac{{\dot \phi_0}^4 }{384 r^4} - \frac{A_3(v)}{32 r^4}\right) + \frac{1}{R^2}\left(\frac{A_1(v)}{96 r^4} + \frac{A_2(v)}{120 r^5}\right) + {\cal O}\left(\frac{1}{R}\right)^4\\
g_4(r,v)&= -\frac{C_4(v)}{r} +\frac{3A_3(v) - {\dot \phi_0}^4}{24 r^2} +\frac{1}{48 r^3}\left(3 \dot A_3(v) - 4 {\dot\phi_0}^3{\ddot \phi_0}\right)\\
&- \frac{1}{R^2}\bigg[\frac{A_1(v)}{24 r^2} + \frac{\dot A_1(v)}{48 r^3} + \frac{15 A_3(v) + 4 A_2(v) - \dot \phi_0^4}{240 r^4}\bigg]   + {\cal O}\left(\frac{1}{R}\right)^4\\
\end{split}
\end{equation}
Where
\begin{equation}\label{sangya}
\begin{split}
K(v)&= \int_{-\infty}^v dt~\dot\phi_0\left(-C_2(t) + \dot\phi_0\ddot\phi_0\right)\\
A_1(v) &= \dot\phi_0(v)\int_{-\infty}^{v}dt_1\int_{-\infty}^{t_1} dt_2\left( -3 K(t_2) + {\dot\phi_0}^3(t_2)\right)\\
A_2(v) &= \dot\phi_0(v)\int_{-\infty}^v dt\left( -3 K(t) + {\dot\phi_0}^3(t)\right)\\
A_3(v) &= \dot\phi_0 K(v)\\
\end{split}
\end{equation}
\begin{equation}\label{shakti}
\begin{split}
C_2(v) &= -\frac{1}{2}\int_{-\infty}^v dt~\dot\phi_0(t)\left(\frac{\dot\phi_0(t)}{R^2} + \dddot\phi_0(t)\right)\\
C_4(v) &= -\frac{3}{8}\int_{-\infty}^v dt~\dot\phi_0(t)\left(K(t) - \dot\phi_0(t)^3\right)\\
& -\frac{1}{8 R^2} \int_{-\infty}^v dt_1~\dot\phi_0(t_1)\int_{-\infty}^{t_1} dt_2\int_{-\infty}^{t_2}dt_3 \left(3 K(t_3) - {\dot\phi_0}(t_3)^3\right) + {\cal O}\left(\frac{1}{R}\right)^4
\end{split}
\end{equation}

\subsection{The solution at late times}

As in previous sections, our solution is normalizable (unforced) for 
$v > \delta t$. Naive perturbation theory reliably establishes
the initial conditions for this unforced evolution at $v=\delta t$. To leading
order, this evolution is given by global AdS black hole metric with 
$M= C_2(\delta t)$ (see \eqref{set}), perturbed by 
$\phi(\delta t) =\frac{K(\delta t)}{r^3}$ see \eqref{expsolglo}. 
As in the previous two subsections, the qualitatively important point is that 
this represents a small perturbation about the black hole background. 
Moreover, it follows on general grounds that perturbations in a black hole 
background in $AdS$ space never grow unboundedly (in fact they decay) with time.
Consequently, we may reliably conclude that our spacetime takes the 
Vaidya form \eqref{los} at all times to leading order in the amplitude 
expansion.  

In order to determine an explicit expression for the subsequent dilaton 
evolution, one needs to solve for the linear, minimally coupled, 
 evolution of a $\frac{1}{r^3}$ initial condition in the background of global 
$AdS$ with a Schwarzschild black hole of arbitrary mass. As in the previous 
two sections, the linear differential equation one needs to solve appears 
to be analytically intractable, but could easily be solved numerically. 
We will not, however, attempt this evaluation in this paper.

\section{Discussion}\label{disc}

In this paper we have used the $AdS/CFT$ correspondence to determine 
the response of a conformal field 
theory, initially in its vacuum, to a low amplitude perturbation by a source
coupled to a marginal operator. When the CFT in question lives on $R^{d-1,1}$
it responds to the perturbation by the source by thermalizing into a plasma
type phase. On the other hand, when the CFT in question lives on a sphere
it either thermalizes into a plasma type phase or settles down into a glueball 
type phase depending on the details of the perturbation procedure.  
In this paper we have demonstrated that, to leading order in the amplitude 
expansion, the dual description of the thermalization into a plasma type 
phase  is a spacetime of the Vaidya form. In odd boundary field theory 
dimensions the Vaidya metric reduces exactly to the uniform black brane 
metric in the causal future $v>\delta t$ at the boundary. As was discussed 
in detail in section \ref{introtrans}, for many purposes our system 
behaves as if it has thermalized {\it instantaneously}.

In this paper we have only studied solutions with a high degree of symmetry; 
for instance, solutions that maintain spatial translational invariance. 
The solutions of this paper may prove to be a useful starting point 
in describing the response of the field theory to a forcing function that 
breaks this symmetry, provided the scale of spatial variation of the forcing 
function is large compared to the inverse temperature of the black brane 
that is set up in our solutions. Consider, for example, the Einstein dilaton 
system studied in section \ref{transdil} perturbed by the non normalizable 
part of a small amplitude dilaton field that takes the form 
$\phi_0(v, {\vec x})$. Let us further assume that the length scales for 
spatial variation $L({\vec x})$ in $\phi_0$ are all large compared to 
$\frac{\delta t({\vec x})}
{\epsilon^\frac{2}{3}}$ (the inverse of the temperature of the black brane
that is eventually formed). As $\epsilon \ll 1$ this implies, in particular, 
that $L({\vec x}) \gg \delta t({\vec x})$. We expect the resultant 
thermalization process to be described by a dual metric of the form  
\begin{equation}\label{derexp}  
ds^2=2 dr dv -\left(r^2-\frac{M(v, {\vec x})}{r^{d-3}}\right)dv^2+r^2 dx_i^2 .
\end{equation}
where 
\begin{equation}\label{eninpl} \begin{split}
M(v, {\vec x})& =C_2(v, {\vec x} )+ {\cal O}(\epsilon^4) \\
C_2(v, {\vec x})&= 
-\frac{1}{2}\int_{-\infty}^v dt {\dot \phi_0(t, {\vec x}) } {\dddot \phi_0}(t, 
{\vec x}) \\
C_2({\vec x})&=
\frac{1}{2}\int_{-\infty}^\infty dt {\ddot \phi_0(t, {\vec x}) } {\ddot \phi_0}(t, 
{\vec x}) \\
\end{split}
\end{equation}
i.e. to be approximated tubewise by the solutions described in this paper. 
The metric \eqref{derexp} will then be corrected in a power series expansion
in two variables; $\epsilon$ (as in this paper) and a spatial derivative 
expansion weighted by $\frac{\delta t}{L \epsilon^\frac{2}{3}}$. The last 
expansion should reduce to the fluid dynamical expansion at late times.
Indeed, at $t=\delta t$, the metric described in \eqref{derexp} is dual to a locally thermalized conformal fluid, everywhere at rest, but with a space varying energy density $C_2({\vec x})$. The evolution of this fluid after $v=\delta t$ will 
simply be governed by the Navier Stokes equations of fluid dynamics; the metric
dual to the corresponding flow was determined in 
\cite{Bhattacharyya:2008jc, Baier:2007ix, 
VanRaamsdonk:2008fp, Loganayagam:2008is, Bhattacharyya:2008xc, Dutta:2008gf, 
Bhattacharyya:2008ji, Haack:2008cp, Bhattacharyya:2008mz, Erdmenger:2008rm, 
Bhattacharyya:2008mz, Fouxon:2008tb,  Bhattacharyya:2008kq, 
Haack:2008xx, Gupta:2008th, Hansen:2008tq, Fouxon:2008ik, 
Kanitscheider:2009as, David:2009np, Springer:2009wj}. Provided we can 
solve the relevant fluid dynamical equations,  we have a complete
description of the evolution of our spacetime for all $v$. 

The gravitational solutions presented in this paper appear to be qualitatively
different, in several ways, for odd and even $d$ (see Appendix \ref{arbdim}). 
This suggests that the equilibration process at strong coupling is 
qualitatively different in odd and even dimensional field theories. 
At leading order in amplitudes, equilibration takes place faster in 
in field theories with odd spacetime dimensions as compared to their even
dimensional counterpart. It would be interesting to find a direct field theory
explanation of this fact. 

In this paper we have investigated the response of an $AdS$ space to a 
marginal, non normalizable deformation of small amplitude. It would be 
natural to extend our work to determine the response of the same space 
to a relevant or irrelevant non normalizable marginal deformation of 
small amplitude. More ambitiously, one could also hope to explore the 
response of the system to large amplitude deformations, perhaps using 
a combination of analytic and numerical techniques (see \cite{Chesler:2008hg}). 
We leave these issues to the future.

In this paper we have constructed several solutions to 
bulk equations in a perturbative expansion. It is natural to wonder whether 
the dynamical processes we have constructed in this paper are 
stable to small fluctuations, when embedded into familiar 
examples of the $AdS/CFT$ correspondence. We will not address this question 
in detail in this paper; in these paragraphs we simply address the question 
of when the end point of the collapse processes, studied in this paper, 
are stable. 

In this paper we have described time evolutions that end up in big black holes, small black holes (big and small compared to the $AdS$ radius) and a thermal 
gas in $AdS$. To the best of our knowledge, large uncharged 
$AdS$ black holes are stable solutions in every familiar example of the
$AdS/CFT$ correspondence. The $AdS$ thermal gas has a potential instability, 
the  Jeans instability, which is triggered at energies at or larger than a 
critical density of order unity \cite{Page:1985em}, in the units of our paper. However the 
collapse situations described in this paper, that end up in a 
thermal gas, do so at energies of order $\frac{\epsilon^2}{x^d} 
\ll \epsilon^\frac{d-2}{d-1} \ll 1$. We conclude that the thermal gases 
produced as the end point of collapse, in our paper, are also stable.

Small black holes in $AdS_{d+1} \times X$, on the other hand, are usually 
unstable to a Gregory Laflamme type clumping on the internal 
manifold $X$ \cite{Hubeny:2002xn}. Consequently, the evolutions leading 
to small $AdS$ black holes, constructed in this paper, are necessarily 
unstable when embedded into familiar examples of the $AdS/CFT$ 
correspondence\footnote{We thank O. Aharony and B. Kol for discussions on this 
point.}. Note, however, that the time scale associated with this Gregory 
Laflamme instability  is $R^2 r_H$ where $r_H$ is the Schwarszchild radius of 
the small black hole. Assuming $R r_H  \ll 1$ (so that the black hole 
that is formed is really small), $\frac{\delta t}{r_H R^2}
\sim \left(\frac{x^{d-1}}{\epsilon} \right)^\frac{2}{d-2}$. It follows that, 
in the limit $x^{d-1} \ll \epsilon$, studied in this paper, the black hole 
formation processes discussed in this paper occur over a time scale much 
smaller than that of Gregory Laflamme instability. In other words 
the thermalization to small black holes (in the perturbative regime described
in this paper) is described by a two stage process. In the first stage 
the solution is well described by the Vaidya metric (plus corrections) 
described in this paper. The second stage describes the evolution of an almost 
completely formed small $AdS$ black hole perturbed by the 
Gregory Laflamme instability. This black hole will then undergo the 
Gregory Laflamme type transition in the usual manner. In other words the 
perturbative solutions presented in this paper correctly describe the 
process of small black hole formation even when embedded in $AdS/CFT$ type 
situations in which this black hole is unstable. 

As we have explained above, a CFT on the sphere can respond to a forcing 
function by settling down into either of its two available phases. It appears
that the space of possible forcings is divided into two regions, by 
a critical surface of unit codimension. On either side of this critical 
surface, the forcing drives the system into different phases. This critical 
surface occurs at $x^{d-1} \sim \epsilon^2$. Ignoring the issue of the 
Gregory Laflamme instability for a moment, the local (small $r$) form 
of this gravitational solution precisely on this critical surface is simply 
Choptuik's critical solution in $R^{d-1, 1}$, which is known to display 
surprisingly robust universal behavior characterized by a universal, 
nakedly singular solution and universal critical exponents. Note, however, 
that the Gregory Laflamme instability generically 
cannot be ignored in the neighbourhood 
of the critical surface. In the neighbourhood of this transition 
black holes that are formed near criticality have arbitrarily small 
Schwarzscild radius, and so trigger the Gregory Laflamme instability over 
very short time scales. Consequently, in order to access the universal 
$R^{d-1,1}$ Choptuik behavior in field theory one would 
have to tune initial data with exponential accuracy.
 
It is of course true on general grounds that the solutions described in 
this paper fall into two classes distinguished by an order 
parameter ( the presence of a horizon at late times). An important question
about the transition between these two behaviours is whether it is 
continuous or discontinuous over time scales small compared to 
$\frac{1}{R}$. In situations in which Gregory Laflamme 
instability occurs there is a very special submanifold of this transition 
manifold; the submanifold on which Gregory Laflamme instabilities are 
precisely tuned away. On this submanifold we know that the relevant 
gravitational solutions are those of Choptuik collapse in $R^{d-1,1}$, 
and so are continuous (second order) and singular. We do not konw if this
singular second order behaviour persists away from this special point. 
The investigation of these issues, as well as the study of 
the smoothing out of singularities on this manifold by finite 
$N$ fluctuations, is potentially interesting area of future research.

\subsection*{Acknowledgements}

We would like especially to thank O. Aharony, B. Kol and S. Trivedi for 
several interesting discussions. We would also like to thank 
J.Bhattacharya, S. Dutta, 
R. Loganayagam, P. Joshi, G. Mandal, S. Raju and 
S. Wadia and all the participants of `String Theory and Fundamental Physics' 
at the Kanha Tiger sanctuary for useful discussions. 
We also thank O. Aharony, J. Bhattacharya, S. Das, G. Horowitz, V. Hubeny, B. Kol, R. Loganayagam, G. Mandal, 
T. Wiseman, M. Rangamani for detailed comments on a preliminary version 
of this manuscript. We thank H. Antia and M. Mahato
for helping in the numerical solution of the differential equation 
in section \ref{transdil}. We would also like to 
acknowledge useful discussion with the students in the TIFR theory room. 
The work of S.M. was supported in part by Swarnajayanti Fellowship. We would 
all also like to acknowledge our debt to the people of India for their 
generous and steady support to research in the basic sciences.


\section{Appendices}
\appendix


\section{Translationally invariant graviton collapse}\label{grav}

In sections \ref{transdil} and \ref{global} above we have studied the collapse
triggered by a minimally coupled scalar wave in an asymptotically $AdS$ 
background. Our study was, in large part, motivated by potential applications
to CFT dynamics via the $AdS/CFT$ correspondence. From this 
point of view the starting point of our analyses in e.g. section \ref{transdil}
has a drawback as not every bulk system that arises 
in the study of the $AdS/CFT$ correspondence, admits a consistent truncation 
to the theory of gravity coupled to a minimally coupled massless scalar field. 

On the other hand, every two derivative theory of gravity that admits $AdS$ 
space as a solution admits a consistent truncation to Einstein gravity 
with a negative cosmological constant. Consequently, any results that 
may be derived using the graviton instead of dilaton waves, applies 
universally to all examples of the AdS/CFT correspondence with two derivative
gravity duals. In this section we study a situation very analogous to 
the set up of section \ref{transdil}, with, however, a transverse graviton 
playing the place of the dilaton field of section \ref{transdil}. All the 
calculations of this section apply universally to any CFT that admits a 
two derivative gravitational dual. 

While the equations that describe the propagation of gravity waves are more 
complicated in detail than those that describe the propagation of a massless 
minimally coupled scalar field, it turns out that the final results of the 
calculations presented in this subsection are extremely similar 
to those of section \ref{transdil}. We take this to suggest that 
all the qualitative results of sections \ref{transdil} and \ref{global} would 
continue to qualitatively apply to the most general 
approximately translationally invariant gravitational 
perturbations of Poincare Patch $AdS$ space or approximately spherically 
symmetric gravitational perturbation of global $AdS$ space. If this 
guess is correct, it suggests that the qualitative lessons 
learnt in this paper have a wide degree of applicability. 

In this section we restrict our attention to the simplest dimension 
$d=3$. It should we possible, with some additional effort, to extend our 
results at least to all odd $d$, and also to work out the corresponding 
results for even $d$. We leave this extension to future work.

The set up of this Appendix is very closely analogous to that employed by 
Yaffe and Chesler in \cite{Chesler:2008hg}. The main differences are as follows. 
Yaffe and Chesler worked in $d=4$; they numerically studied the effect of 
a specific large amplitude non normalizable deformation on the gravitational 
bulk. We work in $d=3$, and analytically study the the effect of the 
arbitrary small amplitude deformation on the gravitational bulk.

\subsection{The set up and summary of results}

In this section we study solutions to pure Einstein 
gravity with a negative cosmological constant. We study solutions 
that preserve an $R^2 \times Z_2 \times Z_2$ symmetry. Here $R^2$ denotes the 
symmetry of translations in spatial field theory directions, while the two 
$Z_2$s respectively denote the spatial parity flip and the discrete
exchange symmetry between the two Cartesian spatial boundary coordinates 
$x$ and $y$. 

As in section \ref{transdil}, our symmetry requirements determine our metric 
up to three unknown functions of $v$ and $r$. With the same choice of 
gauge as in section \ref{transdil}, our metric takes the form
\begin{equation}
 ds^2 = -2~dv~dr +  g(r,v)~dv^2 + f^2(r,v) (dx^2 + dy^2) + 2 r^2 h(r,v) dx~dy
\end{equation}
The boundary conditions on all fields are given by \eqref{bcsm}
under the replacement 
$\phi(r,v) \rightarrow h(r,v)$ and $\phi_0(v) \rightarrow h_0(v)$.  
Here $h_0(v)$ gives the boundary conditions on the off diagonal mode, 
$g_{xy}$, of the boundary metric. $h_0(v)$ is 
taken to be of order $\epsilon$. Physically, our boundary conditions set up a graviton wave, with polarization 
parallel to the spatial directions of the brane.

As in section \ref{transdil}, in order to solve Einstein's equations with 
the symmetries above, it turns 
out to be sufficient to solve the three equations $E_C^2, E_C^1$ and 
$E_{xy}$ (see \eqref{eqnset}) (plus the energy conservation condition 
$r E_{ec}$ at one $r$). 

As in section \ref{transdil} it is possible to solve these equations 
order by order in $\epsilon$. We present our solution later in this section. 
To end this subsection, 
we list the principal qualitative results of this section. We are able 
to show that

\begin{itemize}
\item The boundary conditions described above result in black brane 
formation for an arbitrary (small amplitude) source function $h_0(v)$. 
\item Outside the event horizon of our spacetime, we find an explicit 
analytic form for the metric as a function of $h_0(v)$. Our metric 
is accurate at leading order in the $\epsilon$ expansion, and takes 
the Vaidya form \eqref{lo} with a mass function 
\begin{equation}\label{consq}
M(v)= - \frac{1}{2}   \int_{-\infty}^v dt {\dot h}_0 {\dddot h_0}
\end{equation}
\item  In particular, we find that the energy density of resultant 
black brane is given by 
\begin{equation}\label{ctgrav}
M \approx -E_2= \frac{1}{2} \int_{-\infty}^\infty dt {\ddot h_0 }^2 
\end{equation}
Note that $E_2\sim \frac{\epsilon^2}{(\delta t)^3}$.
\item As this leading order metric is of the same form as that in the 
previous subsection, the analysis of the event horizons presented above 
continues to apply. In particular it follows that singularities 
formed in the process of black brane formation are always shielded by a 
regular event horizon at small $\epsilon$. 
\item  Going beyond leading order, perturbation theory in the amplitude 
$\epsilon$ yields systematic corrections to this metric at higher orders 
in $\epsilon$. We unravel the structure of this perturbation expansion in 
detail and work out this perturbation theory explicitly to fifth order at small 
times.  
\end{itemize}

\subsection{The energy conservation equation}

As we have explained above, the equations of motion for our system 
include the energy conservation relation, in addition to the one 
dynamical and two constraint equations. The form of the dynamical and 
constraint equations is easily determined using Mathematica-6; these 
equations turn out to be rather lengthy and we do not present them here. 
In this section we content ourselves with presenting 
an explicit form for the energy conservation 
equation.
As in section \ref{transdil}, it is possible to solve for the functions 
$\frac{f}{r}$, $\frac{g}{r^2}$ and $h$ in a power series in $\frac{1}{r}$. 
This solution is simply the Graham Fefferman expansion. To order 
$\frac{1}{r^3}$ (relative to the leading result) we find 
\begin{equation}\label{gfgr}\begin{split}
f(r,v)&= r\left(1 + \frac{\frac{\left[ {\dot h}_0\right]^2}
{8\left(1 - h_0^2\right)} }{r^2} + \frac{\frac{1}{2}h_0\sigma(v)}{r^3}  
+ \cdots\right)\\
 g(r,v)&= r^2\left(1 + \frac{\frac{1}{4\left(-1 + h_0^2\right)^2}\bigg[{\left(1 + 3 h_0^2\right)}\left[{\dot h}_0 \right]^2-4 h_0 \left(-1 + h_0^2\right) 
\partial_v^2 h_0 \bigg]}{r^2} - \frac{M(v)}{r^3}  + \cdots\right)\\
h(r,v)&= \left(h_0 + \frac{ {\dot h_0}}{r} + \frac{\frac{h_0 
{\dot h_0}^2 }{4\left(-1 + h_0^2\right)}}{r^2} + \frac{\sigma(v)}{r^3} + \cdots\right)\\
\end{split}
\end{equation}
where the parameters $M$ and $\sigma$ are constrained by the energy 
conservation equation
\begin{equation}\label{conserv}
\begin{split}
{\dot M} &= -\frac{ {\dot h}_0}{2\left(-1 + h_0^2\right)^4}
\bigg[+ 3 M(v) h_0\left(-1 + h_0^2\right)^3
-3\left(-1 + h_0^2\right)^3 \sigma \\
 &- 4\left(-1 + h_0^2\right) h_0  {\dot h}_0 \partial_v^2h_0
 + \left(-1 + h_0^2\right)^2 \partial_v^3h_0 + \left(1 + 3 h_0^2\right) 
\left[ {\dot h}_0\right]^3\bigg]
\end{split}
\end{equation}
\footnote{The stress tensor is given by
\begin{equation}\label{sttensor}
 \begin{split}
  T_{tt} &= M\\
T_{xx}&=T_{yy}= -\frac{M}{2}\\
T_{xy}&=-\frac{1}{2\left(-1 + h_0^2\right)^3}\bigg[-3\left(-1 + h_0^2\right)^3 
\sigma(v) - 4\left(-1 + h_0^2\right) h_0  {\dot h}_0 \partial_v^2 
h_0\\
& + \left(-1 + h_0^2\right)^2 h_0^3 + \left(1 + 3 h_0^2\right)\left[ {\dot h}_0\right]^3\bigg]
 \end{split}
\end{equation} 
Using these relations, it may be verified that 
\eqref{conserv} is simply a statement of the conservation 
of the stress tensor.}

In the perturbative solution we list below, we will find that 
$\sigma \sim {\cal O}(\epsilon^3)$. It follows that, to order 
${\cal O}(\epsilon^2)$, the function $M(v)$ is given by \eqref{consq}. 

\subsection{Structure of the amplitude expansion}
As in subsection \ref{transdil} we set up a naive amplitude expansion 
by  formally replacing $h_0$ with $\epsilon h_0$ 
and then solving our equations in a power series in $\epsilon$. We expand 
\begin{equation}\label{ampexpg} \begin{split}
f(r,v)&=\sum_{n=0}^\infty \epsilon^{n} f_{n}(r,v) \\
g(r,v)&= \sum_{n=0}^\infty  \epsilon^{n} g_{n}(r, v)  \\
h(r,v)&= \sum_{n=0}^\infty \epsilon^{n} h_{n}(r,v) \\
\end{split}
\end{equation}
with 
\begin{equation}\label{zerordgrav}
f_0(r,v)=r, ~~~~ g_0(r,v)=r^2,~~~ h_0(r,v)=0.
\end{equation}
The formal structure of this expansion is identical to that described 
in section \ref{transdilampexp}; in particular $f_n$ and $g_n$ are nonzero 
only for even $n$  while $h_n$ is nonzero only for odd $n$. 
 At first order we find   
\begin{equation}\label{phiograv}
h_1(r,v)=h_0(r,v)+ \frac{{\dot h_0(r,v)}}{r}
\end{equation}
which then leads to simple expressions (see below) for 
$f_2$ and $g_2$. In particular 
$h_1$ and $f_2$ vanish for $v \geq \delta t$ while $g_2=M/r$ for 
$v \geq \delta t$.

Turning to higher orders in the perturbative expansion, it is 
possible to inductively demonstrate that for $n \geq 1$ 
\begin{itemize} 
\item{1.} The functions $h_n$, $g_n$ and 
$f_n$ have the following analytic structure in the variable $r$
\begin{equation}\label{art}\begin{split}
h_{2n+1}(r, v)&=\sum_{k=2}^{2n+1} \frac{\phi_n^k(v)}{r^k}\\
f_{2n+2}(r,v)&=r \sum_{k=2}^{2n+2} \frac{f_n^k(v)}{r^k}\\
g_{2n+2}(r,v)&=r \sum_{k=1}^n \frac{g_n^k(v)}{r^k}\\
\end{split}
\end{equation}
\item{2.} The functions $h_{2n+1}^k(v)$, $f_{2n+2}^k(v)$ and $g_{2n+2}^k(v)$ 
are each functionals of $h_0(v)$ that scale like $\lambda^{-k}$ under 
the scaling $v \rightarrow \lambda v$.
\item{3.} For $v>\delta t$ 
these functions are all polynomials in $v$ of a degree 
that grows with $n$. For example, the degree of $h_{2n+1}^k$ is of 
at most $3n-k$.  
\end{itemize}

As in the section \ref{transdil}, this structure ensures that 
naive perturbation theory is good for times $v \ll M^\frac{1}{3}$, but 
fails for later times. As in section \eqref{transdil}, the correct perturbative
expansion uses the Vaidya metric \eqref{lo} as the zero order solution.

\subsection{Explicit results up to 5th order}

At leading order we have
\begin{equation}\label{ao}
\begin{split}
 h_1(r,v) &= h_0(v) + \frac{{\dot h}_0}{r}\\
f_2(r,v)&=\frac{\left[{\dot h}_0\right]^2}{8 r}\\
g_2(r,v) &= \frac{E_2(v)}{r}+\frac{1}{4}\left[{\dot h}_0\right]^2 +{\dot h}_0
\partial_v^2 h_0\\
\end{split}
\end{equation}

At next order 
\begin{equation}\label{at} \begin{split}
h_3(r,v)&=\frac{1}{4r^3}\left\{\int_{-\infty}^v E_2(x)\partial_x h_0~dx - r~ h_0 \left[ {\dot h}_0\right]^2\right\}\\
\\
f_4(r,v)&=\frac{h_0^2(v)\left[ {\dot h}_0\right]^2}{8r} + \frac{D(v)h_0(v)}{8 r^2} - \frac{ {\dot h}_0}{128 r^3}\left(-12 D(v) + \left[ {\dot h}_0\right]^3\right) \\
g_4(r,v)&=\frac{E4(v)}{r} + \frac{5}{4}h_0(v)^2\left[ {\dot h}_0\right]^2 + h_0(v)^3 \partial_v^2 h0\\
&+ \frac{ {\dot h}_0}{8 r^2}\bigg[D(v) + 4 E_2(v) h_0(v)\bigg]  + \frac{1}{16 r^3}\left(E_2(v)\left[ {\dot h}_0\right]^2 + D(v) \partial_v^2 h_0\right) \\
h_4(r,v)&=0\\
\text{where}~~& D(v) = \int_{-\infty}^v E_2(x) \partial_x h_0~dx\\
\end{split}
\end{equation}

Finally at the next order 
\begin{equation}\label{ath} \begin{split}
h_5(r,v)&=\frac{D_1(v)}{2 r^2} \\
&+\frac{1}{24 r^3}\bigg[6\int_{-\infty}^v D_2(x)~dx 
+ 5\left\{\int^v_{-\infty}dz\int^z_{-\infty}dy\int^y_{-\infty}D_4(x)~dx\right\}\\
&~~~~~~~~~~~ + 4\left\{ \int^v_{-\infty}dy\int^y_{-\infty}D_3(x)~dx\right\}\bigg]\\
&+\frac{1}{r^4}\bigg[\frac{5}{24}\left\{\int^v_{-\infty}dy\int^y_{-\infty}D_4(x)~dx\right\} + \frac{1}{6}\left\{\int^v_{-\infty}D_3(x)~dx\right\}\bigg]\\
& + \frac{1}{8 r^5}\left[\int^v_{-\infty}D_4(x)~dx\right]
\end{split}
\end{equation}
where
\begin{equation}\label{defla} \begin{split}
D_1(x)&=-h_0(x)^3\left[\partial_x h_0\right]^2\\
D_2(x)&=E_4(x) \partial_x h_0 + \frac{1}{4}D(x) h_0(x) \partial_x h_0 + E_2(x)h_0(x)^2 \partial_x h_0\\
D_3(x)&=\frac{1}{8}\bigg[5 D(x)\left[\partial_x h_0\right]^2 + 15 E_2(x) h_0(x) \left[\partial_x h_0\right]^2 + 15 D(x) h_0(x) \partial_x^2h_0\bigg]\\
D_4(x)&= \frac{1}{8}\bigg[18 D(x) E_2(x) + 5 E_2(x) \left[\partial_x h_0\right]^3 +7 D(x) h_0(x) \partial_x^2h_0\bigg]
\end{split}
\end{equation}
and (this follows from energy conservation) 
\begin{equation}\label{gravconser}
 \begin{split}
  {\dot E_2} &=\frac{1}{2}{\dot h}_0 \partial_v^3h_0\\
{\dot E_4} &=\frac{3}{8}D(v){\dot h}_0 + \frac{{\dot h}_0}{2}\bigg[3 E_2(v)h_0(v) + \left[{\dot h}_0\right]^3 + 4 h_0(v) {\dot h}_0 \partial_v^2 h_0 
+ 2 h_0^2 \partial_v^3 h_0 \bigg]
 \end{split}
\end{equation}

It follows in particular that the the `initial' condition for normalizable 
evolution at $v=\delta t$ is given, to leading order,  by 
\begin{equation}\label{incondgrav}
h(r, \delta t)=\frac{1}{8r^3}\int_{-\infty}^v \left(\int_{-\infty}^x dy
\left( \partial_y h_0 \partial_y^3 h_0 \right) 
\partial_x h_0(x) dx \right)
\end{equation}
This initial condition is of order $\frac{\epsilon^3}{(\delta t)^3 r^3}$
i.e. of order $\frac{\epsilon}{{\tilde r}^3}$ where 
${\tilde r}=\frac{r}{E_2}$. This demonstrates that, for $v>\delta t$, 
our solution is a small perturbation about the black brane of energy 
density $E_2$.  

\subsection{Late Times Resummed perturbation theory}

To leading order, the initial condition for the normalizable evolution of 
resummed perturbation theory for the field $h(r, v)$ is given by 
$$h(\delta t)=\frac{1}{4r^3}\left(\int_{-\infty}^{\delta t} E_2(x)\partial_x h_0~dx \right) \equiv \frac{h^{0}_3(\delta t)}{r^3}$$
Now, at the linearized level the equation of motion for the function $h$ is 
simply the minimally coupled scalar equation. It follows that the subsequent 
evolution of the field $h$ is simply given by 
\begin{equation}\label{logth}
h= \frac{h^0_3(\delta t)}{M} \psi( \frac{r}{M^{\frac{1}{3}}}, 
(v-\delta t)M^{\frac{1}{3}})
\end{equation}
where the universal function $\psi$ was defined in section \ref{transdil}.
As in section \ref{transdil}, this perturbation is small initially, and at 
all subsequent times, justifying the resummed perturbation procedure.

\section{Generalization to Arbitrary Dimension}\label{arb}

\subsection{Translationally Invariant Scalar Collapse in Arbitrary Dimension}
\label{arbdim}

In this subsection we will investigate how the results of section 
\ref{transdil}, which were worked out for the special case $d=3$, generalize
to $d \geq 3$. The mathematical problem 
we will investigate in this Appendix was already 
set up in general $d$ in subsection \ref{transdilsetup}. 
It turns out that the dynamical details of collapse processes in odd and 
even dynamics are substantially different, so we will deal with those two 
cases separately.

\subsubsection{Odd $d$}

The general structure of the solutions that describe 
collapse in odd $d \geq 5$ is similar in many ways to the solution reported
in section \ref{transdil}. The energy conservation 
equations may be studied via a large $r$ Graham Fefferman expansion 
closely analogous to that 
described in section \ref{transdil}. The functions 
$\phi$ $f$ and $g$  may be expanded at large $r$ as 
\begin{equation} \label{strucfgpo} \begin{split}
\phi(r,v)&=\sum_{n=0}^\infty \frac{A_\phi^n(v)}{r^n}\\
f(r,v)&=r \left(\sum_{n=0}^\infty \frac{A_f^n(v)}{r^n}  \right)\\
g(r,v)&=r^2 \left(\sum_{n=0}^\infty \frac{A_g^n(v)}{r^n} \right)\\
\end{split}
\end{equation}
For $n\leq d-1$ the equations of motion locally determine 
$A_\phi^n(v)$, $A_f^n(v)$ and $A_g^n(v)$ in terms of $\phi_0(v)$. Each 
of these functions is a local expression (of $n^{th}$ order in $v$ derivatives)
of $\phi_0(v)$. However local analysis does not determine  
$A_g^d(v) \equiv M(v)$ and $A_\phi^d(v) \equiv L(v)$ in terms of 
$\phi_0(v)$. $M(v)$ and $L(v)$ are however constrained to obey 
an energy conservation equation that takes the form 
\begin{equation}\label{encong}
{\dot M}= k {\dot \phi} L(v) +{\rm local}
\end{equation}
where $k$ is a constant and `local' represents the a set of terms built out 
of products of derivatives of $\phi_0(v)$ that we will return to below. 
 As in $d=3$, $L(v)= {\cal O}(\epsilon^3)$, so the first term in 
\eqref{encong} does not contribute at lowest order of the amplitude 
expansion of interest to this paper. The local terms in this equation
\eqref{encong} are easily worked out at lowest order, 
${\cal O}(\epsilon^2)$, in the amplitude expansion, and we find
$M(v)=C_2(v) +{\cal O}(\epsilon^4)$ with  
\begin{equation}\label{massgend}
\begin{split}
C_2(v)&= -\frac{2^{d-2}}{(d-2)} 
\left( \frac{(\frac{d-1}{2})!}{(d-1)!} \right)^2 
\int_{-\infty}^v dt\bigg[ \left(\partial_t^\frac{d+3}{2}\phi_0\right)
\left(\partial_t^{\frac{d-1}{2}} \phi_0\right)
-\frac{d-3}{d-1} 
\left( \partial_t^{\frac{d+1}{2}}\phi_0
\right)^2 \bigg]
\end{split}
\end{equation} 

\begin{equation}\label{ct}
C_2=\frac{2^{d-1}}{(d-1)} 
\left( \frac{(\frac{d-1}{2})!}{(d-1)!} \right)^2 
\int_{-\infty}^\infty dt
\left( \partial_t^{\frac{d+1}{2}}\phi_0(t)
\right)^2 \sim \frac{\epsilon^2}{(\delta t)^d},
\end{equation}
the generalization of \eqref{eninp} and \eqref{eninpc} 
to arbitrary odd $d$. \eqref{ct} gives the leading order expression for the 
mass of the black brane that is eventually formed at the end of the 
thermalization process. 

Let us now turn to the naive amplitude expansion in arbitrary odd $d$. 
The first term in this expansion, $\phi_1$ is easily determined and  we find 
\begin{equation}\label{phiod}
\phi_1(r,v)= \sum_{k=0}^{\frac{d-1}{2}} \frac{2^k}{k!} 
\frac{ \left(\frac{d-1}{2}
\right)!}{(d-1)!} \frac{(d-1-k)!}{(\frac{d-1-2k}{2})!}
\frac{\partial_v^k \phi_0}{r^k} 
\end{equation}
Equations \eqref{maineqs} then immediately determine $f_2$ and $g_2$. 
Turning to higher orders, it is possible to demonstrate that

\begin{itemize} 
\item{1.} The  functions $\phi_{2n+1}$, $g_{2n}$ and 
$f_{2n}$ have the following analytic structure in the variable $r$
\begin{equation}\label{ar}\begin{split}
\phi_{2n+1}(r, v)&=\sum_{k=0}^{\frac{(2n+1)(d-1)}{2}-p(n)}
\frac{\phi_{2n+1}^k(v)}{r^{\frac{(2n+1)(d-1)}{2} -k}}\\
f_{2n}(r,v)&=r \left(\sum_{k=0}^{n(d-1)-f(n)} \frac{f_{2n}^k(v)}{r^{n(d-1)-k}}
\right)\\
g_{2n}(r,v)&=- \frac{C_{2n}(v)}{r^{d-2}}+ 
 r \left( \sum_{k=0}^{n(d-1)-g(n)} \frac{g_{2n}^k(v)}{r^{n(d-1)-k}} \right)\\
\end{split}
\end{equation}
where 
$$p(n)=d, ~~~(2n+1 \geq d),~~~ p(n)=2n+1~~~(2n+1 \leq d),$$ 
$$f(n)=d, ~~~(2n \geq d),~~~ f(n)=2n~~~(2n \leq d),$$ 
$$g(n)=d-1, ~~~(2n \geq d-1),~~~ g(n)=2n-1~~~(2n \leq d).$$
\item{2.} The functions $\phi_{2n+1}^k(v)$, $f_{2n}^k(v)$ and 
$g_{2n}^k(v)$ 
are each functionals of $\phi_0(v)$ that scale like $\lambda^{-k}$ under 
the scaling $v \rightarrow \lambda v$.
\item{3.} For $v>\delta t$ $f_2=f_4=0$, $g_2= -\frac{C_2}{r^{d-2}}$ and $g_4=\frac{-C_4}{r^{d-2}}$. Further, effectively, $p(n)=d$, 
 $f(n)=2d$ and $g(n)=2d-1$ 
for $v>\delta t$ (all additional terms present in \eqref{ar} vanish at these 
late times). Moreover the functions $\phi_{2n+1}^k(v)$, $f_{2n}^k(v)$ 
and $g_{2n}^k(v)$  are all polynomials in $v$ whose degrees are bounded 
from above by $n+k-1$, $n+k-3$ and $n+k-4$ 
respectively. 
\end{itemize}
 
As in $d=3$, the polynomial growth in $v$ of the coefficients of the naive 
perturbative expansion invalidates this expansion for large enough $v$.
More specifically, the sums over $k$ and $n$ in the expressions above 
are weighted by $r v$ and $\frac{\epsilon^2 v}{r^{d-1}}$ respectively. 
In the neighborhood of the horizon, $r\sim r_H\sim T \sim 
\frac{\epsilon^\frac{2}{d}}{\delta t}$ each of these sums is effectively 
weighted by the factor  $v T$. Consequently, naive perturbation theory 
fails at times large compared to the inverse temperature of the brane. 
At times of order $\delta t$ and for $r\sim r_H$
 the sum over $k$ and $n$ are each weighted 
effectively by $\epsilon^\frac{2}{d}$. More generally, naive perturbation 
theory is good at times of order $\delta t$ provided 
$r \delta t \gg \epsilon^{\frac{2}{d-1}}$, a condition that is 
satisfied at the event horizon.

As in $d=3$ the IR divergence of the naive perturbation expansion has a simple 
explanation. Even within the validity of the naive perturbation expansion, 
the spacetime is not well approximated by empty $AdS$ space, but rather 
by  the Vaidya metric \eqref{lo}. The naive expansion, which may be 
carried out with comparative ease up to $v=\delta t$, may be used to 
supply initial conditions for the subsequent  unforced normalizable evolution 
for resummed perturbation theory. For $v\geq \delta t$, the 
spacetime metric is given, to leading order,  by the Vaidya form 
\eqref{lo}, with $C_2(v)$ given by the constant $C_2$ listed in \eqref{ct}

Consequently, the spacetime metric for $v \geq \delta t$ is the black 
brane metric with temperature of order $\frac{\epsilon^{\frac{2}{d}}}
{\delta t}$, perturbed by a propagating $\phi$ field and consequent 
spacetime ripples. The initial conditions at $v=\delta t$,  
that determine these perturbations at later times,  are given to leading 
order in $\epsilon$ (read off from the most small $r$ singular term in 
$\phi_3$) as 
\begin{equation}\label{incondgend}
\begin{split}
 \phi(r,v) &= \frac{A}{r^{\frac{3(d-1)}{2}}}\\
\text{where}&\\
A &= \frac{(d-1)^2}{2(d-2)}\int_{-\infty}^\infty dt \bigg[(d-2)\left(2^{\frac{d-1}{2}}\frac{\left(\frac{d-1}{2}\right)!}{(d-1)!}\right) C_2(t)\left(\partial_t^\frac{d-1}{2}\phi_0\right)\\
 &-\left(2^{\frac{d-1}{2}}\frac{\left(\frac{d-1}{2}\right)!}{(d-1)!}\right)^3\left(\partial_t^\frac{d-1}{2}\phi_0\right)^2\left(\partial_t^\frac{d+1}{2}\phi_0\right)\bigg]
\end{split}
\end{equation}

In terms of the 
normalized variable $x=\frac{r}{M^{\frac{1}{d}}}$  and $y=v M^{\frac{1}{d}}$ this initial 
condition takes the form 
\begin{equation}\label{incd}
\phi(x) \sim \frac{\epsilon^{\frac{3}{d}}}
{x^{\frac{3(d-1)}{2}}}
\end{equation}

It follows that the solution at $v \geq \delta t$ is (in the appropriate 
$x,y$ coordinates) an order $\epsilon^{\frac{3}{d}}$ 
perturbation about the uniform black brane. The coefficient of this 
perturbation is bounded for all $y$, and decays exponentially
for large $y$ over a time scale of order unity in that variable. The explicit form of the solution for $\phi$, for 
$v>\delta t$, may be obtained in terms of a universal function,
$\psi_d(x,y)$ as in 
section \ref{transdil}.  The equation that we need to solve is 
\begin{equation} \label{gendeq}
\partial_x\left(x^{d+1}\left(1-\frac{1}{x^d} \right) \partial_x \psi_d \right)
+ 2 x^\frac{d-1}{2} \partial_x \partial_y \left( x^\frac{d-1}{2} \psi_d 
\right)=0
\end{equation}

\begin{figure}[here]
\centering
\includegraphics[scale=1.0]{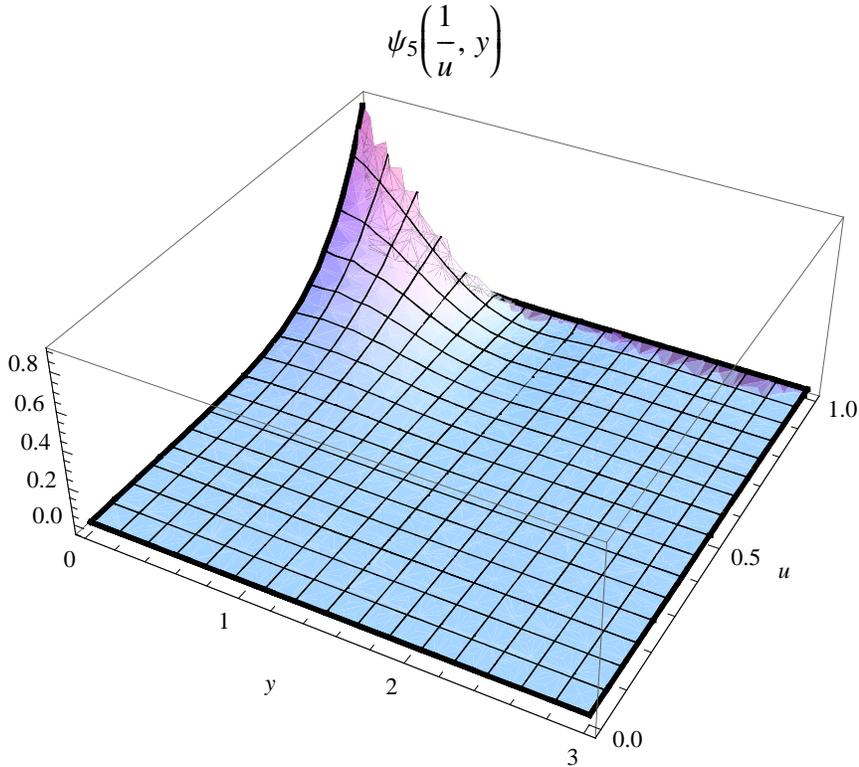}
\caption{Numerical solution for the dilaton at late time in $d = 5$}
\label{fig:plot6}
\end{figure}

As in section \ref{transdil}, this universal function
appears to be difficult to obtain analytically, but is easily 
evaluated numerically.
As an example in Figure~\ref{fig:plot6} we present a numerical plot of this function in $d=5$. As in section \ref{transdil} we find it convenient to display 
the numerical output for the function $\psi_5(\frac{1}{x}, y)$ over the 
full exterior of the event horizon,  $u \in (0, 1)$.

\begin{figure}[here!]
\centering
\includegraphics[scale=1.0]{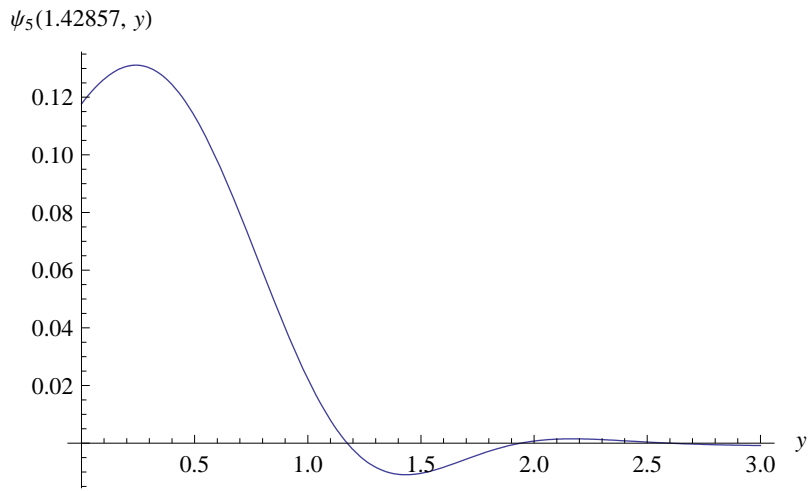}
\caption{A plot of $\psi_5(\frac{1}{0.7}, y)$ as a function of $y$}
\label{fig:pt6}
\end{figure}

\footnote{In order to obtain this plot, as in \ref{transdil}, we worked 
with the 
redefined field $\chi_5(u,y)= (1-u) \psi_5(\frac{1}{u}, y)$ and imposed 
Dirichlet boundary conditions 
on this field at $u=0$ and $u=0.999999$. We also imposed the initial conditions 
$\chi_5=(0.999999-u)u^6$. The Figure~\ref{fig:plot6} was outputted by Mathematica-6's partial differential equation solver, with a step size of 0.0005 and 
an accuracy goal of 0.001.  }
In figure \ref{fig:pt6} we present a 
graph of $\psi_5(\frac{1}{0.7}, y)$ 
(i.e. as a function of time at a fixed radial location)
Notice that this graph decays, roughly exponentially for $v>0.5$ and that 
this exponential decay is dressed with a sinusodial osciallation, as
expected for quasinormal type behavior. A very 
very rough estimate of this decay constant is provided by the equation 
$\omega_I$ using the equation 
$\frac{\psi_5(\frac{1}{0.7}, 1)}{\psi_5(\frac{1}{0.7}, .5)}= 
e^{-0.5 \omega_I}$ 
which gives $\omega_I \approx 8.2 T$ (here $T$ is the temperature of our 
black brane given by $T=\frac{4\pi}{5}$). This number is the same ballpark 
as the decay constants for the first quasi normal mode of the uniform
black brane reported in \cite{Horowitz:1999jd} (unfortunately those 
authors have not reported the precise numerical value for $d=5$) .

\subsubsection{Even $d$}\label{ed}

In our analyses above we have so far focused attention on odd $d$ (recall 
that $d$ is the spacetime dimension of the dual field theory). In this 
subsection we will study how our results generalize to even $d$. While all the broad qualitative conclusions of the odd $d$ 
analysis plausibly continue to apply, several intermediate details are quite 
different.The analysis of all equations is more difficult in even than 
in odd dimensions. In this appendix we aim only to initiate a serious analysis of these equations, and to carry this analysis far enough to have a plausible 
guess for the behavior of our system. We leave a systematic analysis of these 
equations to future work. 

The qualitative differences between even and odd $d$ show themselves already 
in the Graham Fefferman expansion. We illustrate this by working out this 
expansion in $d=4$. In this dimension the expansion of $f, g, \phi$ at large 
$r$ take the form

\begin{equation}\label{largerf}\begin{split}
f(r, v)&= r-\frac{({\dot \phi_0})^2}{12 r}-\frac{{\ddot \phi_0}{\dot \phi_0}}{36 r^2}+\frac{-3 ({\dot \phi_0})^4+2 {\dddot \phi_0} {\dot \phi_0}-(\partial_v^2 \phi _0)^2}{288 r^3}\\
&+\frac{-19 {\ddot \phi_0} ({\dot \phi_0})^3-1440 L(v) {\dot \phi_0}-18 \partial_v^4\phi_0 \partial_v\phi_0+45 {\ddot \phi_0}{\dddot \phi_0}}{21600 r^4}\\
&-\frac{\log (r) {\dot \phi_0} \left(\partial_v^4\phi_0-2 ({\dot \phi_0})^2 \partial_v^2 \phi _0\right)}{240 r^4} + \ldots \\
g(r, v)& =r^2-\frac{5}{12} ({\dot \phi_0})^2-\frac{M(v)}{r^2} + \frac{\log (r) \left(-({\dot \phi_0})^4+2 {\dddot \phi_0} {\dot \phi_0}-(\partial_v^2 \phi _0)^2\right)}{24 r^2} +\ldots \\
\phi(r,v)& =\phi _0+\frac{{\dot \phi_0}}{r}+\frac{ \partial_v^2 \phi _0}{4 r^2}+\frac{\frac{5}{36} ({\dot \phi_0})^3-\frac{1}{12}
   {\dddot \phi_0}}{r^3} +\frac{L}{r^4} + \ldots \\
&+ \frac{\log (r) \left(\partial_v^4\phi_0-2 ({\dot \phi_0})^2 \partial_v^2\phi _0\right)}{16 r^4}
\end{split}
\end{equation}
The energy conservation equation is 
\begin{equation} \label{eomf}
 {\dot M}=\frac{1}{144} \left(40 {\ddot \phi_0}({\dot \phi_0})^3-192 L(v) {\dot \phi_0}-17 \partial_v^4\phi_0 {\dot \phi_0}+6 
{\ddot \phi_0} {\dddot \phi_0}\right)
\end{equation}
and at quadratic order in $\epsilon$ we have 
\begin{equation}\label{mvfa} \begin{split}
M(v)&=C_2(v)+{\cal O}(\epsilon^4)\\
C_2(v)&=\frac{1}{144}\int_{-\infty}^v dt 
\left(-192 L(t) {\dot \phi_0}-17 {\ddddot \phi_0} {\dot \phi_0}+6 
\partial_t^2 \phi_0 {\dddot \phi_0}\right)\
\end{split}
\end{equation}

Unlike in even dimensions, it turns out that in odd dimensions $L(v)$ is 
nonzero at order $\epsilon$. This is fortunate, as all the local terms in 
\eqref{mvfa} are total derivatives, and so vanish when $v$ is taken to be 
larger than $\delta t$. The full contribution to the mass of the black brane 
that is eventually formed from our collapse process arises from the 
term in \eqref{mvfa} that is proportional to $L(v)$. As a consequence, 
the mass of the eventual black brane is not determined simply by Graham 
Fefferman analysis, but requires the details of the full dynamical process. 
These details may be worked out at lowest order in the $\epsilon$ expansion, 
(see below) and we will find  
\begin{equation}
 L(v) = \left(\frac{-7 + 12 \log 2}{192}\right)\partial_v^4 \phi_0 + \frac{1}{16}\int_{-\infty}^v dt \log(v-t)~\partial_t^5\phi_0(t) +{\cal O}(\epsilon^3)
\end{equation}
Plugging into \eqref{mvfa} we find that 
$C_2(v)$ reduces to the constant $C_2$ for $v>\delta t$, and we have 
\begin{equation}\label{cteven}
 C_2=-\frac{1}{12}\int_{-\infty}^\infty dt_1 dt_2 
\left( \partial_{t_1}^3\phi_0(t_1) \log(t_1 - t_2)\Theta(t_1 - t_2) \partial_{t_2}^3 \phi_0(t_2)\right)\
\end{equation}

Let us now turn to the amplitude expansion of our solutions. We will work 
this expansion out only at leading order; already the leading order 
solution turns out to have qualitative differences (and to be much harder 
to determine and manipulate) than the corresponding solution in odd $d$. 

Recall that $\phi_1$ \eqref{phiod} is extremely simple when 
$d$ was odd. To start with,  the 
solution is local in time, i.e. $\phi_1(r,v_0)$ is completely determined 
by the  value, and a finite number of derivatives, of $\phi_0(v_0)$. Relatedly
$\phi(r,v)$ has a very simple analytic expression in $r$; 
it is a polynomial in $\frac{1}{r}$ of degree $\frac{d-1}{2}$. 
In even $d$, on the other hand the dependence of  $\phi_1(r,v)$ on $\phi_0(v)$
is not local in time. Relatedly, the expansion of $\phi_1(r,v)$ in a power 
series in $\frac{1}{r}$ has terms of every order in $\frac{1}{r}$. Explicitly 
we find 
\begin{equation}\label{phio}
 \begin{split}
\phi_1(r,v)&= \int_{0}^\infty \partial_v^{d+1} \phi_0(v -t) \left( 
 \frac{ h(r t)}{r^d} \right) dt  \\
h(x)&=\int_0^x dy \frac{ \left(y(y+2) \right)^{\frac{d-1}{2}}}
{(d-1)!}\\
&= (-1)^{\frac{d}{2}}
{d\choose \frac{d}{2}} \frac{\theta}{2^d}
+ \frac{1}{2^{d-1}} \sum_{k=0}^{\frac{d}{2}-1}
\frac{(-1)^k}{d-2k} { d \choose k}  \sinh \left( (d-2k) \theta \right)\\
&\text{where} ~~\cosh \theta = 1+x
\end{split}
\end{equation}

Note that the  function $h(x)$ admits the following large $x$ expansion
\begin{equation}\label{hlx} \begin{split}
h(x)&= \frac{x^d}{(d-1)!}+ 
\sum_{k=1}^{d-1} \frac{x^{d-k}}{(d-k) k!(d-1)!} 
\left( \prod_{m=1}^k (d-2m-1) \right) \\
&+\frac{(-1)^{\frac{d}{2}+1} (d)!}{(d-1)! 2^d 
((\frac{d}{2})!)^2}\left(\sum_{p=0}^{\frac{d}{2}-1}\frac{1}{(d-2p)(d-2p-1)} 
\right)
+ \frac{(-1)^{\frac{d}{2}} (d)}{2^{d}\left(\frac{d}{2}! \right)^2 }
\ln (2x) +{\cal O}(\frac{\ln x}{x})
\end{split}
\end{equation}
The fact that $h(x)$ grows (rather than decays) with $x$ may cause the 
reader to worry that $\phi_(r,v)$ blows up at large $v$. That this is not 
the case may be seen by noting that $v^k \partial_v^{d+1}\phi_0$  may be 
rewritten as a sum of total derivatives when $k \leq d+1$ and so integrates 
to zero when  
$v>\delta t$ (in general it integrates to a simple local expression even for 
$v<\delta t$). Explicitly, plugging \eqref{hlx}   
into \eqref{phio} and integrating by parts we find that $\phi_1(r,v)$ has 
the following large $r t$ behavior  
\begin{equation}\label{phiolr} \begin{split}
\phi_1(r,v)& =\sum_{i=0}^d \frac{A_i(v)}{r^i} + \frac{ B(v) \ln(r)}{r^d}
+{\cal O}(\frac{\ln r}{r^{d+1}})\\
&=\phi_0(v)\\
&+ \sum_{k=1}^{d-1} \frac{\partial_v^k \phi_0(v)}{r^k}
\left[\frac{(d-k-1)!}{k!(d-1)!} 
\left( \prod_{m=1}^k (d-2m-1) \right) \right] \\
&+  \frac{\partial_v^d \phi_0(v)}{r^d}
\left[\frac{(-1)^{\frac{d}{2}+1} (d)!}{(d-1)! 2^d 
((\frac{d}{2})!)^2}\left(\sum_{p=0}^{\frac{d}{2}-1}\frac{1}{(d-2p)(d-2p-1)} 
\right) \right]\\
&+ \int_0^\infty dt \frac{\partial_v^{d+1} \phi_0(v-t)}{r^d}  \ln (2 r t ) 
\left[\frac{(-1)^{\frac{d}{2}} (d)}{2^{d}\left(\frac{d}{2}! \right)^2 }
\right]
\\
&+{\cal O}(\frac{\ln(r)}r^{d+1})
\end{split}
\end{equation}
(where the functions $A_i(v)$ and $B(v)$ are defined by this equation).
On the other hand at small $x$ we have 
\begin{equation}\label{smallxh}
h(x)=\frac{(2x)^{\frac{d+1}{2}}}{(d+1)(d-1)!} \left(1+{\cal O}(x) \right)
\end{equation}
from which it follows that 
\begin{equation}\label{phis}
\phi_1(r,v)= \frac{1}{r^\frac{d-1}{2}} \frac{1}{(d+1)(d-1)!}
\int_{-\infty}^v dt (2(v-t))^{\frac{d+1}{2}} \partial_t^{d+1} \phi_0(t)
+{\cal O}(\frac{1}{r^{\frac{d-3}{2}}}),
\end{equation}
an expression that is valid at small $r v$.
Note, in particular, that for $\delta t \ll v $,  \eqref{phis}
reduces to 
\begin{equation}\label{phiss}
\phi_1(r,v)= \frac{2^\frac{d+1}{2} \int_0^{\delta t} \phi_0(t) dt}
{r^\frac{d-1}{2} v^\frac{d+1}{2}} \frac{1}{(d+1)(d-1)!}
+{\cal O}(\frac{1}{r^{\frac{d-3}{2}}}) +{\cal O}(\frac{1}{t^{\frac{d+3}{2}}})
\end{equation}
In particular this formula determines the behavior of the field 
$\phi_1$ in the neighborhood of the event horizon $r_H\sim T$ for times
that are large compared to $\delta t$ but small compared to $T^{-1}$. 

The functions $f_2$ and $g_2$ are easily expressed in terms of 
the function $\phi_0$. We find 
\begin{equation}\label{fgt} \begin{split}
f_2(r,v)&=-\frac{1}{2(d-1)} \left[ r \int_r^\infty 
(\partial_\rho \phi_1)^2 d \rho - 
\int_r^\infty \rho^2 (\partial_\rho \phi_1)^2 d \rho \right]\\
g_2(r,v)&=-\left(2 \partial_v f_2(r,v) + (d-2)r f_2(r,v) + r^2\partial_rf_2(r,v)\right)\\
& +\frac{d(d-1)}{r^{d-2}}
\int_0^r \rho^{d-2} f_2(\rho, v) d \rho -\frac{D_2(v)}{r^{d-2}}
\end{split}
\end{equation}
The function $D_2(v)$ is determined by the requirement that the 
coefficient of $\frac{1}{r^{d-2}}$, in the large $r$ expansion 
of $g_2(r,v)$ is $-C_2(v)$ (see \eqref{mvfa}); in particular, 
for $v>\delta t$, $D_2(v)=C_2(v)$. At small $r$ and for $v>\delta t$
\begin{equation}\label{gt} \begin{split}
f_2(r,v)&=-\frac{ K^2(v)}{2(d-1)(d-2)(d-3) r^{d-2}} + {\cal O}(\frac{1}{r^{d-3}})\\ 
g_2(r,v)&=-\frac{C_2}{r^{d-2}}+\frac{\partial_v K^2(v)}{(d-1)(d-2)(d-3) r^{d-2}} +
{\cal O}(\frac{1}{r^{d-3}})\\
K(v)&=  \frac{1}{(d+1)(d-1)!}
\int_{-\infty}^v dt (2(v-t))^{\frac{d+1}{2}} \partial_t^{d+1} \phi_0(t)\\
&\approx  \frac{2^\frac{d+1}{2} \int_0^{\delta t} \phi_0(t) dt}
{ v^\frac{d+1}{2}} \frac{1}{(d+1)(d-1)!}~~~(v \gg \delta t)
\end{split}
\end{equation}

We would like to draw attention to several aspects of these results. 
First note that $\phi_1(r,v)$ is small provided 
$(r \delta t)^\frac{d-1}{2} \gg \epsilon$. Consequently, we expect
a perturbative analysis to correctly capture the dynamics of our situation 
over this range of coordinates; note that this is exactly the same estimate 
as for odd $d$. Next note that the maximal singularity, at small $r$, 
in the functions $f_2$ and $g_2$, are both of order $\frac{1}{r^{d-2}}$; this 
is the same as the maximal singularity in the analogous functions in odd 
$d$ (see the previous subsection). As the function $g_0(r,v)=r^2$, it follows, 
as in the previous function, that our spacetime metric is not uniformly well
approximated by the empty $AdS$ space over the full range of validity of 
perturbation theory. Over this entire range, however, it is well approximated
by a Vaidya type metric, where the mass function for this metric is given
at leading order by the coefficient of $-\frac{1}{r^{d-2}}$ in $g_2(r,v)$ 
above. 

Unlike the situation in odd dimensions, the leading order 
mass function $M(v)$,  in the effective Vaidya metric,  
is not given simply by $C_2(v)$. In particular, when 
$v \gg \delta t$ we have from \eqref{gt} that 
$$\frac{C_2-M(v)}{C_2}\sim \left(\frac{\delta t}{v} \right)^{d+2}.$$
In other words, the leading order metric for the thermalization process, 
in even $d$, is not given precisely by the metric of the uniform black
brane for $v>\delta t$. However it decays, in a power law fashion, 
 to the black brane metric at times larger than $\delta t$. As a consequence
at times $\delta t \ll v \ll T^{-1}$ the leading order metric that captures 
the thermalization process is arbitrarily well approximated by the metric 
of a uniform black brane. It follows that, while the spacetime described 
in this subsection does not capture the dual of instantaneous field theory 
thermalization (as was the case in odd $d$), it yields the dual of 
a thermalization process that occurs over the time scale of the forcing 
function rather than the much longer linear response time scale of the 
inverse temperature. 

We will not, in this paper, continue the perturbative expansion to higher 
orders in $\epsilon$. We suspect, however, that the computation of 
$\phi_3$ when carried through will yield a term proportional to
$\frac{\epsilon^3}{r^\frac{3(d-1)}{2}}$ that is constant in time. This term 
will dominate the decaying tail of $\phi_1(r,v)$ at a time intermediate between
$\delta t$ and $T^{-1}$ and will set the initial condition for the late 
time decay of the $\phi$ field (over time scale $T^{-1}$) 
as was the case in odd dimensions. It would be very interesting to 
verify or correct this guess.   

\subsection{Spherically Symmetric flat space collapse in arbitrary dimension}
\label{arbflat}

\subsubsection{Odd $d$}

The discussion of section \ref{flat} also extends to the study of 
spherically symmetric collapse in a space that is asymptotically 
flat $R^{d,1}$ for arbitrary odd $d$. In this section we will very briefly 
explain how this works, focussing on the limit $y=\frac{r_H}{\delta t}\gg 1$. 

To lowest order in the amplitude expansion we find 
\begin{equation}\label{pholo}
\phi_1(r, v)= \sum_{m}^{\frac{d-3}{2}}2^{\frac{d-3}{2} -m}
\frac{(-1)^m}{ m!} \frac{ \left( \frac{d-3}{2}+m \right)!}
{ \left( \frac{d-3}{2}-m \right)!}  \frac{\partial_v^{\frac{d-3}{2}-m} 
\psi(v)}{r^{\frac{d-1}{2}+m}}
\end{equation}
Here $\psi(v)$ is a function of time that we take, as usual, 
 to vanish outside $v \in (0,\delta t)$, and be of order 
$\epsilon_f (\delta t)^\frac{d-1}{2}$, where $\epsilon_f$ is a dimensionless
number such that $\epsilon_f \gg 1$. As in section \ref{flat} the parameter
that will justify the amplitude expansion will be $\frac{1}{\epsilon_f}$.  

\eqref{pholo} together with constraint equations 
immediately yields an expression for 
the functions $f_2$ and $g_2$. In particular, the leading large $r$ 
approximation to $g_2$ is given by 
\begin{equation}\label{gtgd} \begin{split}
g_2(r,v)&=-\frac{M(v)}{r^{d-2}}\\
M(v)&= -\frac{2^{(d-4)}}{d-1}\int_{-\infty}^vdt~\bigg[\left(\partial_t^{\frac{(d-3)}{2}}\psi(t)\right)\left(\partial_t^{\frac{(d+1)}{2}}\psi(t)\right) - \frac{d-3}{d-2}\left(\partial_t^{\frac{(d-1)}{2}}\psi(t)\right)^2\bigg]
\end{split}
\end{equation}

Note that $\phi_1 \ll 1 $ whenever $ r^\frac{d-1}{2} \ll 
(\delta t)^\frac{d-1}{2} \epsilon_f $ so we expect the amplitude 
expansion to reliably describe dynamics over this range of parameters. 
As in section \ref{flat}, however, $g_2$ cannot be ignored in comparison 
to $g_0=1$  throughout this parameter regime. As in section \ref{flat}, this 
implies that our spacetime is well approximated by a Vaidya type metric 
rather than empty flat space even at arbitrarily small $\frac{1}{\epsilon_f}$. 
The mass function of this Vaidya metric is given by $M(v)$ in 
\eqref{gtgd}. 

As in section \ref{flat} one may ignore this complication at early times 
$v \ll r_H$ over which the solution is well approximated by a naive 
perturbation expansion that uses empty flat space as its starting point. 
It is possible to demonstrate that this naive expansion has the following 
analytic structure in the variables $r$ and $v$
\begin{itemize} 
\item{1.} The functions $\Phi_{2n+1}$, $F_{2n}$ and 
$G_{2n}$ have the following analytic structure in the variable $r$
\begin{equation}\label{arfgend}\begin{split}
\Phi_{2n+1}(r, v)&=\sum_{m=0}^{\infty} \frac{\Phi_{2n+1}^m(v)}
{r^{(2n+1)\frac{d-1}{2}+m}}\\
F_{2n}(r,v)&=r \sum_{m=0}^{\infty} \frac{F_{2n}^m(v)}
{r^{n(d-1)+m}}\\
G_{2n}(r,v)&=- \delta_{n, 1} \frac{ M(v)}{r^{d-2}} + 
r \sum_{m=0}^\infty \frac{G_{2n}^m(v)}{r^{n(d-1)+m}}\\
\end{split}
\end{equation}
\item{2.} The functions $\Phi_{2n+1}^m(v)$, $F_{2n}^m(v)$ and 
$G_{2n}^m(v)$ 
are each functionals of $\psi(v)$ that scale like $\lambda^{m-(2n+1)
\frac{d-3}{2}}$ $\lambda^{m -n(d-3)}$ and $\lambda^{m -n(d-3)-1}$ 
under the the scaling $v \rightarrow \lambda v$. $M(v)$ scales like 
$\lambda^{2-d}$ under the same scaling. 
\item{3.} For $v>\delta t$ the 
$\Phi_{2n+1}^m(v)$ is polynomials in $v$ of degree $\leq n+m-1$; 
$F_{2n}^m(v)$ and $G_{2n}^m$ are polynomials in $v$ of degree 
$\leq n+m-3$ and $n+m-4$ respectively.    
\end{itemize}

It follows that, say, $\phi(r,v)$, is given by a double sum
$$\phi(r,v)=\sum_{n}\Phi_{2n+1}(r, v)=\sum_{n,m=0}^{\infty} 
\frac{\Phi_{2n+1}^m(v)}{r^{(2n+1)\frac{d-1}{2}+m}}.$$
Now sums over $m$ and $n$  are controlled by the effective 
expansion parameters $\sim \frac{v}{r}$ (for $m$) and 
$\frac{\psi^2 v  }{(\delta t)^{d-2} r^{d-1}} 
\sim \frac{v}{\delta t \epsilon_f^\frac{2}{d-2}} 
\sim \frac{v}{r_H}$ (for $n$; recall that in the neighborhood of the horizon 
$r_H^{d-2} \sim (\delta t)^{d-2} \epsilon_f^2$).

As in section \ref{flat}, it follows that the naive perturbation expansion 
breaks down for times $v \gg r_H$. However this expansion is valid everwhere
outside the event horizon at times of order $\delta t$, and so may be used 
to set the initial conditions for a resummed perturbation expansion that uses 
the Vaidya metric as its starting point. For $v>\delta t$ the mass function 
of the Vaidya metric reduces to a constant. At long times our solution 
is given by a small perturbation around a black hole of mass $M$. This 
perturbation is best analyzed in the coordinates $x=\frac{r}{M^\frac{1}{d-2}}$ 
and $y=\frac{v}{M^\frac{1}{d-2}}$. In these coordinates the leading 
order tail of $\phi$, at long times, is given by motion about a black hole 
of unit Schwarzschild radius perturbed by the $\phi$ field with initial 
condition 
$$\phi(x,0)=\frac{\phi_3^0(\delta t)}{M^\frac{3(d-1)}{2(d-2)} 
x^{\frac{3(d-1)}{2}}} \sim  \frac{1}{\epsilon_f^\frac{3}{d-2}}$$
The smallness of this perturbation justifies linearized treatment of 
the subsequent dynamics. 

\subsubsection{Even $d$}

We will not, in this paper, attempt an analysis of the spherically 
symmetric collapse to form a black hole asymptotically $R^{d,1}$ for 
even $d$. Here we simply note that the leading order large 
$\epsilon_f$ solution for $\phi_1(v)$ may formally be expressed as 
\begin{equation}\label{evd}
 \phi_1(r,v)= \int d\omega \left( q(\omega)  e^{i \omega (v-r)} 
\frac{H^{(1)}_{\frac{d-2}{2}}(r \omega)}{ r^{\frac{d-2}{2}}} \right)
\end{equation}
for any function $q(\omega)$ 
where $H_n(x)$ is the $n^{th}$ Hankel function of the first kind, i.e. 
$$H^{(1)}_n(x) \approx  \sqrt{\frac{2}{\pi x}} \left(
e^{i(x -\frac{\pi}{4} -\frac{n \pi}{2})}+{\cal O}(\frac{1}{x}) \right) $$
Using this expansion, it is easily verified that $\phi_1(r,v)$ reduces, 
at large $r$, to an incoming wave that takes the form 
$\frac{\psi(v)}{r^{\frac{d-1}{2}}}$. The evolution of this wave to small $r$ is 
implicitly given by \eqref{evd}. It should be possible to mimic the analysis 
of subsubsection \ref{ed} to explicitly express $\phi_1(r,v)$ as a spacetime 
dependent Kernel function convoluted against $\psi(v)$. In analogy with 
subsection \ref{ed} it should also be possible to expand $g_2(r,v)$
about small $r$. It is tempting to guess that such an analysis would 
reveal that the leading singularity in $g_2(r,v)$ scales like 
$\frac{1}{r^{d-2}}$, so that the metric is well approximated by a spacetime 
of the Vaidya form. We leave the verification of these guesses to future work.

\subsection{Spherically symmetric asymptotically $AdS$ collapse in 
arbitrary dimension}

It should be straightforward to generalize the analysis of section \ref{global}
to arbitrary odd $d$, and perhaps also to arbitrary even $d$. We do not 
explicitly carry out this generalization in this paper. However it is  
a simple matter to infer the various scales that will appear in this 
generalization using the intuition and results of subsections \ref{arbdim}
and \ref{arbflat}, and the fact that the results of global spherically 
symmetric $AdS$ collapse must reduce to Poincare patch collapse in one limit 
and flat space collapse in another. We have reported these scales in 
the introductionto section \ref{global}.

\bibliographystyle{JHEP}
\bibliography{cp}

\end{document}